\documentclass[12pt]{article}
\input epsfig.sty
\def\laq{\raise 0.4ex\hbox{$<$}\kern -0.8em\lower 0.62
ex\hbox{$\sim$}}
\def\gaq{\raise 0.4ex\hbox{$>$}\kern -0.7em\lower 0.62
ex\hbox{$\sim$}}

\begin{document}

\begin{titlepage}
\begin{flushright}
CERN-PH-TH/2007-110
\end{flushright}
\vspace*{1cm}

\begin{center}
{\LARGE {\bf Semi-analytical approach \\
\vskip0.5cm
to magnetized temperature autocorrelations}}
\vskip2.cm
\large{Massimo Giovannini \footnote{e-mail address: massimo.giovannini@cern.ch}}
\vskip 1.cm
{\it Centro ``Enrico Fermi",  Via Panisperna 89/A, 00184 Rome, Italy}
\vskip 0.5cm
{\it Department of Physics, Theory Division, CERN, 1211 Geneva 23, Switzerland}

\end{center}
\begin{abstract}
The cosmic microwave background (CMB) temperature autocorrelations,
induced by a magnetized adiabatic mode of curvature 
inhomogeneities, are computed with semi-analytical methods. 
As suggested by the latest CMB data, a nearly scale-invariant spectrum for the adiabatic mode 
is consistently assumed. 
In this situation, the effects of a fully inhomogeneous magnetic field are scrutinized and constrained
with particular attention to harmonics which are relevant for the region of Doppler oscillations.  
Depending on the parameters of the stochastic magnetic field a hump may replace the second 
peak of the angular power spectrum. Detectable effects on the Doppler region are then expected only 
if the magnetic power spectra have quasi-flat slopes and typical amplitude (smoothed over a comoving scale of Mpc size and redshifted to the epoch of gravitational collapse of the protogalaxy) 
exceeding 0.1 nG.  If the magnetic energy spectra are bluer (i.e. steeper in frequency) the allowed value of the smoothed amplitude becomes, comparatively, larger (in the range of 20 nG).
The implications of this investigation for the origin of large-scale magnetic fields in the Universe are discussed.
Connections with forthcoming experimental observations of CMB temperature fluctuations are also suggested and partially explored.
\end{abstract}

\end{titlepage}

\newpage
\renewcommand{\theequation}{1.\arabic{equation}}
\setcounter{equation}{0}
\section{Formulation of the problem}
\label{sec1}
Since the Cosmic Microwave Background (CMB) is extremely isotropic in nearly all angular 
scales, it is rather plausible to infer that the Universe was quite homogeneous 
(and isotropic) at the moment when the ionization fraction dropped 
significantly and the photon mean free path became, almost suddenly, comparable with the present 
Hubble radius.  

The inhomogeneities present for length-scales larger than the Hubble radius right before recombination are believed to be, ultimately, the seeds of structure formation and they can be studied by looking at the temperature 
autocorrelations which are customarily illustrated in terms of the angular power spectrum. 
The distinctive features of the angular power spectrum (like the Doppler peaks) can be phenomenologically reproduced by assuming the presence, before recombination, of a primordial 
adiabatic \footnote{The terminology adiabatic (and non-adiabatic) is used to classify the initial 
conditions of curvature perturbations in the pre-equality plasma. A solution is said to be 
adiabatic if the fluctuations of the specific entropy vanish at large length-scales. The opposite is true 
for a non-adiabatic solution. See section \ref{sec3} and discussions therein.}mode arising in a spatially flat Universe \cite{WMAP1,WMAP2,WMAP3,WMAP4,WMAP5}. 
Possible deviations from this working hypothesis  can also be 
bounded: they include, for instance, the plausible presence of non-adiabatic modes (see \cite{nonad1,nonad2,nonad3} and references 
therein), or even features in the power-spectrum that could be attributed either to the pre-inflationary stage of expansion or to 
the effective modification of the dispersion relations (see  \cite{features1,features2,features3,features4} and references therein).
For a pedagogical introduction to the physics of CMB anisotropies see, for instance, Ref. \cite{MAXG1}.
In short the purpose of the present paper is to show that CMB temperature autocorrelations may also 
be a source of valuable informations on large-scale magnetic fields whose possible presence 
prior to recombination sheds precious light on the origin of the largest magnetized 
structures we see today  in the sky such as galaxies, clusters of galaxies and  even some 
supercluster.

In fact, spiral galaxies and rich clusters possess a large-scale magnetic field that ranges 
 from $500$ nG  \cite{carilli,max1} (in the case of Abell clusters) to few $\mu$ G in the case 
of spiral galaxies \cite{beck}. Elliptical galaxies have 
also magnetic fields in the $\mu$G range but with correlation scales of the order of $10$--$100$ pc (i.e. 
much smaller than in the spirals where typical correlation lengths are of the order of $30$ kpc, as in the case 
of the Milky Way). The existence of large-scale magnetic fields in superclusters, still 
debatable because of ambiguities in the determination of the column density of electrons along 
the line of sight, would be rather intriguing. Recently plausible indications of the existence 
of magnetized structures in Hercules and Perseus-Pisces superclusters have been 
reported \cite{kronberg} (see also \cite{kronberg2}): 
the typical correlation scales of the fields whould be $0.5$ Mpc and the intensity 300 nG.

While there exist various ideas put forward throught the years, it is 
fair to say that the origin of these (pretty large) fields is still matter of debate \cite{max1,rees}. 
Even if they are,roughly, one millionth of a typical planetary magnetic field (such as the one of the earth) 
these fields are pretty large for a cosmological standard since their 
energy density is comparable both with energy density of the CMB photons 
(i.e. $T_{\mathrm{CMB}}^4$) and with the cosmic ray pressure. The 
very presence of large scale magnetic fields in diffuse astrophysical plasmas and with 
large  correlation scales (as large of, at least, $30$ kpc) 
seems to point towards a possible primordial origin \cite{max1}. 
At the same time, the efficiency of dynamo amplification can be questioned in different ways 
so that, at the onset of the gravitational collapse of the protogalaxy it seems rather 
plausible that only magnetic fields with intensities\footnote{In the present paper by $B_{\mathrm{L}}$ we denote 
the magnetic field smoothed over a typical comoving scale $\mathrm{L} = 2 \pi/k_{\mathrm{L}}$ 
with $k_{\mathrm{L}} = \mathrm{Mpc}^{-1}$. This choice is purely conventional and refers 
to the occurrence that the gravitational collapse of the protogalaxy occurs over a typical comoving scale 
of the Mpc. The usefulness of this convention will become clear later on.}$B_{\mathrm{L}} > 10^{-14}$ nG may be, eventually, amplified at an observable level \cite{subra1,kulsrud}. 

As emphasized many years ago by Harrison \cite{harrison1,harrison2,harrison3}, this situation 
is a bit reminiscent of what happened  
with the problem of justifying the presence of a flat spectrum of curvature perturbations that could eventually 
seed the structure formation paradigm. Today a possibility along this direction
 is provided by inflationary models in one of their various incarnations. 

It seems therefore appropriate, especially in view of forthcoming satellite missions (like PLANCK Explorer \cite{Planck}), to discuss the effects 
of large-scale magnetic fields on CMB physics. In fact, all along the next decade dramatic improvements in the quality and quantity of CMB data 
can be expected. On the radio-astronomical side, the 
next generation of radio-telescopes such as Square Kilometre Array (SKA) 
\cite{SKA} might be able to provide us with 
unprecedented accuracy in the full sky survey of Faraday Rotation measurements at frequencies 
that may be so large to be, roughly, comparable with \footnote{As the name of the instrument suggests 
the collecting area of SKA will be of $10^{6}\, \mathrm{m}^2$. The angular resolution of SKA is designed 
to be of $0.1$ arcsec at $1.4$ GHz. The frequency capability of the instrument will presumably 
be between $0.1$ and $25$ GHz. While the frequency range may be optimistic, it is 
certainly inspiring to think that $25$ GHz is not so far from the $30$ GHz of the low-frequency 
channel of the PLANCK Explorer \cite{Planck}. This occurrence might have a relevant experimental impact for the 
possible analysis of Faraday rotated CMB polarization, as recently emphasized \cite{mg2}.}
(even if always smaller than) the lower frequency channel of the PlANCK Explorer (i.e. 
about 30 GHz). The  question before us today is, therefore, the following: 
is CMB itself able to provide
compelling bounds on the strength of large-scale magnetic fields prior to hydrogen recombination?
In fact, all the arguments connecting the present strength of magnetic field to their primordial value (say before recombination)
suffer undeniable ambiguities. These ambiguities are  related to the evolution of the Universe 
through the dark ages (i.e. approximately, between photon decoupling and galaxy formation). 
So, even if it is very reasonable to presume  that during the stage of galaxy formation the magnetic flux and helicity are, according 
to Alfv\'en theorems, approximately conserved, the strengths of the fields prior to gravitational collapse is unknown and it is only predictable within a specific model for the origin of large-scale magnetic fields. 
In general terms, the magnetic fields produced in the early Universe may have different features.
They may be helical or not, they may have different spectral slopes and different intensities. 
There are, however, aspects that are common to diverse mechanisms like the stochastic nature 
of the produced field. Furthermore, since as we go back in time the conductivity increases 
with the temperature, it can be expected that the flux freezing and the helicity conservation
are better and better verified as the Universe heats up say from few eV to few MeV.

Along the past decade some studies addressed the 
analysis of vector and tensor modes induced by large-scale magnetic fields \cite{vt1,vt2,vt3,vt4}. There have 
been also investigations within a covariant approach to perturbation theory \cite{tsb1,tsb2}. Only recently the 
analysis of the scalar modes has been undertaken \cite{mg1,mg3,mg4,t1,tsb3}.
The set-up of the aforementioned analyses is provided by an effective one-fluid 
description of the plasma which is essentially the curved space analog of magnetohydrodynamics (MHD).
This approach is motivated since the typical length-scales of the problem are much larger 
of the Debye length. However, it should be borne in mind that the treatment of Faraday 
rotation is a typical two-fluid phenomenon. So if we would like to ask the question on how 
the polarization plane of the CMB is rotated by the presence of a uniform magnetic field
a two-fluid description would be mandatory (see section 2 and references therein).

In the framework described in the previous paragraph, it
 has been shown that the magnetic fields affect the scalar modes in a threefold way. In the first
place the magnetic energy density and pressure gravitate inducing a computable modification 
of the large-scale adiabatic solution. Moreover, the anisotropic stress and the divergence 
of the Lorentz force affect the evolution of the baryon-lepton fluid. Since, prior to decoupling, photons 
and baryons are tightly coupled the net effect will also be a modification of the temperature 
autocorrelations at angular scales smaller than the ones relevant for the ordinary SW contribution 
(i.e. $\ell > 30$). 

In the present paper, elaborating on the formalism developed in \cite{mg1,mg3,mg4},
a semi-analtytical approach for the calculation of the temperature autocorrelations 
is proposed. Such a framework allows the estimate 
of the angular power spectrum also for angular scales compatible with the first 
Doppler peak. A gravitating magnetic field will be included from the very beginning 
and its effects discussed both at large angular scales and small angular scales. 
The main theme of the present paper can then be phrased by saying that large-scale magnetic fields 
affect the geometry and the evolution of the (scalar) sources. We ought to compute how all these 
effects combine in the final power spectra of the temperature autocorrelations.
It should be remarked, incidentally, that the vector and the tensor modes 
are only partially coupled to the evolution of the various plasma quantities while the
treatment of the scalar modes necessarily requires a consistent inclusion of large-scale 
magnetic fields in the equations governing the evolution of the gravitational perturbations.

The plan of the present paper will therefore be the following. In section 2 the typical scales of the problem will be discussed. In section 3 the attention will be focused on the large-scale evolution of the curvature 
perturbations with particular attention to the magnetized contribution, i.e. the contribution associated with 
the gravitating magnetic fields. In section 4 the evolution at smaller angular scales 
will be investigated accounting, in an approximate manner, for the finite thickness effects of the last-scattering 
surface. In section 5 the estimates of the angular power spectra of the temperature autocorrelations will be presented. Section 6 contains the concluding remarks.
Some of the relevant theoretical tools  needed for the 
discussion of the problem have been collected in the appendix 
with the sole aim to make the overall presentation more self-contained. The material presented in the 
appendix collects the main equations whose solutions are reported and discussed in section 3 and 4.

\renewcommand{\theequation}{2.\arabic{equation}}
\setcounter{equation}{0}
\section{Typical scales of the problem}
\label{sec2}
The analysis starts by defining all the relevant physical scales of the problem. 
These scales stem directly from the evolution equations of the gravitational 
perturbations in the presence of a stochastic magnetic field. The interested reader may 
also consult appendix \ref{APPA} where some relevant technical aspects are briefly summarized.
\subsection{Equality and recombination}
According to the present understanding of the Doppler oscillations the
space-time geometry is well described by a conformally flat line element 
of Friedmann-Robertson-Walker (FRW) type
\begin{equation}
ds^2 = a^2(\tau) [d\tau^2 - d\vec{x}^2],
\label{LEL}
\end{equation}
where $\tau$ is the conformal time coordinate. In the present paper 
the general scheme will be to introduce the magnetic fields in the 
standard lore where the space-time geometry is spatially flat. This is the first 
important assumption which is supported by current experimental data 
including the joined analysis of, at least, three sets of data stemming, respectively
from large-scale structure, from Type Ia supernovae and from the three 
year WMAP data (eventually combined with other CMB experiments). 
For the interpretation of the data a specific model must also be adopted.    
The framework of the present analysis 
will be the one provided  by the $\Lambda$CDM model. This is 
probably the simplest 
case where the effects of magnetic fields can be included. Of course one may also
ask the same question within a different underlying model (such as the open CDM model
or the $\Lambda$CDM model with sizable contribution from the tensor modes and so on and so forth). 
While the calculational scheme will of course be a bit different, the main logic will remain the same. 
More details 
on the typical values of cosmological parameters inferred 
in the framework of the $\Lambda$CDM model can be found at the beginning of section \ref{sec5}.

In the geometry given by Eq. (\ref{LEL}) the scale factor for the radiation-matter transition can be smoothly parametrized as 
\begin{equation}
a(\tau) = a_{\mathrm{eq}} \biggl[ \biggl(\frac{\tau}{\tau_{1}}\biggr)^2 + 2 \biggl(\frac{\tau}{\tau_{1}}\biggr)\biggr], \qquad 
\tau_{1} = \frac{2}{H_{0}} \sqrt{\frac{a_{\mathrm{eq}}}{\Omega_{\mathrm{M}  0}}}  \simeq 288 \,\, \biggl(\frac{h_{0}^2 \Omega_{\mathrm{M}0}}{0.134}\biggr)^{-1}\, \mathrm{Mpc}.
\label{SCF}
\end{equation}
Concerning Eqs. (\ref{LEL}) and (\ref{SCF}) few comments 
are in order:
\begin{itemize}
\item{} the conformal time coordinate is rather useful 
for the treatment of the evolution of magnetized curvature perturbations
and is extensively employed in the appendix \ref{APPA};
\item{} $H_{0}$ is the present value of the Hubble constant and $\Omega_{\mathrm{M}0}$ is the 
present critical fraction in non-relativistic matter, i.e. $\Omega_{\mathrm{M}0} = \Omega_{\mathrm{b}0}+ \Omega_{\mathrm{c}0}$, given by the sum of the CDM component and of the baryonic component;
\item{} in the notation of Eq. (\ref{SCF}) the equality time (i.e. the time at which the 
radiation contrribution equals the contribution of dusty matter) is easily determined to be $\tau_{\mathrm{eq}} = (\sqrt{2} -1) \tau_{1}$, i.e. 
roughly, $\tau_{\mathrm{eq}} \simeq \tau_{1}/2$.
\end{itemize}
Equation (\ref{SCF}) is a solution of the Friedmann-Lema\^itre equations
whose specific form is
\begin{eqnarray}
&&{\mathcal H}^2 = \frac{8\pi G}{3} a^2 \rho_{\mathrm{t}},
\label{F1}\\
&&{\mathcal H}^2- {\mathcal H}' = 4\pi G a^2 (\rho_{\mathrm{t}} + p_{\mathrm{t}}),
\label{F2}\\
&& \rho_{\mathrm{t}}' + 3 {\mathcal H} (\rho_{\mathrm{t}} + p_{\mathrm{t}}) =0,
\label{F3}
\end{eqnarray}
where ${\mathcal H} =a' / a $ and the prime will denote, throughout the paper, a derivation with 
respect to $\tau$.  Equation (\ref{SCF}) is indeed solution of Eqs. (\ref{F1}), (\ref{F2}) and (\ref{F3}) when the total 
energy density  $\rho_{\mathrm{t}}$ is given by the sum of the matter density $\rho_{\mathrm{M}}$ and of 
the radiation density $\rho_{\mathrm{R}}$ (similarly $p_{\mathrm{t}} = p_{\mathrm{R}} + p_{\mathrm{M}}$). 

Often, for notational convenience, the rescaled time coordinate 
$x = \tau/\tau_{1}$ will be used. Within this $x$ parametrization the critical fractions of radiation and 
dusty matter become
\begin{equation}
\Omega_{\mathrm{R}}(x) = \frac{1}{\alpha(x) +1}= \frac{1}{(x+ 1)^2},\qquad 
\Omega_{\mathrm{M}}(x) = \frac{\alpha(x)}{\alpha(x) +1} = \frac{x^2 + 
2 x}{(x+ 1)^2}.
\label{OMEX}
\end{equation}
The redshift to equality is given, from Eq. (\ref{SCF}), by 
\begin{equation}
z_{\mathrm{eq}} + 1\simeq \frac{\rho_{\mathrm{M}0}}{\rho_{\mathrm{R}0}}=
\frac{h_{0}^2 \Omega_{\mathrm{M}0}}{h_{0}^2 \Omega_{\mathrm{R}0}}= 3228.91 \biggl(\frac{h_{0}^2 \Omega_{\mathrm{M}0}}{0.134}\biggr).
\label{SCF2}
\end{equation}
The redshift to recombination $z_{\mathrm{rec}}$ is, approximately, between $1050$ and $1150$. 
From this hierarchy of scales, i.e. $z_{\mathrm{dec}} > z_{\mathrm{rec}}$, it appears that 
recombination takes place when the Universe is already dominated by matter. Furthermore, a decrease 
in the fraction of dusty matter delays the onset of the matter dominated epoch.

If the recombination 
happens suddenly, the ionization fraction $x_{\mathrm{e}}$ 
drops abruptly from $1$ to $10^{-5}$. Prior to recombination 
the photons interact with protons and electrons via Thompson 
scattering so that the relevant mean free path is, approximately, 
\begin{equation}
\lambda_{\mathrm{T}}(z_{\mathrm{rec}}) \simeq \frac{1.8}{x_{\mathrm{e}}} \biggl(\frac{0.023}{h_{0}^2 \Omega_{\mathrm{b}0}}\biggr) \biggl(\frac{1100}{1 + z_{\mathrm{rec}}}\biggr)^2 \biggl(\frac{0.88}{1 - Y_{\mathrm{p}}/2}\biggr)\,\,\mathrm{Mpc},
\label{lambdaT}
\end{equation}
where $Y_{\mathrm{p}} \simeq 0.24$ is the abundance of $^{4} \mathrm{He}$. Since 
$m_{\mathrm{p}} =0.938$ GeV and $m_{\mathrm{e}} = 0.510$ MeV, 
 the mean free path of the photons will be essentially determined by the electrons because the Thompson 
 cross section is smaller for protons than for electrons. 
 Furthermore the protons and the electrons 
are even more tightly coupled, among them, by Coulomb scattering 
whose rate is larger than the Thompson rate of interaction. When the 
ionization fraction drops the photon mean free path gets as large as $10^{4}$
Mpc. For the purposes of this investigation it will be also important 
to take into account, at least approximately, the finite thickness 
of the last scattering surface. This can be done by approximating the 
visibility function with a Gaussian profile 
\cite{pavel1,pavel2,seljak,hu1,hu2} (see also \cite{wein,muk}) with 
finite width. We recall that the visibility function 
simply gives the probability that a photon was last scattered between 
$\tau$ and $\tau + d \tau$ (see section \ref{sec4}).
The scale factor (\ref{SCF}) can be used to express the ratios 
of two typical time-scales in terms of the ratio between the corresponding 
redshifts. So, for instance, 
\begin{equation}
x_{\mathrm{rec}}=\frac{\tau_{\mathrm{rec}}}{\tau_{1}}  = \sqrt{ \frac{z_{\mathrm{eq}} +1}{z_{\mathrm{rec}} + 1} + 1 } -1,
\label{RED1}
\end{equation}
which implies that, for $z_{\mathrm{rec}}$ and $h_{0}^2 \Omega_{\mathrm{M}0} = 0.134$, $\tau_{\mathrm{rec}}= 1.01 \tau_{1}$.

There is another typical scale that plays an important role  in the discussion 
of the Doppler oscillations. It is the baryon to photon ratio and it is 
defined as 
\begin{equation}
R_{\mathrm{b}}(z) = \frac{3}{4} \frac{\rho_{\mathrm{b}}}{\rho_{\gamma}}  = 0.664 \biggl(\frac{h_{0}^2 \Omega_{\mathrm{b}0}}{0.023}\biggr) \biggl(\frac{1051}{z + 1}\biggr).
\label{Rbdef}
\end{equation}
In the treatment of the angular power spectrum at intermediate 
angular scales $R_{\mathrm{b}}(z)$ appears ubiquitously either 
alone or in the expression of the sound speed of the photon-baryon 
system (see appendix \ref{APPA} for further details)
\begin{equation}
c_{\mathrm{sb}}(z)= \frac{1}{\sqrt{3 (R_{\mathrm{b}}(z) +1)}}.
\label{csbdef}
\end{equation}
In the absence of a magnetized 
contribution, $R_{\mathrm{b}}(z_{\mathrm{rec}})$ sets the height of the first Doppler 
peak as it can be easily argued by solving the evolution of the 
photon density contrast in the WKB approximation (see Eqs. (\ref{TC7}) and
(\ref{TC7a})).

\subsection{Plasma scales}
The Debye scale and the plasma frequency of the electrons 
can be easily computed in terms of the cosmological parameters 
introduced so far. The results are, respectively:
\begin{eqnarray}
&& \lambda_{\mathrm{D}}(z) = \sqrt{\frac{T_{\mathrm{e}}}{8\pi e^2 n_{\mathrm{e}} } }\simeq \frac{4.26}{\sqrt{x_{\mathrm{e}}}} \biggl(\frac{1050}{z + 1}\biggr) 
\biggl(\frac{h_{0}^2 \Omega_{\mathrm{b}0}}{0.023}\biggr)^{-1/2} \, \mathrm{cm}, 
\label{DEB}\\
&& \omega_{\mathrm{p\,e}}(z) = 3.45 \,\biggl(\frac{h_{0}^2 \Omega_{\mathrm{b}0}}{0.023}\biggr)^{1/2} \, \biggl(\frac{1 + z}{1050} \biggr)^{3/2} \,\, \mathrm{MHz}.
\label{PLE}
\end{eqnarray}
By comparing Eqs. (\ref{lambdaT}) and (\ref{DEB}), 
$\lambda_{\mathrm{T}} \gg \lambda_{\mathrm{D}}$ both around 
equality and recombination.  For typical scales comparable with 
the Hubble radius at recombination, therefore, the plasma will be, to an
excellent apprfoximation, globally neutral, i.e. 
\begin{equation}
\vec{\nabla}\cdot \vec{E} = 4 \pi e (n_{\mathrm{p}} - n_{\mathrm{e}}) =0
\label{DIV1}
\end{equation}
where $ \vec{E}(\tau,\vec{x}) = a^2(\tau) \vec{{\mathcal E}}(\tau,\vec{x})$ denote the 
rescaled electric fields and where, by charge neutrality, the electron density equals the proton 
density, i.e.
\begin{equation}
n_{\mathrm{e}}(z) = n_{\mathrm{p}}(z) = x_{\mathrm{e}} \eta_{\mathrm{b}} n_{\gamma}(z),\qquad \eta_{\mathrm{b}} = 6.27 \times 10^{-10} \biggl(\frac{h_{0}^2 \Omega_{\mathrm{b}0}}{0.023}\biggr);
\end{equation}
$\eta_{\mathrm{b}}$ is the ratio between the baryonic charge density and the photon density.
When the ionization fraction drops, the Debye scale 
is still the smallest length of the problem.
From Eq. (\ref{PLE}) the plasma 
frequency for the electrons is, around recombination, in the 
MHz range. The plasma frequency for the ions (essentially protons)
will then be smaller (in the kHz range). Both these frequencies 
are smaller than the maximum of the CMB emission (which is, today,
around $300$ GHz and around $300$ THz around recombination).
Since the main focus of the present investigation will be on frequencies 
$\omega \ll \omega_{\mathrm{p\,e}}$,  
the electromagnetic propagation of disturbances can be safely neglected and this 
implies, in terms of the rescaled electric and magnetic fields, that 
\begin{equation}
\vec{\nabla} \times \vec{B} = 4\pi \vec{J},\qquad \vec{\nabla} \cdot \vec{B}=0\label{DIV2}
\end{equation}
where $\vec{B}(\tau,\vec{x}) = a^2 \vec{{\mathcal B}}(\tau,\vec{x})$ and where 
\begin{equation}
\vec{J} = \sigma_{\mathrm{c}} (\vec{E} + \vec{v} \times \vec{B}),
\label{Ohm}
\end{equation}
is the Ohmic current and $\sigma_{\mathrm{c}} = a(\tau) \overline{\sigma}_{\mathrm{c}}$ defined in terms of the the rescaled conductivity. Since we are in the situation 
where $ T\ll m_{\mathrm{e}}$, $\overline{\sigma}_{\mathrm{c}}= \alpha_{\mathrm{em}}^{-1} T \sqrt{T/m_{\mathrm{e}}}$. By now using the Ohmic 
electric field inside the remaining Maxwell equation, i.e.
\begin{equation}
\vec{\nabla} \times\vec{E} = - \frac{\partial \vec{B}}{\partial \tau}, 
\label{MX1}
\end{equation}
 the magnetic diffusivity equation can be obtained
\begin{equation}
\frac{\partial \vec{B}}{\partial\tau} = \vec{\nabla}\times(\vec{v} \times \vec{B}) + 
\frac{1}{4\pi \sigma_{\mathrm{c}}} \nabla^2 \vec{B}.
\label{magndiff}
\end{equation}
Equation (\ref{magndiff}) together with the previous 
equations introduced in the present subsection are the starting point of the 
magnetohydrodynamical (MHD) description adopted in the present paper. 
They hold for typical frequencies $\omega \ll \omega_{\mathrm{p\,e}}$ 
and for typical length scales much larger than the Debye scale.
In this approximation (see Eq. (\ref{DIV2})) the Ohmic current is solenoidal, i.e. 
$\vec{\nabla} \cdot \vec{J} =0$.

As in the flat-space case, the MHD equations can be obtained 
from a two-fluid description by combining the relevant equations and by 
using global variables. As a consequence of this derivation 
$\vec{J}$ will be the total current and $\vec{v}$ will be the bulk velocity 
of the plasma, i.e. the centre-of-mass 
velocity of the electron-proton system \cite{krall,biskamp}. It should be remembered 
that various phenomena involving the possible existence 
of a primordial magnetic field at recombination should not be treated 
within a single fluid approximation (as it will be done here) but rather 
within a two-fluid (or even kinetic) description. An example 
along this direction is Faraday rotation of the CMB polarization 
\cite{biref} or any other phenomenon where the electromagnetic 
branch of the plasma spectrum is relevant, i.e.
 $\omega > \omega_{\mathrm{p\,e}}$.  In fact, the CMB is linearly 
 polarized. So if a uniform magnetic field is present at recombination 
the polarization plane of the CMB can be rotated. From the 
appropriate dispersion relations (obtainable in the usual two-fluid 
description) the Faraday rotation rate can be computed
bearing in mind that the Larmor frequency of electrons and ions at
recombination, i.e. 
\begin{equation}
 \omega_{\rm Be} =\frac{e B_{\mathrm{L}}(\tau_{\mathrm{rec}})}{m_{\rm e} c} \simeq 
18.08 \biggl(\frac{B_{\mathrm{L}}(\tau_{\mathrm{rec}})}{ 10^{-3}\,\,\, {\rm G}}\biggr)\,\,\,{\rm kHz},
\,\,\,\,\,\,\,\,\,\,\,  \omega_{\rm Bi}=\frac{e B_{\mathrm{L}}(\tau_{\mathrm{rec}})}{m_{\rm i} c}  
\simeq 9.66  
\biggl(\frac{B_{\mathrm{L}}(\tau_{\mathrm{rec}})}{10^{-3}\,\,\, {\rm G}}\biggr)\,\,\,{\rm Hz},
\label{Bei}
\end{equation}
are both smaller than $\omega_{\mathrm{p\,e}}$. In Eq. (\ref{Bei})
 $B_{\mathrm{L}}(\tau_{\mathrm{rec}})$ is the smoothed magnetic field strength at recombination.

It is the moment to spell out clearly two concepts that are central to the 
discussion of the evolution of large-scale magnetic fields in a FRW Universe 
with line element (\ref{LEL}):
\begin{itemize}
\item{} the concept of comoving and physical magnetic fields;
\item{} the concept of stochastic magnetic field.
\end{itemize}
The comoving magnetic field $\vec{B}(\tau,\vec{x})$ is related to the 
physical magnetic field $\vec{{\mathcal B}}(\tau,\vec{x})$ as 
$\vec{B}(\tau,\vec{x}) = a^2(\tau) \vec{{\mathcal B}}(\tau,\vec{x})$. 
We will choose as the reference time the epoch of gravitational 
collapse of the protogalaxy. At this time the comoving and physical 
field coincide. So, for instance, a (physical) magnetic field of nG strength at the onset of gravitational collapse 
will be roughly of the order of the mG (i.e. $10^{-3}$ G) at the epoch of recombination. This 
conclusion stems directly from the fact that the physical magnetic field scales with 
$a^{-2}(\tau)$, i.e. with $z^{2}$ where $z$, as usual is the redshift. This implies, in turn, that 
$\vec{B}$ (i.e. the comoving field) is roughly constant (in time) if the plasma does not have sizable kinetic 
helicity\footnote{The breaking of parity (often related to the turbulent 
nature of the bulk velocity field) is one of the necessary 
conditions for the persistence of the dynamo term in the magnetic 
diffusivity equation. For some classic introductions to dynamo 
theory see \cite{parker,zeldovich}. In this paper it will be assumed 
that the pre-recombination plasma is not turbulent since the values 
of the kinetic and magnetic Reynolds numbers are both small \cite{max1}.
Possible turbulent effects have been contemplated by the literature but 
for much higher temperatures in the life of the Universe but anyway 
always above the threshold of electron-positron annihilation (i.e. $T> \mathrm{MeV}$).}
 (i.e. $\langle \vec{v}\cdot \vec{\nabla}\times \vec{v} \rangle =0$) (see, for instance, \cite{max1,subra1,kulsrud}). In this situation 
Eq. (\ref{magndiff}) dictates that $\vec{B}$ is constant for typical wave-numbers 
$k < k_{\sigma}$ (i.e. for sufficiently large comoving 
length-scales) 
 where $k_{\sigma}$ sets the  magnetic diffusivity scale
whose value, at recombination, is 
\begin{equation}
\frac{1}{k_{\sigma}} \simeq 1.24 \times 10^{-14} 
\biggl(\frac{h_{0}^2 \Omega_{\mathrm{M}0}}{0.134}\biggr)^{-3/4} \,\, \mathrm{Mpc}.
\label{ksigma}
\end{equation}
Equation (\ref{ksigma}) can be compared with the estimate of the diffusive scale associated with Silk damping:
\begin{equation}
\frac{1}{ k_{\mathrm{D}} \tau_{\mathrm{rec}}} = 9.63\times 10^{-3} \biggl(\frac{h_{0}^2\Omega_{\mathrm{b}}}{0.023}\biggr)^{-1/2}
\biggl(\frac{h_{0}^2\Omega_{\mathrm{M}}}{0.134}\biggr)^{1/4} \biggl(\frac{1050}{z_{\mathrm{rec}}}\biggr)^{3/4}.
\label{simplistic}
\end{equation}
Hence, for the typical value of the matter fraction appearing in Eq. (\ref{ksigma}),
$\tau_{\mathrm{rec}} \simeq \tau_{1}$ and, consequently $k_{\sigma} \gg 
k_{\mathrm{D}}$. While finite conductivity effects are rather efficient in washing out the magnetic fields for large wave-numbers, the 
thermal diffusivity effects (related to shear viscosity and, ultimately, to Silk 
damping) affect typical wave-numbers that are much smaller than the ones 
affected  by conductivity. 

Under the conditions of MHD, two (approximate) conservations laws 
may be derived, namely the magnetic flux conservation 
\begin{equation}
\frac{d}{d \tau} \int_{\Sigma} \vec{B} \cdot d\vec{\Sigma}=-
\frac{1}{4\pi \sigma_{\mathrm{c}}} \int_{\Sigma} \vec{\nabla} \times\vec{\nabla}
\times\vec{B}\cdot d\vec{\Sigma},
\label{flux}
\end{equation}
and the magnetic helicity conservation
\begin{equation}
\frac{d}{d\tau} \biggl(\int_{V} d^3 x \vec{A}~\cdot \vec{B}\biggr) = - \frac{1}{4\pi \sigma_{\mathrm{c}}} \int_{V} d^3 x
{}~\vec{B}\cdot\vec{\nabla} \times\vec{B}.
\label{hel}
\end{equation}
In Eq. (\ref{flux})  $\Sigma$ is an arbitrary closed surface 
that moves with the plasma. In Eq. (\ref{hel}) $\vec{A}$ is the vector potential. According to Eq. (\ref{flux}),  in MHD the magnetic field 
has to be always solenoidal (i.e. $\vec{\nabla} \cdot \vec{B} =0$).
Thus, the magnetic flux conservation implies that, in the 
ideal MHD  limit (i.e. $\sigma_{\mathrm{c}} \to \infty$) the 
magnetic flux lines, closed because of the transverse nature of the field, evolve always 
glued together with the plasma element. In this 
approximation, as far as the magnetic field 
evolution is concerned, the plasma is a collection 
of (closed) flux tubes. The theorem of flux conservation 
states then  that the energetical properties of large-scale
magnetic fields are conserved throughout the plasma evolution. 

While 
the flux conservation concerns the 
energetic properties of the magnetic flux lines, the 
magnetic helicity, i.e. Eq. (\ref{hel}), concerns chiefly the 
 topological  properties of the 
magnetic flux lines. In the simplest situation,
 the magnetic flux lines will be closed loops 
evolving independently in the plasma and the helicity 
will vanish. There could be, however, 
more complicated topological situations \cite{mgknot}
where a single magnetic loop is twisted (like some 
kind of M\"obius stripe) or the case where 
the magnetic loops are connected like the rings of a chain:
now the non-vanishing magnetic helicity 
 measures, essentially, the number of links and twists 
in the magnetic flux lines \cite{biskamp}. Furthermore, in the 
superconducting limit, the helicity will not change 
throughout the time evolution. 
The conservation of the magnetic flux and of the magnetic 
helicity is a consequence of the fact that, in ideal 
MHD, the Ohmic electric field is always orthogonal 
both to the bulk velocity field and to the magnetic 
field. In the resistive MHD approximation this conclusion may  not apply.
The quantity at the right-hand-side of Eq. (\ref{hel}), i.e. $\vec{B}\cdot\vec{\nabla}\times \vec{B}$ is called magnetic gyrotropy and it is a gauge-invariant measure 
of the number of contact points in the magnetic flux lines.  As we shall see 
in a moment, the only stochastic fields contributing to the scalar fluctuations 
of the goemetry are the ones for which the magnetic gyrotropy vanishes.

Nearly all 
mechanisms able to generate large scale magnetic fields imply the 
existence of a stochastic background of magnetic disturbances 
\cite{max1} that could be written, 
in Fourier space, as \footnote{For the Fourier transforms we use the following conventions: $B_{i} (\vec{x}) = 
(2\pi)^{-3/2} \,\,\int d^{3}k  e^{- i \vec{k}\cdot\vec{x}} B_{i}(\vec{k})$
and, conversely, $B_{i} (\vec{k}) = (2\pi)^{-3/2}
\int d^{3}x  e^{ i \vec{k}\cdot\vec{x}} B_{i}(\vec{x})$.}
\begin{equation}
\langle B_{i}(\vec{k},\tau) B_{j}(\vec{p},\tau) \rangle =  P_{ij}(k)  \delta^{(3)}(\vec{k} + \vec{p}),
\label{stoch1}
\end{equation}
where 
\begin{equation}
P_{ij}(k) = {\cal Q}(k) \biggl(\delta_{ij} - \frac{k_{i}\, k_{j}}{k^2}\biggr),\qquad 
{\cal Q}(k) = {\cal Q}_{0} k^{m}.
\label{definitions1}
\end{equation}
From Eq. (\ref{definitions1})  the magnetic field configuration 
of Eq. (\ref{stoch1}) depends on the amplitude of the field ${\cal Q}_{0}$ and on the spectral index $m$. 

It is often  useful, in practical estimates, to regularize the two-point function 
by using an appropriate ``windowing". Two popular windows  
are, respectively, the Gaussian and the top-hat functions , i.e.
\begin{equation}
{\cal W}_{\rm g}(k,L) = e^{- \frac{k^2 L^2}{2}},\qquad {\cal W}_{\rm th}(k,L) =
\frac{3}{k L}\,j_{1}( k L).
\end{equation}
For instance, the regularized magnetic energy density with Gaussian 
filter can be obtained from the previous expressions by shifting 
${\cal Q}(k) \to {\cal Q}(k) W_{\rm g}^2 (k, L)$. The result is 
\begin{equation}
\langle B_{i}(\tau, \vec{x})B^{i}(\tau, \vec{x} + \vec{r}) \rangle= \frac{ 2 {\cal Q}_{0}}{( 2\pi)^2} \frac{\Gamma\biggl(\frac{m+3}{2}\biggr)}{L^{3 + m}} F\biggl( \frac{m + 3}{2}, \frac{3}{2}, - \frac{r^2}{4 L^2} \biggr)  ,
\label{trace}
\end{equation}
where $F(a, b, x) \equiv_{1} F_{1}(a,b, x) $ is the confluent hypergeometric function \cite{abr,grad}. Notice that the integral appearing 
in the trace converges for $m > -3$.  The amplitude of the magnetic power 
spectrum ${\cal Q}_{0}$ can be traded for $ B^2_{\mathrm{L}}$ where  
 $ B^2_{\mathrm{L}}$  is by definition the regularized two-point function 
 evaluated at coincident spatial points, i.e. 
 \begin{equation}
B_{\mathrm{L}}^2 =\lim_{r\to 0} \langle B_{i}(\tau, \vec{x}) B^{i}(\tau, \vec{x} +\vec{r})\rangle 
\label{DEFBL}
\end{equation}
Combining  Eq. (\ref{trace})  with Eq. (\ref{DEFBL}) we have that ${\cal Q}_{0}$ becomes 
\begin{equation}
{\cal Q}_{0} = \frac{(2 \pi)^{ m + 5}}{2} \frac{k_{\mathrm{L}}^{-( 3 + m)}}{\Gamma\biggl(
\frac{m + 3}{2}\biggr)} B_{\mathrm{L}}^2,
\label{Q0}
\end{equation}
 where $k_{\mathrm{L}} = 2\pi/L$. The two main parameters that 
 will therefore characterize the magnetic background will be 
 the smoothed amplitude $B_{\mathrm{L}}$ and the spectral slope. For reasons related to the way power spectra are assigned for curvature 
 perturbations, it will be practical to define the magnetic spectral index as 
 $\epsilon = m+ 3$ (see Eqs. (\ref{SW3})--(\ref{SW4}) and comments 
 therein).

In the case of the configuration (\ref{stoch1}) 
the magnetic gyrotropy is vanishing, i.e. 
$\langle \vec{B} \cdot \vec{\nabla} \times \vec{B} \rangle =0$.
There are situations where magnetic fields are produced in a 
state with non-vanishing gyrotropy (or helicity) (see for instance \cite{tk2} and references 
therein). In the latter case, the two point function can be written in the same 
form given in Eq. (\ref{stoch1}) 
\begin{equation}
\langle B_{i}(\vec{k},\tau) B_{j}(\vec{p},\tau) \rangle =  \tilde{P}_{ij}(k)  \delta^{(3)}(\vec{k} + \vec{p}),
\label{stoch2}
\end{equation}
but where now 
\begin{equation}
\tilde{P}_{ij}(k) = {\cal Q}(k) \biggl(\delta_{ij} - \frac{k_{i}\, k_{j}}{k^2}\biggr) + i 
\tilde{{\cal Q}}(k) \epsilon_{i j \ell} \,\frac{k^{\ell}}{k},\qquad 
\tilde{{\cal Q}}(k) = \tilde{{\cal Q}}_{0} k^{\tilde{m}}.
\label{definitions2}
\end{equation}
From Eq. (\ref{definitions2}) we can appreciate that, on top of the parity-invariant 
contribution (already defined in Eqs. (\ref{stoch1}) and (\ref{definitions1})), there 
is a second term proportional to the Levi-Civita $\epsilon_{ij\ell}$.
In Fourier space, the introduction of gyrotropic configurations 
implies also the presence of a second function of the momentum 
$\tilde{{\cal Q}}(k)$. In the case of scalar fluctuations of the geometry this second power 
spectrum will not give any contribution (but it does contribute to the vector modes 
of the geometry as well as in the case of the tensor modes).

The correlators that contribute to the evolution of the scalar 
fluctuations of the geometry will be essentially the ones of magnetic energy density and pressure 
(i.e. $\vec{B}^2/(8\pi)$ and $\vec{B}^2/(24\pi)$) and the one related to the  divergence 
of the MHD Lorentz force (i.e. $\vec{\nabla}\cdot[ \vec{J}\times \vec{B}]$) which appears 
as source term in the evolution equation of the divergence 
of the peculiar velocity of the baryons (see Eqs. (\ref{B2}) and (\ref{TC2}) of the appendix \ref{APPA}).
Since in MHD $4\pi \vec{J} = \vec{\nabla}\times \vec{B}$ the divergence of the 
Lorentz force will be proportional to 
$\vec{\nabla}\cdot [( \vec{\nabla}\times \vec{B}) \times \vec{B}]$.
The magnetic anisotropic stress $\tilde{\Pi}_{i}^{j}$ does also contribute to the scalar problem 
but it can be related, through simple vector identities, to the magnetic energy density and to the 
divergence of the Lorentz force (see Eqs. (\ref{identity}) and (\ref{identity2})).
To specify the effect of the stochastic  background of magnetic 
fields on the scalar  modes of the geometry we shall therefore need 
the correlation functions of two dimensionless quantities denoted, in what 
follows, by $\Omega_{\mathrm{B}}$ and $\sigma_{\mathrm{B}}$, i.e. 
\begin{equation}
\Omega_{\mathrm{B}}(\tau, \vec{x}) = \frac{1}{a^4(\tau)~\rho_{\gamma}(\tau)}\frac{\vec{B}^2(\tau,\vec{x})}{8\pi},\qquad 
\partial_{j}\partial^{i} \tilde{\Pi}_{j}^{i} = (p_{\gamma}+\rho_{\gamma}) \nabla^2 \sigma_{\mathrm{B}}
\end{equation}
where $\rho_{\gamma}$ is the energy density of the photons. Since $\Omega_{\mathrm{B}}$ and $\sigma_{\mathrm{B}}$ are both quadratic in the magnetic field intensity, their corresponding two-point 
functions will be quartic in the magnetic field intensities. Consequently $\Omega_{\mathrm{B}}$ and $\sigma_{\mathrm{B}}$ will have Fourier transforms that are defined as convolutions of the original magnetic fields 
and, more precisely:
\begin{eqnarray}
&&\Omega_{\rm B}(\vec{x}, \tau) = \frac{\delta \rho_{\rm B}( \vec{x},\tau)}{\rho_{\gamma}(\tau)} = \frac{1}{(2\pi)^{3/2}}
\int d^{3} q \,\,\Omega_{\rm B} (\vec{q},\tau) \,e^{- i\vec{q}\cdot \vec{x}}, \qquad \delta p_{\rm B} = \frac{\delta \rho_{\rm B}}{3},
\label{OMFT}\\
&& 
\sigma_{\rm B}(\vec{x},\tau) = \frac{1}{(2\pi)^{3/2}} \int d^{3} q  e^{ - i \vec{q} \cdot\vec{x}} \sigma_{\rm B}(\vec{q},\tau),
\end{eqnarray}
where 
\begin{eqnarray}
&& \Omega_{\rm B}(\vec{q},\tau) = \frac{1}{(2\pi)^{3/2}}  \frac{1}{8\pi 
\overline{\rho}_{\gamma}}  \int d^{3} p B_{i}(\vec{p}) 
B^{i} (\vec{q} - \vec{p}),
\label{OMEGA}\\
&& \sigma_{\rm B}(\vec{q},\tau) = \frac{1}{(2\pi)^{3/2}} 
\frac{1}{16\pi \overline{\rho}_{\gamma} } \frac{1}{q^2} \int d^{3} p \biggl[  3 ( q^{j} - p^{j}) p^{i} B_{j}(\vec{p}) B_{i}(\vec{q} -\vec{p}) - q^2 B_{i}(\vec{q} - \vec{p}) B^{i}(\vec{p})\biggr].
\label{SIGMA}
\end{eqnarray}
having defined, for notational convenience, $\overline{\rho}_{\gamma}=\rho_{\gamma}(\tau) a^4(\tau)$.

\renewcommand{\theequation}{3.\arabic{equation}}
\setcounter{equation}{0}
\section{Large-scale solutions}
\label{sec3}

After equality but before recombination the fluctuations of the geometry 
evolve coupled with the fluctuations of the plasma. The plasma contains
four species: photons, neutrinos (that will be taken to be effectively 
massless at recombination), baryons and cold dark matter (CDM) 
particles. The evolution equations go under the name of Einstein-Boltzmann 
system since they are formed by the perturbed Einstein equations and by the evolution equations of the brightness perturbations. In the case 
of temperature autocorrelations, the relevant Boltzmann hierarchy 
will be the one associate with the $I$ stokes parameter giving the 
intensity of the Thompson scattered radiation field. Furthermore, since neutrinos are collisionless after $1$ MeV, the Boltzmann hierarchy for neutrinos has also to be consistently included. In practice, however, the 
lowest multipoles (i.e. the density contrast, the velocity and the anisotropic stress) will be the most important ones for the problem of setting the 
pre-recombination initial conditions.

Since stochastic magnetic fields are present prior to recombination, the Einstein-Boltzmann system has to be 
appropriately modified. This system has been already derived
in the literature (see Ref. \cite{mg1,mg3}) but since 
it will be heavily used in the present and in the following sections
the main equations have been collected and discussed in
appendix \ref{APPA}.  It is also appropriate to remark, on a more technical 
ground, that the treatment of the curvature perturbations 
demands the analysis of quantities that are 
invariant under infinitesimal coordinate transformations (or, for short, gauge 
invariant). The strategy adopted in the appendix has been to pick up a
specific gauge (i.e. the conformally Newtonian gauge) and to 
derive, in this gauge, the relevant evolution equations for the 
appropriate gauge-invariant quantities such as the density contrast
on uniform density hypersurfaces (denoted, in what follows, by $\zeta$) and 
the curvature perturbations on comiving orthogonal hypersurfaces
(denoted, in what follows, by ${\mathcal R}$).
Defining as $k$ the comoving wave-number of the fluctuations, 
the magnetized Einstein-Boltzmann system can be discussed 
in three complementary regimes:
\begin{itemize}
\item{} the wavelengths that are larger than the Hubble radius 
at recombination, i.e. $k \tau_{\mathrm{rec}} < 1$;
\item{} the wavelengths that crossed the Hubble radius before 
recombination but that were still larger than the Hubble radius at 
equality, i.e. $k\tau_{\mathrm{eq}} <1$;
\item{} the wavelengths that crossed the Hubble radius prior to equality 
and that are, consequently, inside the Hubble radius already at equality (i.e. 
$k\tau_{\mathrm{eq}} >1$).
\end{itemize}
The wavelengths that are larger than the Hubble radius at recombination 
determine the large-scale features of temperature autocorrelations 
and, in particular, the so-called Sachs-Wolfe plateau. The wavelengths 
that crossed the Hubble radius around $\tau_{\mathrm{rec}}$ determine 
the features of the temperature autocorrelations in the region of the 
Doppler oscillations.

The initial conditions of the Einstein-Boltzmann system are 
set in the regime when the relevant wavelengths are
larger than the Hubble radius before equality (i.e. deep in the radiation 
epoch).  The standard unknown is represented, in this context, by the primordial spectrum 
of the metric fluctuations whose amplitude and slope are two essential parameters of the 
$\Lambda$CDM model. To this unknown we shall also add the possible presence 
of a stochastically distributed magnetized background.
In the conventional case, where magnetic fields are not contemplated, 
the system of metric fluctuations admits 
various (physically different) solutions that are 
customarily classified in adiabatic and non-adiabatic modes (see, for 
instance, \cite{nonad1,nonad2} and also \cite{MAXG1}).
For the adiabatic modes the fluctuations of the specific entropy 
vanish at large scales. Conversely, for non-adiabatic (also 
sometimes named isocurvature)  solutions the fluctuations of the specific 
entropy do not vansih. The WMAP 3-year data \cite{WMAP1,WMAP2,WMAP3} suggest that the temperature autocorrelations 
are well fitted by assuming a primordial adiabatic mode of curvature 
perturbations with nearly scale-invariant power spectrum. 
Therefore, the idea will be now to assume 
the presence of an adiabatic mode of curvature perturbations 
and to scrutinize the effects of fully inhomogeneous 
magnetic fields.  It should be again stressed that this is the minimal 
assumption compatible with the standard $\Lambda$CDM paradigm.
As it will be briefly discussed later on, all the non-adiabatic solutions 
in the pre-equality regime can be generalized to include a magnetized background \cite{mg3}.
However, for making the discussion both more cogent and simpler, the attention will be focussed 
on the physical system with the fewer number of extra-parameters, i.e. the case of a magnetized 
adiabatic mode.
\subsection{Curvature perturbations}

Consider the large angular scales that 
were outside the  horizon at recombination.  While 
smaller angular scales (compatible with the first Doppler peak) 
necessarily demand the inclusion of finite thickness effects 
of the last scattering surface, the largest angular scales (corresponding 
to harmonics $\ell \leq 25$) can be safely treated in the approximation 
that the visibility function is a a Dirac delta function centered around 
$\tau_{\mathrm{rec}}$. Moreover, for the modes satisfying the condition 
$k \tau_{\mathrm{rec}} < 1$ the radiation-matter transition 
takes place when the relevant modes have wavelengths still larger 
than the Hubble radius. 

It is practical, for the present purposes, 
 to think the matter-radiation fluid as a unique 
physical entity with time-dependent barotropic index and time-dependent sound speed:
\begin{equation}
w_{\mathrm{t}}(\alpha) = \frac{p_{\mathrm{t}}}{\rho_{\mathrm{t}}}=\frac{1}{3 ( \alpha + 1)},\qquad c_{\mathrm{st}}^2 = 
\frac{p'_{\mathrm{t}}}{\rho'_{\mathrm{t}}} = \frac{4}{3 ( 4 + 3 \alpha)},
\label{wcs}
\end{equation}
where $\alpha = a / a_{\mathrm{eq}}$. According to
 Eq. (\ref{wcs}),  when $a\gg a_{\mathrm{eq}}$ both 
$c_{\mathrm{st}}^2$ and $w_{\mathrm{t}}$ go to zero (as appropriate when matter dominates) while in the opposite 
limit (i.e. $\alpha \ll 1$) $ c_{\mathrm{st}}^2 \simeq w_{\mathrm{t}} \to 1/3$ which is the usual result of the radiation epoch.
Since recombination takes place after equality
 it will be crucial, for the present purposes, to determine the perturbations of the spatial curvature at this moment.
The presence of fully inhomogeneous magnetic fields affects the evolution of the curvature perturbations 
across the radiation-matter transition.  
This issue has been addressed in \cite{mg1} by following, outside the Hubble
radius, the evolution of the gauge-invariant density contrast on uniform density hypersurfaces 
(customarily denoted by $\zeta$):
\begin{equation}
\zeta= - \psi + \frac{{\mathcal H}(\delta\rho_{\mathrm{t}} + \delta \rho_{\mathrm{B}})}{\rho_{\mathrm{t}}'}.
\label{ZETA}
\end{equation}
where $\psi$ is related to the fluctuation of the spatial component of the metric (i.e. $\delta_{\mathrm{s}} g_{ij} = 2 a^2 \psi 
\delta_{ij}$ in the conformally Newtonian gauge) and 
\begin{equation}
\delta \rho_{\mathrm{t}} = \delta \rho_{\gamma} + \delta\rho_{\nu} + \delta \rho_{\mathrm{c}} + \delta\rho_{\mathrm{b}},\qquad 
\delta\rho_{\mathrm{B}} = \frac{B^2(\vec{x})}{8\pi a^4},
\label{DC}
\end{equation}
are, respectively, the total density fluctuation of the fluid sources (i.e. photons, neutrinos, CDM and baryons) 
and the density fluctuations induced by a fully inhomogeneous magnetic field. The gauge-invariant density contrast 
on uniform curvature hypersurfaces is related, via the Hamiltonian constraint (see Eq. (\ref{HAM})), to the curvature perturbations 
on comoving orthogonal hypersufaces customarily denoted by ${\mathcal R}$. Since both ${\mathcal R}$ and $\zeta$
are gauge-invariant, their mutual relation can be worked out in any gauge and, in particular, in the conformally Newtonian 
gauge where ${\mathcal R}$ can be expressed as \cite{MAXG1}
\begin{equation}
{\mathcal R} = - \psi - \frac{2\rho_{\mathrm{t}}}{3 (\rho_{\mathrm{t}} + p_{\mathrm{t}})}
\biggl( \phi + \frac{\psi'}{{\mathcal H}}\biggr),
\label{Rdef}
\end{equation}
 where $\phi$ is defined as the spatial part of the perturbed metric in the conformally Newtonian 
 gauge, i.e. $\delta_{\mathrm{s}} g_{00} = 2 a^2 \phi$.   
 In the same gauge the Hamiltonian constraint reads (see 
 also appendix \ref{APPA} and, in particular, Eq. (\ref{HAM}))
\begin{equation}
\nabla^2 \psi - 3 {\mathcal H}( {\mathcal H}\phi + \psi') = 4\pi G a^2 ( \delta \rho_{\mathrm{t}} + \delta \rho_{\mathrm{B}}).
\label{HAM1}
\end{equation}
Using Eq. (\ref{F3}) inside Eq. (\ref{ZETA}) 
and inserting the obtained equation into Eq. (\ref{HAM1}) we obtain, through Eq. (\ref{Rdef}) the following 
relation
\begin{equation}
\zeta = {\mathcal R} + \frac{\nabla^2 \psi}{12 \pi G a^2 (\rho_{\mathrm{t}} + p_{\mathrm{t}})}.
\end{equation}
implying that \footnote{In the present and in the following sections the we we will often pass from the real space 
to the Fourier space description. We will avoid, though, to write the explicit subscripts referring to the Fourier mode since they might  get confused with the indices labeling the fluctuations of the different species of the plasma.} for $k \tau \ll 1$, ${\mathcal R}(k) \sim \zeta(k) + {\mathcal O}(|k\tau|^2)$. 
From the covariant conservation equation we can easily deduce the evolution 
for $\zeta$:
\begin{equation}
\zeta' = - \frac{{\mathcal H}}{p_{\mathrm{t}} + \rho_{\mathrm{t}}} \delta p_{\mathrm{nad}} + 
\frac{{\mathcal H}}{p_{\mathrm{t}} + \rho_{\mathrm{t}}} \biggl( c_{\mathrm{st}}^2 - \frac{1}{3}\biggr)\delta\rho_{\mathrm{B}} - \frac{\theta_{\mathrm{t}}}{3}.
\label{zetaprime}
\end{equation}
In the case of a CDM-radiation entropy mode we have that 
\begin{equation}
\delta p_{\mathrm{nad}} = \rho_{\mathrm{M}} c^2_{\mathrm{s}} {\mathcal S}_{*},\qquad {\mathcal S}_{*} = \frac{\delta \varsigma}{\varsigma},
\end{equation}
where ${\mathcal S}_{*}$ is the relative fulctuation of the specific entropy $\varsigma= T^{3}/n_{\mathrm{CDM}}$ defined 
in terms of the temperature $T$ and in terms of the CDM concentration $n_{\mathrm{CDM}}$.

\subsection{Magnetized adiabatic mode}

The possible presence of entropic contributions will be neglected since 
 the attention will now be focused 
on the simplest situation which implies solely the presence of an adiabatic mode. It is however 
useful to keep, for a moment, the dependence of the curvature perturbations also upon ${\mathcal S}_{*}$ since the present 
analysis can be easily extended, with some algebra, to the 
case of magnetized non-adiabatic modes.
Recalling now the expression of the total sound speed $c_{\mathrm{st}}^2$ given in Eq.  (\ref{wcs}) 
and noticing that 
\begin{equation}
\frac{\rho_{\mathrm{M}}}{\rho_{\mathrm{t}} + p_{\mathrm{t}}} = \frac{3 \alpha}{3 \alpha + 4},
\qquad \frac{\rho_{\mathrm{R}}}{\rho_{\mathrm{t}} + p_{\mathrm{t}}} = \frac{ 3 }{3\alpha + 4},
\end{equation}
Eq. (\ref{zetaprime}) can be recast in the following useful form
\footnote{In Eq. (\ref{useful}) $R_{\gamma}$ denotes the fraction of photons in the radiation plasma.
Neutrinos are effectively massless prior to equality and their fraction will be denoted by $R_{\nu}$.
Note that $R_{\gamma} + R_{\nu} =1$ and $R_{\nu} = r/(r+1)$ where $r = 0.681 (N_{\nu}/3)$. Consequently
$R_{\gamma} = 0.594$ for three families, i.e. $N_{\nu}=3$.}
\begin{equation}
\frac{d \zeta}{d\alpha} = - \frac{4 {\mathcal S}_{*}}{(3 \alpha + 4)^2} - \frac{3 R_{\gamma} \Omega_{\mathrm{B}}}{(3 \alpha + 4)^2},
\label{useful}
\end{equation}
whose solution is
\begin{equation}
\zeta(k,\tau)= \zeta_{*}(k) - \frac{\alpha [ 4 {\mathcal S}_{*}(k) + 3 R_{\gamma} \Omega_{\mathrm{B}}(k)]}{4 (3 \alpha + 4)},
\label{zetaalpha}
\end{equation}
where $\zeta_{*}(k)$ is the constant value of curvature perturbations implied by the presence of the adiabatic mode; $\Omega_{\mathrm{B}}(k)$ has
been introduced in Eq. (\ref{OMEGA}). The dependence 
upon the Fourier mode $k$ has been explicitly written to remind 
that $\zeta_{*}(k)$ is constant in time but not in space. 
In the two relevant physical limits, i.e. well before and well after equality, Eq. (\ref{zetaalpha}) implies, respectively, 
\begin{eqnarray}
&& \lim_{\alpha\gg 1} \zeta(k) = \zeta_{*}(k) - \frac{{\mathcal S}_{*}(k)}{3} - \frac{R_{\gamma} \Omega_{\mathrm{B}}(k)}{4},
\label{alphagg}\\
&& \lim_{\alpha\ll 1} \zeta = \zeta_{*}(k) - \frac{{\mathcal S}_{*}(k)}{4}\alpha - \frac{3R_{\gamma} \Omega_{\mathrm{B}}(k)}{16}\alpha.
\label{alphall}
\end{eqnarray}
When $\psi = \phi$ we can also obtain the evolution of $\psi$ for the large scales 
\begin{equation}
\frac{ d \psi }{d\alpha} + \frac{5 \alpha + 6}{2\alpha (\alpha + 1)} \psi = - \frac{3 \alpha + 4}{2 \alpha (\alpha + 1)} \zeta_{*} 
+ \frac{4 {\mathcal S}_{*}  + 3 R_{\gamma} \Omega_{\mathrm{B}}}{8(\alpha + 1)}.
\label{evolp}
\end{equation}
Equation (\ref{evolp}) 
can be easily solved by noticing that it can be rewritten as 
\begin{equation}
\frac{\partial}{\partial \alpha} \biggl(\frac{\alpha^3}{\sqrt{\alpha+ 1}} \psi\biggr)  \frac{\sqrt{\alpha + 1}}{{\alpha^3}} = - 
\frac{3 \alpha + 4}{2 \alpha (\alpha + 1)} \zeta_{*} + \frac{4 {\mathcal S}_{*} + 3 R_{\gamma} \Omega_{\mathrm{B}}}{8 (\alpha + 1)},
\label{zetapsi}
\end{equation}
implying that 
\begin{equation}
\frac{\alpha^3}{\sqrt{\alpha + 1}} \psi = - \frac{\zeta_{*}}{2} {\mathcal I}_{1}(\alpha) + \frac{4 {\mathcal S}_{*} + 3 R_{\gamma} \Omega_{\mathrm{B}}}{8} {\mathcal I}_{2}(\alpha),
\label{implsol}
\end{equation}
where 
\begin{equation}
 {\mathcal I}_{1}(\alpha) = \int_{0}^{\alpha} \frac{\beta^2 ( 3 \beta + 4)}{(\beta+ 1 )^{3/2}} d\beta,\qquad{\mathcal I}_{2}(\alpha) = \int_{0}^{\alpha} \frac{\beta^3}{(\beta+ 1)^{3/2}} d\beta.
\end{equation}
By using the obvious change of variables $ y = \beta + 1$ 
both integrals can be calculated with elementary methods with the result that 
\begin{eqnarray}
&& {\mathcal I}_{1}(\alpha) =  \frac{2\{ 16 [ \sqrt{\alpha + 1} -1] + \alpha [ \alpha ( 9\alpha + 2) -8]\}}{15 \sqrt{\alpha + 1}},
\nonumber\\
&& {\mathcal I}_{2}(\alpha) = \frac{2 \{ 16 [ 1 - \sqrt{\alpha + 1}] + \alpha [ 8 + \alpha (\alpha -2)]\}}{5 \sqrt{\alpha + 1}}.
\label{integrals}
\end{eqnarray}
Inserting Eq. (\ref{integrals}) into Eq. (\ref{implsol}) the explicit result for 
$\psi$ can be written as:
\begin{eqnarray}
&& \psi(k,\tau) = - \frac{\zeta_{*}(k)}{15\alpha^3} \{ 16 [ \sqrt{\alpha + 1} -1] + \alpha [ \alpha ( 9\alpha + 2) -8]\} 
\nonumber\\
&&+ \frac{4 {\mathcal S}_{*}(k) + 3 R_{\gamma} \Omega_{\mathrm{B}}(k)}{20 \alpha^3} \{ 16 [ 1 - \sqrt{\alpha + 1}] + \alpha [ 8 + \alpha (\alpha -2)]\}.
\label{expl1}
\end{eqnarray}
Equation (\ref{expl1}) can be evaluated in the two limits mentioned 
above, i.e., respectively, well after and well before equality:
\begin{eqnarray}
&& \lim_{\alpha \gg 1} \psi(k,\tau)= - \frac{3}{5} \zeta_{*}(k) + 
\frac{ 4 {\mathcal S}_{*}(k) + 3 R_{\gamma} \Omega_{\mathrm{B}}(k)}{20},
\nonumber\\
&& \lim_{\alpha \ll 1}\psi(k,\tau) = - \frac{2}{3}  \zeta_{*}(k) 
+ \frac{\alpha}{32} \biggl[ 
\frac{4}{3} \zeta_{*}(k) + 4 {\mathcal S}_{*}(k) + 3 R_{\gamma}\Omega_{\mathrm{B}}(k)\biggr].
\end{eqnarray}
Notice that $\zeta_{*}(k)$ appears also in the correction
which goes as $\alpha= a/a_{\mathrm{eq}}$. 
In this derivation the role of the anisotropic stress has been neglected.
As full numerical solutions of the problem (in the tight coupling approximation) shows \cite{mg3,mg4}  that the magnetic anisotropic stress 
can be neglected close to recombination but it is certainly 
relevant deep in the radiation-dominated regime. 
To address this issue let us solve directly the system provided by the evolution 
equations of the longitudinal fluctuations of the geometry (i.e. Eqs. (\ref{anisstrein}), (\ref{HAM}) and 
(\ref{MOM})--(\ref{psidouble}))coupled with the evolution equations of the matter sources which are reported in appendix \ref{APPA}. The evolution of the background will be the one dictated 
 by Eq. (\ref{SCF}) and by Eq. (\ref{OMEX}). The solution of the Hamiltonian 
 constraint (\ref{HAM}) and of the evolution equations for various density contrasts (i.e. $\delta_{\nu}$, $\delta_{\gamma}$, $\delta_{\mathrm{b}}$ and $\delta_{\mathrm{c}}$) can be written, in the limit $x= \tau/\tau_{1} \ll 1$ as 
\begin{eqnarray}
&& \delta_{\gamma}(k,\tau) =\delta_{\nu}(k,\tau) - 2\phi_{*} - R_{\gamma} \Omega_{\mathrm{B}}(k) + 4 \psi_{1} x,
\nonumber\\
&& \delta_{\mathrm{c}}(k,\tau) = \delta_{\mathrm{c}}(k,\tau) = \frac{3}{4} \delta_{\gamma}(k,\tau).
\label{ans1}
\end{eqnarray}
The Hamiltonian constraint (\ref{HAM}) implies, always for $x  \ll 1$, that the following relation 
must hold among the various constants:
\begin{equation}
\phi_{1}(k) + 3 \psi_{1}(k) + \frac{\phi_{*}(k)}{2} = \frac{3}{4} R_{\gamma} \Omega_{\mathrm{B}}(k).
\label{CONHAM}
\end{equation}
Going on along the same theme we have that Eq. (\ref{psidouble})
is automatically satisfied by Eq. (\ref{ans1}) in the small-$x$ limit. 
The solution of Eq. (\ref{anisstrein}) can be obtained with similar methods and always 
well before equality:
\begin{equation}
\sigma_{\nu}(k,\tau) = - \frac{R_{\gamma}}{R_{\nu}} \sigma_{\mathrm{B}}(k) + \frac{k^2 x^2}{6 R_{\nu}} (\psi_{*}(k) - \phi_{*}(k)) + 
\frac{k^2 x^3}{6 R_{\nu}} [ (\psi_{*}(k) - \phi_{*}(k)) + (\psi_{1}(k) - \phi_{1}(k))],
\end{equation}
where $\sigma_{\nu}(k,\tau)$ is the neutrino anisotropic stress and 
$\sigma_{\mathrm{B}}(k,\tau)$ has been already introduced in Eq. (\ref{SIGMA}).
Notice that, as $\Omega_{\mathrm{B}}(k)$ also $\sigma_{\mathrm{B}}(k)$ is 
approximately constant in time when the flux-freezing condition is verified.

Using Eq. (\ref{ans1}) into  the evolution equations of the peculiar velocities
(i.e. Eqs. (\ref{CDM2}), (\ref{nueq2}) and (\ref{TC2})),
the explicit expressions 
for $\theta_{\mathrm{c}}$,  $\theta_{\nu}$ and $\theta_{\gamma\mathrm{b}}$ can be easily obtained.
In particular, for $\theta_{\mathrm{c}}$ and $\theta_{\nu}$ we have:
\begin{eqnarray}
\theta_{\mathrm{c}}(k,\tau) &=& \frac{k^2}{2}\biggl[ \phi_{*}(k) x + \frac{x^2}{6} ( 4 \phi_{1}(k) - \phi_{*}(k))\biggr],
\label{THC}\\
\theta_{\nu}(k,\tau) &=& \frac{k^2 x }{4} \biggl[ 2 \phi_{*}(k) + \frac{R_{\gamma}}{R_{\nu}} ( 4 \sigma_{\mathrm{B}}(k) - R_{\nu}\Omega_{\mathrm{B}}(k))\biggr] + \frac{k^2 x^2}{2}[\psi_{1}(k) + \phi_{1}(k)].
\label{THNU}
\end{eqnarray}
Finally, from Eq. (\ref{TC2}), the photon-baryon peculiar velocity field 
is determined to be:
\begin{equation}
\theta_{\gamma\mathrm{b}}(k,\tau)  = \frac{k^2 x}{8} [ R_{\nu} \Omega_{\mathrm{B}}(k) - 4 \sigma_{\mathrm{B}}(k) + 2 \phi_{*}(k)] + 
\frac{k^2 x^2}{3} [\psi_{1}(k) + \phi_{1}(k)].
\end{equation}

By solving Eq. (\ref{nueq3})  (bearing in mind Eqs. (\ref{CONHAM}) and (\ref{THNU})) the following relations can be obtained 
\begin{eqnarray}
&& \psi_{*}(k) = \biggl( 1 + \frac{2}{5} R_{\nu}\biggr) \phi_{*}(k)+ 
\frac{R_{\gamma}}{5} [4 \sigma_{\mathrm{B}}(k) - R_{\nu} \Omega_{\mathrm{B}}(k)],
\nonumber\\
&& \psi_{1}(k) - \phi_{1}(k) = \frac{4}{15} R_{\nu} ( \psi_{1}(k) + \phi_{1}(k)) 
- \frac{2}{5} R_{\nu} \phi_{*}(k) - \frac{R_{\gamma}}{5}[4 \sigma_{\mathrm{B}}(k) - R_{\nu} \Omega_{\mathrm{B}}(k)],
\end{eqnarray}
allowing to determine, in conjunction with Eq. (\ref{CONHAM}), the 
explicit form of $\phi_{1}(k)$ and of $\psi_{1}(k)$:
\begin{eqnarray}
&&\psi_{1}(k) = \frac{15 + 4 R_{\nu}}{60 + 8 R_{\nu}} \biggl[ \frac{3}{4} R_{\gamma} \Omega_{\mathrm{B}}(k) - \frac{\phi_{*}(k)}{2} \biggr]
- \frac{1}{60 + 8 R_{\nu}}\biggl[ \frac{2}{5} R_{\nu} \phi_{*}(k) + 
\frac{R_{\gamma}}{5} ( 4 \sigma_{\mathrm{B}}(k) - R_{\nu}\Omega_{\mathrm{B}}(k))\biggr],
\nonumber\\
&& \phi_{1}(k) = \frac{15 - 4 R_{\nu}}{60 + 8 R_{\nu}} \biggl[ \frac{3}{4} R_{\gamma} \Omega_{\mathrm{B}}(k) - \frac{\phi_{*}}{2}\biggr] 
+ \frac{3}{60 + 8 R_{\nu}}\biggl[ \frac{2}{5} R_{\nu} \phi_{*}(k) + 
\frac{R_{\gamma}}{5} ( 4 \sigma_{\mathrm{B}}(k) - R_{\nu}\Omega_{\mathrm{B}}(k))\biggr].
\end{eqnarray}
If $R_{\nu} = \Omega_{\mathrm{B}} =0$ we have that 
\begin{equation}
\phi_{1}(k) = \psi_{1}(k) = - \frac{\phi_{*}(k)}{8},
\label{recov}
\end{equation}
and this result coincides precisely with the result already obtained 
in Eq. (\ref{alphall}). In fact, recalling that $\alpha(x) = x^2 + 2 x$, we have that, in the small-$x$ 
region $\psi(k,\tau) \simeq - (2/3) \zeta_{*}(k) + (x/12)\zeta_{*}(k)$. But recalling now that, in the limit $R_{\nu}\to 0$ 
and $\Omega_{\mathrm{B}} \to 0$, $\zeta_{*}(k) = - (3/2) \psi_{*}(k)$,  Eq. (\ref{recov}) is recovered.
The obtained large-scale solutions will be important both for the explicit 
evaluation of the Sachs-Wolfe plateau as well as for the normalization 
of the solution at smaller $k$ that will be discussed in the forthcoming section.

\subsection{Estimate of the ordinary Sachs-Wolfe contribution}

The ordinary and integrated Sachs-Wolfe contributions can now be computed. Recalling Eq. (\ref{SWISW}) the large-scale 
limit of the brightness perturbation of the radiation field is (see also 
Eqs. (\ref{BRI}) and (\ref{SWISW}) of  the appendix \ref{APPA})
\begin{eqnarray}
&& \Delta_{\mathrm{I}}^{\mathrm{(SW)}}(\vec{k},\tau_{0})
 =- \frac{\zeta_{*}(k)}{5}+ \frac{R_{\gamma} \Omega_{\mathrm{B}}(k)}{20},
\label{SW}\\
&&  \Delta_{\mathrm{I}}^{\mathrm{(ISW)}}(\vec{k},\tau_{0})
 =  \frac{10 - 4 R_{\nu}}{ 5 ( 15 + 4 R_{\nu})} \zeta_{*}(k)
 + \frac{3}{10} R_{\gamma} \Omega_{\mathrm{B}}(k) + \frac{4}{5} 
 \frac{R_{\gamma} ( R_{\nu} + 5) ( 4 \sigma_{\mathrm{B}}(k) - R_{\nu} \Omega_{\mathrm{B}}(k))}{5 ( 15 + 4 R_{\nu})}.
 \label{ISW}
 \end{eqnarray}
As in the standard case, the ISW effect mimics the ordinary SW effect and it actually cancels partially 
the SW contribution at large angular scales. Notice that, in order to derive the explicit form of the ordinary SW 
it is practical to observe that, for wavelengths 
larger than the Hubble radius at recombination
$(\delta_{\gamma} - 4 \psi)' \simeq 0$. This 
observation implies that, clearly,
$ \delta_{\gamma}^{(\mathrm{f})} = 4 ( \psi^{(\mathrm{f})} -  
\psi^{(\mathrm{i})}) + 
\delta_{\gamma}^{(\mathrm{i})}$ where the superscripts f (for final) and i 
(for initial) indicate that the values of the corresponding quantities 
are taken, respectively, well after and well before equality.
The large angular scale expression of the temperature autocorrelations 
are defined as 
\begin{equation}
C^{\mathrm{(SW)}}_{\ell} = \frac{2}{\pi} \int k^3 d\ln{k} 
|\Delta^{\mathrm{(SW)}}_{\mathrm{I}}(k,\tau_{0})|^2.
\label{SW1}
\end{equation}
To evaluate Eq. (\ref{SW1}) in explicit terms we have to mention the 
conventions for the curvature and for the magnetic power spectra.
The correlators of $\zeta_{*}(k)$, $\Omega_{\mathrm{B}}(k)$ and
$\sigma_{\mathrm{B}}(k)$ are defined, respectively, as 
\begin{eqnarray}
&& \langle \zeta_{*}(\vec{k})\zeta_{*} (\vec{p}) \rangle = 
|\zeta_{*}(k)|^2 \delta^{(3)}(\vec{k} + \vec{p}),
\nonumber\\
&& \langle \Omega_{\mathrm{B}}(\vec{k}) \Omega_{\mathrm{B}}(\vec{p}) 
\rangle = |\Omega_{\mathrm{B}}(k)|^2 \delta^{(3)}(\vec{k} + \vec{p}),
\nonumber\\
&&\langle \sigma_{\mathrm{B}}(\vec{k}) \sigma_{\mathrm{B}}(\vec{p}) 
\rangle = |\sigma_{\mathrm{B}}(k)|^2 \delta^{(3)}(\vec{k} + \vec{p}).
\label{CORR}
\end{eqnarray}
In the case of the curvature perturbations we will have that 
\begin{equation}
|\zeta_{*}(k)|^2 = \frac{2\pi^2}{k^3} {\mathcal P}_{\zeta}(k),
\qquad 
{\mathcal P}_{\zeta}(k) = {\mathcal A}_{\zeta} \biggl(\frac{k}{k_{\mathrm{p}}}\biggr)^{n_{\zeta} -1},
\label{SW2}
\end{equation}
where $k_{\mathrm{p}}$ denotes the pivot scale at which the 
spectrum of curvature fluctuations is computed and ${\mathcal A}_{\zeta}$ 
is, by definition, the amplitude of the spectrum at the pivot scale.
In similar terms the magnetized contributions can be written as 
\begin{eqnarray}
&& |\Omega_{\mathrm{B}}(k)|^2 = 
\frac{2\pi^2 }{k^3} {\mathcal P}_{\Omega}(k), \qquad 
{\mathcal P}_{\Omega}(k) = {\mathcal F}(\epsilon) \overline{\Omega}_{\mathrm{BL}}^2 \biggl(\frac{k}{k_{\mathrm{L}}}\biggr)^{2\epsilon},
\label{SW3}\\
&& |\sigma_{\mathrm{B}}(k)|^2 = 
\frac{2\pi^2 }{k^3} {\mathcal P}_{\sigma}(k), \qquad 
{\mathcal P}_{\sigma}(k) = {\mathcal G}(\epsilon) \overline{\Omega}_{\mathrm{BL}}^2 \biggl(\frac{k}{k_{\mathrm{L}}}\biggr)^{2\epsilon},
\label{SW4}
\end{eqnarray}
where $k_{\mathrm{L}}$ (defined in Eq. (\ref{Q0})) denotes, in some sense, the magnetic pivot scale. The spectral index of the magnetic correlator 
defined in Eq. (\ref{definitions2}) is related to $\epsilon$ as $m + 3 = \epsilon$. Notice also that in defining the correlators of $\Omega_{\mathrm{B}}$ and of $\sigma_{\mathrm{B}}$ the same conventions used 
for the curvature perturbations have been adopted. These conventions 
imply that a factor $k^{-3}$ appears at the right hand side of the first 
relation of Eq. (\ref{SW2}).

Since the spectrum of the magnetic energy density implies the calculation 
of a convolution $k_{\mathrm{L}}$ is also related to the smoothing 
scale of the magnetic energy density (see, for instance, \cite{mg3}).
In Eqs. (\ref{SW3}) and (\ref{SW4}) the functions ${\mathcal F}(\epsilon)$ 
and ${\mathcal G}(\epsilon)$ as well as the smoothed amplitude 
$\overline{\Omega}_{\mathrm{BL}}$ are defined as 
\begin{eqnarray}
&& {\mathcal   F}(\epsilon) = \frac{4(6 - \epsilon) ( 2 \pi)^{ 2 \epsilon}}{\epsilon ( 3 - 2 \epsilon)
 \Gamma^2(\epsilon/2)}, \qquad 
 {\mathcal   G}(\epsilon) = 
 \frac{4(188 - 4 \epsilon^2 - 66\epsilon) (2\pi)^{2 \epsilon}}{3 \epsilon ( 3 - \epsilon) ( 2 \epsilon +1) \Gamma^2(\epsilon/2)},
\label{SW5}\\
&& \overline{\Omega}_{{\rm BL}} = \frac{\rho_{\mathrm{BL}}}{\overline{\rho}_{\gamma}}, 
\qquad \rho_{{\rm BL}}=\frac{ B_{\mathrm{L}}^2}{8\pi}, 
\qquad \overline{\rho}_{\gamma} = a^4(\tau)\rho_{\gamma}(\tau).
\label{SW6}
\end{eqnarray}
From Eq. (\ref{SW6}), recalling that $T_{\mathrm{CMB}} = 2.725 \mathrm{K}$
and that $\overline{\rho}_{\gamma} = (\pi^2/15) T_{\mathrm{CMB}}^4$, we can also write, in more explicit terms:
\begin{equation}
\overline{\Omega}_{\mathrm{BL}} = 7.565 \times 10^{-9} \, \biggl(\frac{B_{\mathrm{L}}}{\mathrm{nG}}\biggr)^2.
\label{SW7}
\end{equation}
It should finally be appreciated that the power spectra of the magnetic energy 
density and of the anisotropic stress are proportional since we focus 
our attention to magnetic spectral slopes $\epsilon < 1$ which are 
the most relevant at large length-scales \footnote{Notice, incidentally, that nearly scale-invariant magnetic energy spectra also arise 
in string inspired cosmological models \cite{mv1} as a consequence of the breaking of conformal invariance 
during the pre-big bang phase.}. In principle, the present analysis 
can be also extended to the case when the magnetic power spectra 
are very steep in $k$ (i.e. $\epsilon >1$). In the latter case the power spectra 
are often said to be violet and they are severely  constrained by thermal diffusivity effects \cite{vt3}.

By performing the integration over the comoving wave-number that appears 
in Eq. (\ref{SW1}) the wanted result can be expressed as \footnote{The analytical calculation 
of the integral of Eq. (\ref{SW1}) holds for $ -3 <n_{\zeta} < 3$ and for $\ell < 30$. This means 
that the result is accurate for sufficiently large angular scale. In fact, the angular separation $\vartheta$ 
is approximately equal to $\pi/\ell$. If $\ell < 26$, then $\theta > 7 \mathrm{deg}$. 
This was, for instance, the region 
explored by the COBE team and it is the regime where CMB anisotropies computations may be usefully normalized.}
\begin{equation}
C^{(\mathrm{SW})}_{\ell} = \biggl[ \frac{{\cal A}_{\zeta}}{25} \,{\mathcal Z}_{1}(n_{\zeta},\ell)  +
\frac{1}{400} \, R_{\gamma}^2  \overline{\Omega}^{2}_{{\rm B}\,L} {\mathcal Z}_{2}(\epsilon,\ell) - 
\frac{1}{50} \sqrt{ {\cal A}_{\zeta}} \, R_{\gamma} \,\overline{\Omega}_{{\rm B}\, L}\,{\mathcal Z}_{3} (n_{\zeta},\epsilon, \ell) \cos{\gamma_{br}}\biggr],
\label{SWP}
\end{equation}
where 
\begin{eqnarray}
{\mathcal Z}_{1}(n,\ell) &=& \frac{\pi^2}{4} \biggl(\frac{k_0}{k_{\rm p}}\biggl)^{n-1} 2^{n} \frac{\Gamma( 3 - n) \Gamma\biggl(\ell + 
\frac{ n -1}{2}\biggr)}{\Gamma^2\biggl( 2 - \frac{n}{2}\biggr) \Gamma\biggl( \ell + \frac{5}{2} - \frac{n}{2}\biggr)},
\label{Z1}\\
{\mathcal Z}_{2}(\epsilon,\ell) &=& \frac{\pi^2}{2} 2^{2\epsilon} {\cal F}(\epsilon) \biggl( \frac{k_{0}}{k_{L}}\biggr)^{ 2 \epsilon} \frac{ \Gamma( 2 - 2 \epsilon) 
\Gamma(\ell + \epsilon)}{\Gamma^2\biggl(\frac{3}{2} - \epsilon\biggr) \Gamma(\ell + 2 -\epsilon)},
\label{Z2}\\
{\mathcal Z}_{3}(n,\epsilon,\ell) &=&\frac{\pi^2}{4} 2^{\epsilon} 2^{\frac{n +1}{2}} \,\sqrt{{\cal F}(\epsilon)}\, \biggl(\frac{k_{0}}{k_{L}}\biggr)^{\epsilon} \biggl(\frac{k_{0}}{k_{\rm p}}\biggr)^{\frac{n + 1}{2}} \frac{ \Gamma\biggl(\frac{5}{2} - \epsilon - \frac{n}{2}\biggr) \Gamma\biggl( \ell + 
\frac{\epsilon}{2} + \frac{n}{4} - \frac{1}{4}\biggr)}{\Gamma^2\biggl(\frac{7}{4} - \frac{{\epsilon}}{2} - \frac{n}{4}\biggr)
\Gamma\biggl( \frac{9}{4} + \ell - \frac{\epsilon}{2} - \frac{n}{4} \biggr)}.
\label{Z3}
\end{eqnarray}
In Eq. (\ref{Z3}) $\gamma_{\mathrm{br}}$ is the correlation angle that has been included to keep 
the expressions as general as possible. In what follows the main focus will however be on the case where 
the adiabatic mode of curvature perturbations is not correlated with the magnetized 
contribution (i.e. $\gamma_{\mathrm{br}} = \pi/2$).
The various pivot scales appearing in Eqs. (\ref{Z1}), (\ref{Z2}) and (\ref{Z3}) 
will now be defined:
\begin{equation}
k_{\mathrm{p}} = 0.002\, \mathrm{Mpc}^{-1},
\qquad k_{0}=\frac{h_{0}}{3000} \, \mathrm{Mpc}^{-1},
\qquad k_{\mathrm{L}}= 1\,\mathrm{Mpc}^{-1}.
\label{kscales}
\end{equation}
Let us now consider some simplified limits. The first one 
is to posit\footnote{In section \ref{sec5}  the scalar spectral index will be denoted by $n_{\zeta}$, stressing, in this way, that we refer to the spectral 
index appearing in the power spectrum of $\zeta$.}  $n_{\mathrm{\zeta}} =1$ and $\epsilon < 1$. 
We will have that the functions ${\mathcal Z}$ will be simplified.
They become:
\begin{eqnarray}
 {\mathcal Z}_{1}(1,\ell) &=& \frac{2\pi}{\ell (\ell + 1)},
\label{Z1a}\\
{\mathcal Z}_{2}(\epsilon,\ell) &=& \frac{\pi^2}{2} 2^{2\epsilon} {\cal F}(\epsilon) \biggl( \frac{k_{0}}{k_{L}}\biggr)^{ 2 \epsilon} \frac{ \Gamma( 2 - 2 \epsilon) 
\Gamma(\ell + \epsilon)}{\Gamma^2\biggl(\frac{3}{2} - \epsilon\biggr) \Gamma(\ell + 2 -\epsilon)},
\label{Z2a}\\
{\mathcal Z}_{3}(1,\epsilon,\ell) &=&\frac{\pi^2}{2} 2^{\epsilon} \,\sqrt{{\cal F}(\epsilon)}\, \biggl(\frac{k_{0}}{k_{L}}\biggr)^{\epsilon} \biggl(\frac{k_{0}}{k_{\rm p}}\biggr)\frac{ \Gamma\biggl(2 - \epsilon) \Gamma\biggl( \ell + 
\frac{\epsilon}{2}\biggr)}{\Gamma^2\biggl(\frac{3}{2} - \frac{{\epsilon}}{2}\biggr)
\Gamma\biggl(  \ell +2  - \frac{\epsilon}{2} \biggr)}.
\label{Z3a}
\end{eqnarray}
We now can enforce the normalization at large scales by assuming a dominant 
adiabatic mode. A preliminary manipulation is the following. We can write the previous 
expression as 
\begin{equation}
\frac{\ell (\ell +1)}{2\pi} C^{(\mathrm{SW})}_{\ell} = \frac{{\mathcal A}_{\zeta}}{25}\biggl[ 1 +
\biggl(\frac{R_{\gamma}^2}{16 {\mathcal A}_{\zeta}}\overline{\Omega}_{\mathrm{BL}}^2\biggr) \frac{\ell(\ell +1)}{2\pi} {\mathcal Z}_{2}(\epsilon,\ell) - \biggl(\frac{R_{\gamma}}{2 \sqrt{{\mathcal A}_{\zeta}}} \overline{\Omega}_{\mathrm{BL}}\biggr)
\frac{\ell(\ell + 1)}{2\pi} {\mathcal Z}_{3}(1,\epsilon,\ell) \cos{\gamma_{\mathrm{br}}}\biggr].
\end{equation}
We can now expand the relevant terms in powers of $\epsilon$. We do get 
\begin{equation}
\frac{\ell (\ell +1)}{2\pi} C^{(\mathrm{SW})}_{\ell} =\frac{{\mathcal A}_{\zeta}}{25}\biggl[ 1 + {\mathcal Q}_{1}(\ell)\biggl(\frac{R_{\gamma}^2}{16 {\mathcal A}_{\zeta}}\overline{\Omega}_{\mathrm{BL}}^2\biggr) {\mathcal F}(\epsilon) \biggl(\frac{k_{0}}{k_{\mathrm{L}}}\biggr)^{2\epsilon} - {\mathcal Q}_{2}(\ell)\biggl(\frac{R_{\gamma}}{2 \sqrt{{\mathcal A}_{\zeta}}} \overline{\Omega}_{\mathrm{BL}}\biggr)\sqrt{{\mathcal F}(\epsilon)} \biggl(\frac{k_{0}}{k_{\mathrm{L}}}\biggr)^{\epsilon}  \biggl(\frac{k_{0}}{k_{\mathrm{p}}}\biggr)\biggr],
\end{equation}
where
\begin{eqnarray}
&& {\mathcal Q}_{1}(\ell) = 1 + \biggl( 2 + \frac{1}{\ell} + \frac{1}{\ell + 1} - 2\ln{2} + 2 \frac{\Gamma'(\ell)}{\Gamma(\ell)}\biggr)\epsilon + {\mathcal O}(\epsilon^2),
\nonumber\\
&& Q_{2}(\ell) = \frac{1}{2} {\mathcal Q}_{1}(\ell).
\end{eqnarray}
We can then compute the various pieces. They will set the scale of the numerical 
results. In particular, it is easy to argue that the presence of the cross correlation 
enhances the results at smaller scales. 
As a final comment it is relevant to remark that the large-scale solutions 
are not only important per se but they will be used to deduce the appropriate 
normalization for the results arising at smaller angular scales.

\renewcommand{\theequation}{4.\arabic{equation}}
\setcounter{equation}{0}
\section{Intermediate scales}
\label{sec4}
From Eqs. (\ref{TC5}), (\ref{TC6}) and (\ref{TC7}) 
the photon density contrast can be determined 
under the assumption that the entropic contribution is absent.  Thus, 
if only the magnetic fields and the 
adiabatic mode are present, Eqs. (\ref{TC7}), (\ref{TC7a}) and (\ref{TC7b})
lead to the following solution
\begin{equation}
\delta_{\gamma}(k,\tau) = - \frac{4}{3 c_{\mathrm{sb}}^2} \psi(k,\tau) + [ 4 \sigma_{\mathrm{B}}(k) - \Omega_{\mathrm{B}}(k)] + \sqrt{c_{\mathrm{sb}}} \, A_{1}(k)\, \cos{[\alpha(\tau,k)]}\,\, e^{- \frac{k^2}{k_{\mathrm{D}}^2}},
\label{DG1}
\end{equation}
where $\psi(k,\tau)$ is assumed to be slowly varying in time and where, recalling Eq. (\ref{lambdaT})
\begin{equation}
\alpha(\tau,k) =k \int_{0}^{\tau_{\mathrm{rec}}} c_{\mathrm{sb}}(\tau) d\tau,\qquad 
\frac{1}{k_{\mathrm{D}}^2} = \frac{2}{5} \int \lambda_{\mathrm{T}}(\tau) c_{\mathrm{sb}}^2(\tau) d\tau.
\label{DG2}
\end{equation}
The constant $A_{1}(k)$ can be determined by matching the 
solution to the large-scale (i.e. super-Hubble) behaviour 
of the fluctuations, i.e. 
\begin{equation}
\frac{\delta_{\gamma}(k,\tau)}{4} + \psi(k,\tau) \to \frac{\psi_{\mathrm{m}}(k)}{3} = - \frac{\zeta_{*}(k)}{5} 
+ \frac{R_{\gamma}}{20} \Omega_{\mathrm{B}}(k),
\end{equation}
where $\psi_{\mathrm{m}}$ denotes the value of $\psi(k)$ after equality and for $k\tau <1$.
From the solution of the evolution equation of 
$\delta_{\gamma}(k,\tau)$ also 
$\theta_{\gamma\mathrm{b}}(k,\tau)$ can be easily obtained (see, in particular, Eq. 
(\ref{TC1}) of appendix \ref{APPA}).
The final result can be expressed, for the present purposes, as\footnote{Notice that, in this paper,
the natural logarithm will be denoted by $\ln$ while the base-$10$ logarithms will be denoted by $\log$.}
\begin{eqnarray}
&& 
\biggl[\frac{\delta_{\gamma}(k,\tau)}{4} + \psi(k,\tau) \biggr] = {\mathcal L}_{\zeta}(k,\tau) + 
[{\mathcal M}_{\zeta}(k,\tau) {\mathcal D}_{\zeta}(k) + {\mathcal M}_{\mathrm{B}}(k,\tau) {\mathcal D}_{\mathrm{B}}(k)]
\sqrt{c_{\mathrm{sb}}} \cos{[\alpha(\tau,k)]},
\label{gen1}\\
&& \theta_{\gamma\mathrm{b}}(k,\tau) = 3 \,c_{\mathrm{sb}}^{3/2}[ {\mathcal M}_{\zeta}(k,\tau)
{\mathcal D}_{\zeta}(k)  + {\mathcal M}_{\mathrm{B}}(k,\tau) {\mathcal D}_{\mathrm{B}}(k)] \sin{[\alpha(\tau,k)]}.
 \label{gen2}
 \end{eqnarray}
The functions ${\mathcal L}_{\zeta}(k,\tau)$ and ${\mathcal M}_{\zeta}(k,\tau)$ are directly related to the curvature perturbations and can be 
determined by interpolating the large-scale behaviour with the
small-scale solutions. In the present case they can be written as
\begin{eqnarray}
&& {\mathcal L}_{\zeta}(k,\tau) = - \frac{\zeta_{*}(k)}{6} \biggl(1 - \frac{1}{3 c_{\mathrm{sb}}^2}\biggr) \ln{\biggl[\frac{14}{w \ell} \frac{\tau_{0}}{\tau_{\mathrm{eq}}}\biggr]},
\label{ELL1}\\
&& {\mathcal M}_{\zeta}(k,\tau) = - \frac{6}{25} \zeta_{*}(k)\ln{\biggl[\frac{14}{25} w \ell \frac{\tau_{\mathrm{eq}}}{\tau_{0}}\biggr]}.
\label{ELL2}
\end{eqnarray}
In Eqs. (\ref{ELL1}) and (\ref{ELL2}) the variable $w = k\tau_{0}/\ell$ has been introduced. This way of writing may seem, at the moment, obscure. However, the variable $w$ will appear as integration variable in the angular power spectrum, so it is practical, as early as possible, to express the integrands directly in terms of $w$.
Finally the functions 
${\mathcal L}_{\mathrm{B}}(k,\tau)$ and ${\mathcal M}_{\mathrm{B}}(k,\tau)$ are 
determined in similar terms and they can be written as 
\begin{eqnarray}
&&{\mathcal L}_{\mathrm{B}}(k,\tau)= \frac{3 R_{\gamma}}{20} \Omega_{\mathrm{B}}(k)\biggl(1 - \frac{1}{3 c_{\mathrm{sb}}^2}\biggr) +
\biggl[\sigma_{\mathrm{B}}(k) - \frac{\Omega_{\mathrm{B}}(k)}{4}\biggr],
\label{ELLB1}\\
&& {\mathcal M}_{\mathrm{B}}(k,\tau)= 3^{1/4} \biggl\{\biggl[\frac{\Omega_{\mathrm{B}}}{4}(k) - \sigma_{\mathrm{B}}(k)\biggr] + \frac{R_{\gamma}}{20} \Omega_{\mathrm{B}}(k)\biggr\}.
\label{ELLB2}
\end{eqnarray}
The functions ${\mathcal D}_{\zeta}(k)$ and 
${\mathcal D}_{\mathrm{B}}(k)$ encode the informations 
related to the diffusivity wave-number:
\begin{equation}
{\mathcal D}_{\zeta}(k) = e^{-\frac{k^2}{k_{\mathrm{D}}^2}},\qquad 
{\mathcal D}_{\mathrm{B}}(k) = e^{-\frac{k^2}{k_{\mathrm{B}}^2}}.
\end{equation}
As introduced before $k_{\mathrm{D}}$ is the thermal diffusivity scale
(i.e. shear viscosity). The quantity named $k_{\mathrm{B}}$ is the smallest 
momentum between the ones defined by magnetic diffusivity, by Alfv\'en 
diffusivity and by thermal diffusivity. The magnetic diffusivity has been 
already introduced in Eq. (\ref{ksigma}) and it arises 
because of the finite value of the conductivity. The Alfv\'en 
diffusivity arises when the magnetic field supports Alfv\'en waves that are 
subsequently damped for typical length-scales that are a bit smaller 
than the Silk damping scale (see \cite{vt3} and, in particular, \cite{kan1}). Now, if the magnetic field 
is fully inhomogeneous (as in the present case) the dominant 
source of diffusivity is represented by the Silk length scale since 
it is larger than the magnetic diffusivity length and than the 
Alfv\'en diffusivity length \cite{vt3}. 
For the purpose of simplifying the integrals to be evaluated numerically 
it is practical to introduce the following rescaled quantities:
\begin{eqnarray}
&&{\mathcal L}_{\zeta}(k,\tau) = \zeta_{*}(k) L_{\zeta}(k,\tau), \qquad {\mathcal L}_{\mathrm{B}}(k,\tau) = \Omega_{\mathrm{B}}(k)
 L_{\mathrm{B}}(k,\tau),
 \nonumber\\
&& {\mathcal M}_{\zeta}(k,\tau) = \zeta_{*}(k) M_{\zeta}(k,\tau),\qquad 
{\mathcal M}_{\mathrm{B}}(k,\tau) = \Omega_{\mathrm{B}}(k)
 M_{\mathrm{B}}(k,\tau),
\end{eqnarray}
after some algebra the angular power spectrum can be written as 
the sum of four integrals, i.e. 
\begin{equation}
C_{\ell} = {\mathcal U}_{1}(\ell) +  {\mathcal U}_{2}(\ell) +  {\mathcal U}_{3}(\ell) 
+  {\mathcal U}_{4}(\ell),
\label{U}
\end{equation}
where:
\begin{eqnarray}
{\mathcal U}_{1}(\ell) &=& 4\pi \int_{0}^{\infty} \frac{d w}{w} \overline{{\mathcal U}}_{1}(\ell,w){\mathcal K}^2(\ell,\ell_{\mathrm{t}},w) j_{\ell}^2 (\ell w),
\label{U1a}\\
 {\mathcal U}_{2}(\ell) &=& 2\pi c_{\mathrm{sb}} \int_{0}^{\infty} 
\frac{ dw}{w^3}[ w^2 + 9 c_{\mathrm{sb}}^2 ( w^2 -1)]   
\overline{{\mathcal U}}_{2}(\ell,w){\mathcal K}^2(\ell,\ell_{\mathrm{t}},w) j_{\ell}^2 (\ell w),
\label{U2a}\\
{\mathcal U}_{3}(\ell) &=&  2 \pi c_{\mathrm{sb}} \int_{0}^{\infty} 
\frac{dw}{w^3} [w^2 - 9 c_{\mathrm{sb}}^2 (w^2 -1)]
\overline{{\mathcal U}}_{3}(\ell,w)
\cos{(2 \ell \gamma w)}{\mathcal K}^2(\ell,\ell_{\mathrm{t}},w) j_{\ell}^2 (\ell w), 
\label{U3a}\\
{\mathcal U}_{4}(\ell) &=& 8 \pi \sqrt{c_{\mathrm{sb}}} \int_{0}^{\infty} \frac{dw}{w} \cos{(\ell \gamma w)} \overline{{\mathcal U}}_{4}(\ell,w){\mathcal K}^2(\ell,\ell_{\mathrm{t}},w) j_{\ell}^2 (\ell w),
\label{U4a}
\end{eqnarray}
where $j_{\ell}(y)$ denote the spherical Bessel functions of the first kind \cite{abr,grad} which 
are related to the ordinary Bessel functions of the first kind as $j_{\ell}= \sqrt{\pi/(2 y)} J_{\ell +1/2}(y)$.
The various functions appearing in Eqs. (\ref{U1a}), (\ref{U2a}), (\ref{U3a}) 
and (\ref{U4a}) are:
\begin{eqnarray} 
\overline{{\mathcal U}}_{1}(\ell,w) &=&  {\mathcal P}_{\zeta}(w,\ell) L_{\zeta}^2(\ell,w) + {\mathcal P}_{\Omega}(w,\ell) L_{\mathrm{B}}^2(\ell,w) 
\nonumber\\
&+& 2 L_{\zeta}(\ell,w) L_{\mathrm{B}}(\ell,w) 
\sqrt{{\mathcal P}_{\zeta}(w,\ell)}\sqrt{{\mathcal P}_{\Omega}(w,\ell)},
\label{U1b}\\
\overline{{\mathcal U}}_{2}(\ell,w) &=&
{\mathcal P}_{\zeta}(w,\ell) M_{\zeta}^2(\ell,w) {\mathcal D}_{\zeta}^2(\ell,\ell_{\mathrm{S}},w)
+ {\mathcal P}_{\Omega}(w) M_{\mathrm{B}}^2(\ell,w)
{\mathcal D}_{\mathrm{B}}^2(\ell,\ell_{\mathrm{A}},w)  
\nonumber\\
&+ &2 \sqrt{{\mathcal P}_{\zeta}(w,\ell)}\sqrt{{\mathcal P}_{\Omega}(w,\ell)} M_{\zeta}(\ell,w) M_{\mathrm{B}}(\ell,w) {\mathcal D}_{\zeta}(\ell,\ell_{\mathrm{S}},w) {\mathcal D}_{\mathrm{B}}(\ell,\ell_{\mathrm{A}},w),
\label{U2b}\\
\overline{{\mathcal U}}_{3}(\ell,w) &=& {\mathcal P}_{\zeta}(w,\ell)  M_{\zeta}^2(\ell,w) {\mathcal D}_{\zeta}^2(\ell,\ell_{\mathrm{t}},w)
+ {\mathcal P}_{\Omega}(w,\ell)(\ell,\ell_{\mathrm{A}},w) M_{\mathrm{B}}^2(\ell,w) {\mathcal D}_{\mathrm{B}}^2
\nonumber\\
&+& 2 \sqrt{{\mathcal P}_{\zeta}(w,\ell)}\sqrt{{\mathcal P}_{\Omega}(w,\ell)}
M_{\zeta}(\ell,w) M_{\mathrm{B}}(\ell,w) {\mathcal D}_{\zeta}(\ell,\ell_{\mathrm{S}},w) {\mathcal D}_{\mathrm{B}}(\ell, \ell_{\mathrm{A}},w),
\label{U3b}\\
\overline{{\mathcal U}}_{4}(\ell,w) &=&  {\mathcal P}_{\zeta}(w,\ell) L_{\zeta}(\ell,w)
M_{\zeta}(\ell,w) {\mathcal D}_{\zeta}(\ell,\ell_{\mathrm{S}},w) 
+
{\mathcal P}_{\Omega}(w,\ell) L_{\mathrm{B}}(\ell,w) M_{\mathrm{B}}(\ell,w) {\mathcal D}_{\mathrm{B}}(\ell,\ell_{\mathrm{A}},w)
\nonumber\\
&+&\sqrt{{\mathcal P}_{\zeta}(w,\ell)}\sqrt{{\mathcal P}_{\Omega}(w,\ell)}L_{\zeta}(\ell,w) M_{\mathrm{B}}(\ell,w) {\mathcal D}_{\mathrm{B}}(\ell,\ell_{\mathrm{A}},w) 
\nonumber\\
&+& \sqrt{{\mathcal P}_{\zeta}(w,\ell)}\sqrt{{\mathcal P}_{\Omega}(w,\ell)} L_{\mathrm{B}}(\ell,w) M_{\zeta}(\ell,w) {\mathcal D}_{\zeta}(\ell,\ell_{\mathrm{S}},w),
\label{U4b}
\end{eqnarray}
where $\ell_{\mathrm{S}}$ and $\ell_{\mathrm{A}}$ denote respectively the typical Silk multipole 
and the typical multipole associated with Alfv\'en diffusivity. They will be defined explicitly 
in a moment.
In Eqs. (\ref{U3a}) and (\ref{U4a}) the oscillatory terms arising, originally, 
in the full expression of the angular power spectrum have been simplified.
The two oscillatory contributions in Eqs. (\ref{U3a}) and (\ref{U4a}) 
go, respectively, as $\cos{(2 \gamma \ell w)}$ and as $\cos{(\gamma \ell w)}$.
The definition of $\gamma$ can be easily deduced from the original 
parametrization of the oscillatory contribution in Eqs. (\ref{DG1}) and 
(\ref{DG2}). In fact we can write $\alpha(k,\tau_{\mathrm{rec}}) = \gamma(\tau_{\mathrm{rec}}) \ell w$. Recalling that 
$w = k \tau_{0}/\ell$, and defining, for notational convenience, $\gamma\equiv \gamma(\tau_{\mathrm{rec}})$,
the following expression for $\gamma$ can be easily obtained
\begin{equation}
\gamma = \frac{1}{\tau_{0}} \int_{0}^{\tau_{\mathrm{rec}}} c_{\mathrm{sb}}(\tau) d\tau = \frac{\tau_{1}}{\sqrt{3}\tau_{0}}\int_{0}^{\tau_{\mathrm{rec}}/\tau_{1}} \frac{d x }{\sqrt{1 + \nu_1 ( x^2 + 2 x)}}.
\label{gamma1}
\end{equation}
In Eq. (\ref{gamma1}) the first equality is simply the definition of $\gamma$ while the second equality can be deduced by inserting in the definition 
the explicit expression of the scale factor of Eq. (\ref{SCF}).
By doing so the constant $\nu_{1}$ is just $R_{\mathrm{b}} z_{\mathrm{rec}}/z_{\mathrm{eq}}$. The expression of $\gamma$ can be made 
even more explicit by performing the integral appearing in Eq. (\ref{gamma1}):
\begin{equation}
\gamma= \frac{1}{\sqrt{3}} \biggl(\frac{\tau_{1}}{\tau_{0}}\biggr) \sqrt{\frac{z_{\mathrm{eq}}}{z_{\mathrm{rec}}}} \ln{\biggl[ 
\frac{\nu_{1} ( 1 + x_{\mathrm{rec}}) + \sqrt{1 + \nu_{1} x_{\mathrm{rec}}(x_{\mathrm{rec}}+2)}}{\sqrt{\nu_{1}} +1}\biggr]},\qquad x_{\mathrm{rec}} = \frac{\tau_{\mathrm{rec}}}{\tau_{1}}.
\label{gamma2}
\end{equation}
It is now practical to recall that the ratio between $\tau_{1}$ and $\tau_{0}$ 
depends upon the critical fraction of the dark energy. So the scale 
factor (\ref{SCF}) must be complemented, at late times, by the contribution of the dark energy. This standard calculation leads to the following estimate
for a spatially flat Universe
\begin{equation}
\frac{\tau_{1}}{\tau_{0}} = \frac{\tau_{1}}{\tau_{\mathrm{rec}}} \frac{\tau_{\mathrm{rec}}}{\tau_{0}}= \frac{{\mathcal I}_{\Lambda}}{\sqrt{z_{\mathrm{eq}}}},
\qquad {\mathcal I}_{\Lambda} = |1 - \Omega_{\Lambda 0}|^{-\nu_{2}}, 
\label{gamma3}
\end{equation}
where $\Omega_{\Lambda 0}$ is the present critical fraction 
of dark energy (parametrized in terms of a cosmological constant in a 
$\Lambda$CDM framework) 
and where $\nu_{2} = 0.0858$. Inserting Eq. (\ref{RED1}) into Eq. (\ref{gamma3}) and recalling the explicit expression of $\nu _{1}$ we will have finally
\begin{equation}
\gamma = \frac{{\mathcal I}_{\Lambda}}{\sqrt{3 R_{\mathrm{b}} z_{\mathrm{rec}}}} \ln{\biggl[ 
\frac{\sqrt{(1 + z_{\mathrm{rec}}/z_{\mathrm{eq}}) R_{\mathrm{b}}} + \sqrt{1 + R_{\mathrm{b}}}}{
1 + \sqrt{ R_{\mathrm{b}} z_{\mathrm{rec}}/z_{\mathrm{eq}}}}\biggr]}.
\label{gamma4}
\end{equation}

At this point the spherical Bessel functions appearing in the above expressions can be evaluated in the limit of large $\ell$ with the result that the above expressions can be made more explicit. In particular, focussing the attention 
on $j_{\ell}(\ell w)$, we have that  \cite{abr,grad}
\begin{equation}
j_{\ell}^2( \ell w) \simeq \frac{\cos^2{[\beta(w,\ell)]}}{\ell w \sqrt{ \ell^2 w^2 - \ell^2}}, \qquad w>1.
\label{besselexp}
\end{equation}
Note that the expansion (\ref{besselexp}) has been used consistently by other authors (see, in particular, 
\cite{wein,muk} and also \cite{pavel1,seljak,hu1,hu2}).
The result expressed by Eq. (\ref{besselexp})  allows to write 
the integrals of Eqs. (\ref{U1a}), (\ref{U2a}), (\ref{U3a}) 
and (\ref{U4a}) as 
\begin{equation}
C_{\ell} = \frac{1}{\ell^2}[{\cal C}_{1}(\ell) + {\cal C}_{2}(\ell) + {\cal C}_{3}(\ell)
+ {\cal C}_{4}(\ell) ],
\label{CELLFIN}
\end{equation}
where 
\begin{eqnarray}
&& {\mathcal C}_{1}(\ell) = \int_{1}^{\infty} I_{1}(w,\ell) 
\overline{{\mathcal U}}_{1}(\ell,w)\, dw,
\label{C1}\\
&& {\mathcal C}_{2}(\ell) = \int_{1}^{\infty} I_{2}(w,\ell) 
\overline{{\mathcal U}}_{2}(\ell,w)\, dw,
\label{C2}\\
&& {\mathcal C}_{3}(\ell) = \int_{1}^{\infty} I_{3}(w,\ell)
\overline{{\mathcal U}}_{3}(\ell,w)\, dw ,
\label{C3}\\
&& {\mathcal C}_{4}(\ell) = \int_{1}^{\infty} I_{4}(w,\ell) 
\overline{{\mathcal U}}_{4}(\ell,w)\, dw,
\label{C4}
\end{eqnarray}
where
\begin{eqnarray}
I_{1}(w,\ell) &=& \frac{4\pi \cos^2{[\beta(w,\ell)]}}{w^2 
\sqrt{w^2 -1}} e^{-2\frac{\ell^2}{\ell_{{\mathrm{t}}}^2} w^2},
\label{I1}\\
I_{2}(w,\ell) &=&  2\pi c_{\mathrm{sb}}
\frac{w^2 + 9 c_{\mathrm{sb}}^2 ( w^2 -1)}{w^4 \sqrt{w^2 -1}} 
 \cos^2{[\beta(w,\ell)] }e^{-2\frac{\ell^2}{\ell_{{\mathrm{t}}}^2} w^2},
 \label{I2}\\
 I_{3}(w,\ell) &=& 2\pi c_{\mathrm{s}}
\frac{w^2 - 9 c_{\mathrm{sb}}^2 ( w^2 -1)}{w^4 \sqrt{w^2 -1}} \cos{(2 \ell \gamma w)} \cos^2{[\beta(w,\ell)]} e^{-2\frac{\ell^2}{\ell_{{\mathrm{t}}}^2} w^2},
\label{I3}\\
 I_{4}(w,\ell) &=& 8\pi \sqrt{c_{\mathrm{sb}}}
  \frac{\cos{(\ell\gamma w)}}{w^2 \sqrt{w^2 -1}} \cos^2{[\beta(w,\ell)]} e^{-2\frac{\ell^2}{\ell_{{\mathrm{t}}}^2} w^2}.
 \label{I4}
\end{eqnarray}
Concerning Eqs. (\ref{C1})--(\ref{C4}) and (\ref{I1})--(\ref{I4}) the following comments are in order:
\begin{itemize}
\item{} the lower limit of integration over $w$ is $1$ in  Eqs. (\ref{C1})--(\ref{C4}) since the asymptotic expansion of Bessel functions implies that 
$k\tau_{0} \geq \ell$, i.e. $w \geq 1$;
\item{} the obtained expressions will be valid for the angular power spectrum 
will be applicable for sufficiently large $\ell$; in practice, as we shall see 
the obtained results are in good agreement with the data in the Doppler region;
\item{} the function $\beta(w,\ell) = \ell \sqrt{w^2 -1} - \ell\arccos{(w^{-1})} - \frac{\pi}{4}$ leads to a rapidly oscillating argument whose effect will be to 
 slow down the convergence of the numerical integration;  
 it is practical, for the present purposes, to replace $\cos^2[\beta(w,\ell)]$ 
 by its average (i.e. $1/2$).
\end{itemize}

In the integrals (\ref{I1}), (\ref{I2}), (\ref{I3}) and (\ref{I4}) the scale $\ell_{\mathrm{t}}$ stems from the finite thickness of the last scattering surface and it is defined as 
\begin{equation}
\ell_{\mathrm{t}}^2 = \frac{1}{4 \sigma^2} \biggl(\frac{\tau_{0}}{\tau_{\mathrm{rec}}}\biggr)^2, \qquad \sigma= 1.49 \times 10^{-2} \frac{\sqrt{ z_{\mathrm{rec}} + z_{\mathrm{eq}}} + \sqrt{z_{\mathrm{rec}}}}{\sqrt{z_{\mathrm{rec}} + 
z_{\mathrm{eq}}}}.
\label{ELLT}
\end{equation}
Furthermore, within the present approximations, 
\begin{equation}
\frac{\ell^2_{\mathrm{t}}}{\ell^2_{\mathrm{S}}} = \frac{\sigma^2 k_{\mathrm{D}}^2 \tau_{\mathrm{rec}}^2}{\sigma^2 k_{\mathrm{D}}^2 \tau_{\mathrm{rec}}^2 + 1}, \qquad \ell_{\mathrm{A}} \simeq \ell_{\mathrm{S}}
\end{equation}
To simplify further the obtained expressions we can also change 
variable in some of the integrals. Consider, as an example, the 
integrals appearing in the expression of $C_{1}(\ell)$ (see Eq. (\ref{C1})).
Changing the variable of integration as $w = y^2 + 1$ we will have that 
\begin{equation}
C_{1}(\ell) = \int_{0}^{\infty} \overline{I}_{1}(y,\ell) \overline{{\mathcal U}}_{1}(\ell,y) dy,
\label{EXIN1}
\end{equation}
where, in explicit terms and after the change of variables,
\begin{equation}
\overline{I}_{1}(y,\ell) = \frac{4\pi}{(y^2 + 1)^2 \sqrt{y^2 + 2}} e^{- 2\frac{\ell^2}{\ell_{\mathrm{t}}^2} (y^2 + 1)^2},
\label{EXIN2}
\end{equation}
and 
\begin{eqnarray}
\overline{{\mathcal U}}_{1}(\ell, y) &=& {\mathcal A}_{\zeta} \biggl(\frac{k_{0}}{k_{\mathrm{p}}}\biggr)^{n -1}\, \ell^{n-1}L_{\zeta}^{2}(\ell,y) (y^2 +1 )^{n -1} + \overline{\Omega}^2_{\mathrm{B L}} 
\biggl(\frac{k_{0}}{k_{\mathrm{L}}}\biggr)^{2\epsilon} \,\ell^{2\epsilon} {\mathcal F}(\epsilon)L_{\mathrm{B}}^{2}(\ell,y) (y^2 +1 )^{2\epsilon} 
\nonumber\\
&+& 2 L_{\mathrm{B}}(\ell,y)L_{\zeta}(\ell,y) \sqrt{{\mathcal A}_{\zeta}}\, \overline{\Omega}_{\mathrm{B L}}\,
\ell^{(n -1)/2 + \epsilon}
\sqrt{{\mathcal F}(\epsilon)}
\biggl(\frac{k_{0}}{k_{\mathrm{p}}}\biggr)^{(n -1)/2} \biggl(\frac{k_{0}}{k_{\mathrm{L}}}\biggr)^{\epsilon}  (y^2 +1 )^{(n -1)/2 + \epsilon}.
\label{EXIN3}
\end{eqnarray}
In Eq. (\ref{EXIN3}) the explicit dependence of the functions 
$L_{\mathrm{B}}(\ell,y)$ and $L_{\zeta}(\ell,y)$ upon $y$ can be simply deduced from the analog expressions in terms of $w$:
\begin{eqnarray}
&& L_{\zeta}(\ell,y) = - \frac{1}{6} \biggl(1 - \frac{1}{3c_{\mathrm{sb}}^2}\biggr)
\ln{\biggl[\frac{14}{\ell (y^2 + 1)} \frac{\tau_{0}}{\tau_{\mathrm{eq}}}\biggr]},
\nonumber\\
&& L_{\mathrm{B}}(\ell,y)  = \frac{3 R_{\gamma}}{20} \biggl(1 - \frac{1}{3 c_{\mathrm{sb}}^2} \biggr) + \biggl[ \frac{{\mathcal G}(\epsilon)}{{\mathcal F}(\epsilon)} - \frac{1}{4} \biggr].
\label{EXIN4}
\end{eqnarray}
With similar manipulations it is possible to transform also all the other 
integrands appearing in Eqs. (\ref{C2}), (\ref{C3}) and (\ref{C4}).

\renewcommand{\theequation}{5.\arabic{equation}}
\setcounter{equation}{0}
\section{Calculation of the temperature autocorrelations}
\label{sec5}
So far the necessary ingredients for the estimate of the magnetized
temperature autocorrelations have been sorted out. 
In particular the angular power spectrum has been computed semi-analytically 
in the two relevant regions, i.e. the Sachs-Wolfe regime (corresponding 
to large angular scales and $\ell \leq 30$) and the Doppler region, i.e.
$\ell > 100$. Furthermore,  for the nature of the approximations made we do not expect 
the greatest accuracy of the algorithm in the intermediate region (i.e. $30 <\ell < 100$).
Indeed, it was recognized 
already in the absence of magnetic fields that it is somehow necessary to 
smooth the joining of the two regimes by assuming an interpolating form 
of the metric fluctuations that depends upon two fitting parameters \cite{hu1,hu2}.
We prefer here to stress that this method is inaccurate in the matching regime 
since the spherical Bessel functions have been approximated for large $\ell$. Therefore, the 
comparison with experimental data should be preferentially conducted, for the present purposes,
in the Doppler region.
The strategy adopted in the present section is, therefore, the following:
\begin{itemize}
\item{} by taking a concordance model as a starting point, 
the shape and amplitude of the Doppler oscillations will be analyzed when 
the amplitude and spectral slope of the stochastic field are allowed 
to vary;
\item{} constraints can then be derived  from the temperature autocorrelations induced 
by the simultaneous presence of the standard adiabatic mode 
and of the stochastic magnetic field.
\end{itemize}

Before plunging into the discussion, it is appropriate to comment on the choice 
of the cosmological parameters that will be employed throughout 
this section. The WMAP 3-year \cite{WMAP1} 
data have been combined, so far, 
with various sets of data. These data sets include the 2dF 
Galaxy Redshift Survey \cite{2dF}, the combination of Boomerang
and ACBAR data \cite{AC1,AC2}, the combination of CBI and VSA 
data \cite{CB1,CB2}. Furthermore the WMAP 3-year 
data can be also combined with the Hubble Space Telescope 
Key Project (HSTKP) data \cite{HSTKP} as well as with 
the Sloan Digital Sky Survey (SDSS) \cite{SDSS1,SDSS2} data.
Finally, the WMAP 3-year data can be also usefully combined  with 
the weak lensing data \cite{WL1,WL2} and with theobservations of  type Ia supernovae
\footnote{In particular the data of the Supernova Legacy Survey (SNLS) \cite{SNLS} and 
the so-called Supernova "Gold Sample" (SNGS) \cite{SNGS1,SNGS2}.}(SNIa).  
\begin{figure}
\begin{center}
\begin{tabular}{|c|c|}
      \hline
      \hbox{\epsfxsize = 7.6 cm  \epsffile{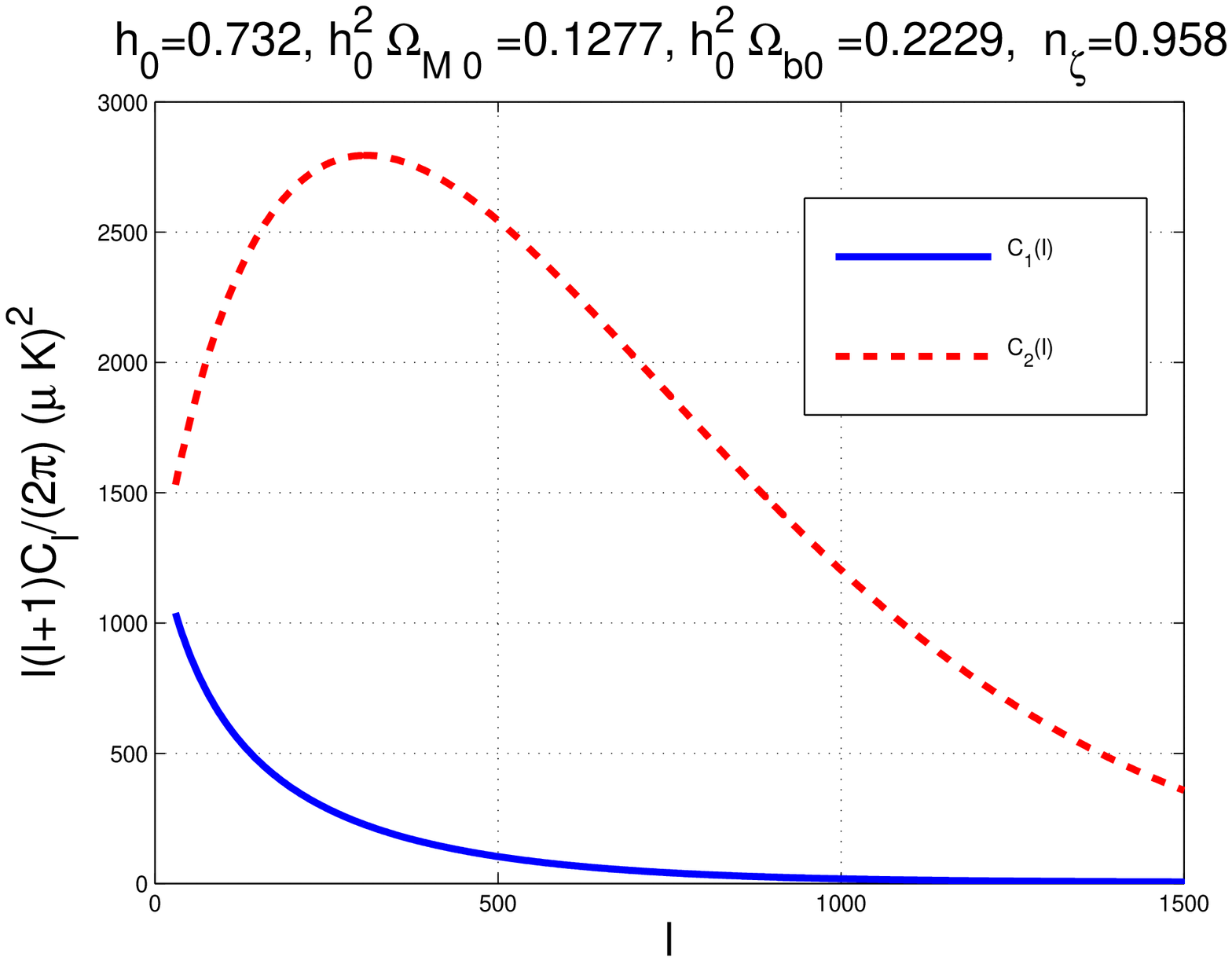}} &
      \hbox{\epsfxsize = 7.6 cm  \epsffile{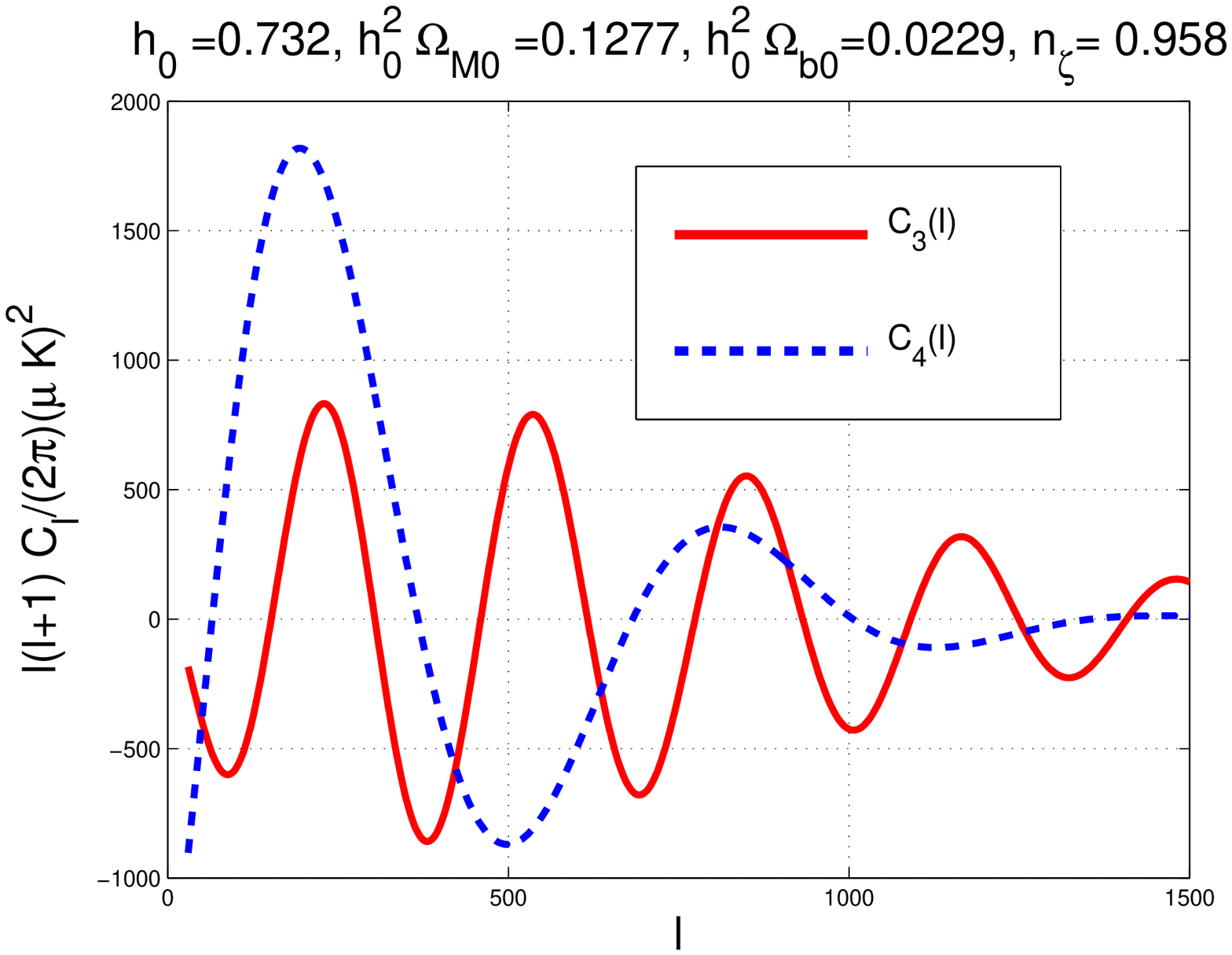}}\\
      \hline
\end{tabular}
\end{center}
\caption[a]{The contribution of each of the integrals giving the 
temperature autocorrelations is reported (see Eq. (\ref{CELLFIN})). The parameters 
are chosen in such a way to match the best fit to the experimental points when only the WMAP data 
are included and the contribution of the tensor modes is assumed to be vanishing. }
\label{Figure1}
\end{figure}
Each of the data sets mentioned in the previous paragraph can be 
analyzed within different frameworks.
The minimal $\Lambda$CDM model with no cut-off in the 
primordial spectrum of the adiabatic mode and with 
vanishing contribution of tensor modes is the simplest concordance 
framework. This is the one that has been adopted in this paper.
Diverse completions of this minimal model 
are possible: they include the addition of the tensor modes, a sharp 
cut-off in the spectrum and so on and so forth.  One of the conclusions
of the present study is that the observational cosmologists 
may also want to include, in their analyses, the possibility 
of pre-recombination large-scale magnetic fields.

All these sets of data (combined with different theoretical models) 
lead necessarily to slightly 
different determinations of the relevant cosmological parameters
To have an idea of the range of variations of the parameters 
the following examples are useful\footnote{The values quoted for all 
the cosmological observables always refer to the case 
of a spatially flat Universe where the semi-analytical calculation 
has been performed.}:
\begin{itemize}
\item{} the WMAP 3-year data alone \cite{WMAP1} (in a $\Lambda$CDM 
framework) seem to favour a slightly 
 smaller value $h_{0}^2 \Omega_{\mathrm{M}0}= 0.127$;
\item{} if the WMAP 3-year data are combined with the "gold" sample
of SNIa \cite{SNGS1} (see also \cite{SNGS2}) the favoured value is $h_{0}^2 \Omega_{\mathrm{M}0}$ is of the order of $0.134$; if 
the WMAP 3-year data are combined with {\em all} the data sets 
$h_{0}^2 \Omega_{\mathrm{M}0} = 0.1324$.
\item{}  similarly, if the WMAP data alone are considered, 
the preferred value of $h_{0}^2 \Omega_{\mathrm{b}0}$ 
is $0.02229$ while this value decreases to $0.02186$ if the WMAP 
data are combined with all the other data sets.
\end{itemize}
The aforementioned list of statements refers to the case of a pure 
$\Lambda$CDM model. If, for instance, tensors are included, then
the WMAP 3-year data combined with CBI and VSA increase a bit 
the value of $h_{0}^2 \Omega_{\mathrm{b}0}$ which becomes, in this 
case closer to $0.023$. 
While in the future it might be interesting to include pre-recombination 
magnetic fields also in non-minimal $\Lambda$CDM scenarios, here 
the logic will be to take a best fit model to the WMAP data alone, compare it 
with the numerical scheme proposed in this paper, and, consequently, 
assess the accuracy of the semi-analytical method. Once this step will be concluded the effects 
stemming from the presence of the magnetic fields will be carefully analyzed.
\begin{figure}
\begin{center}
\begin{tabular}{|c|c|}
      \hline
      \hbox{\epsfxsize = 7.6 cm  \epsffile{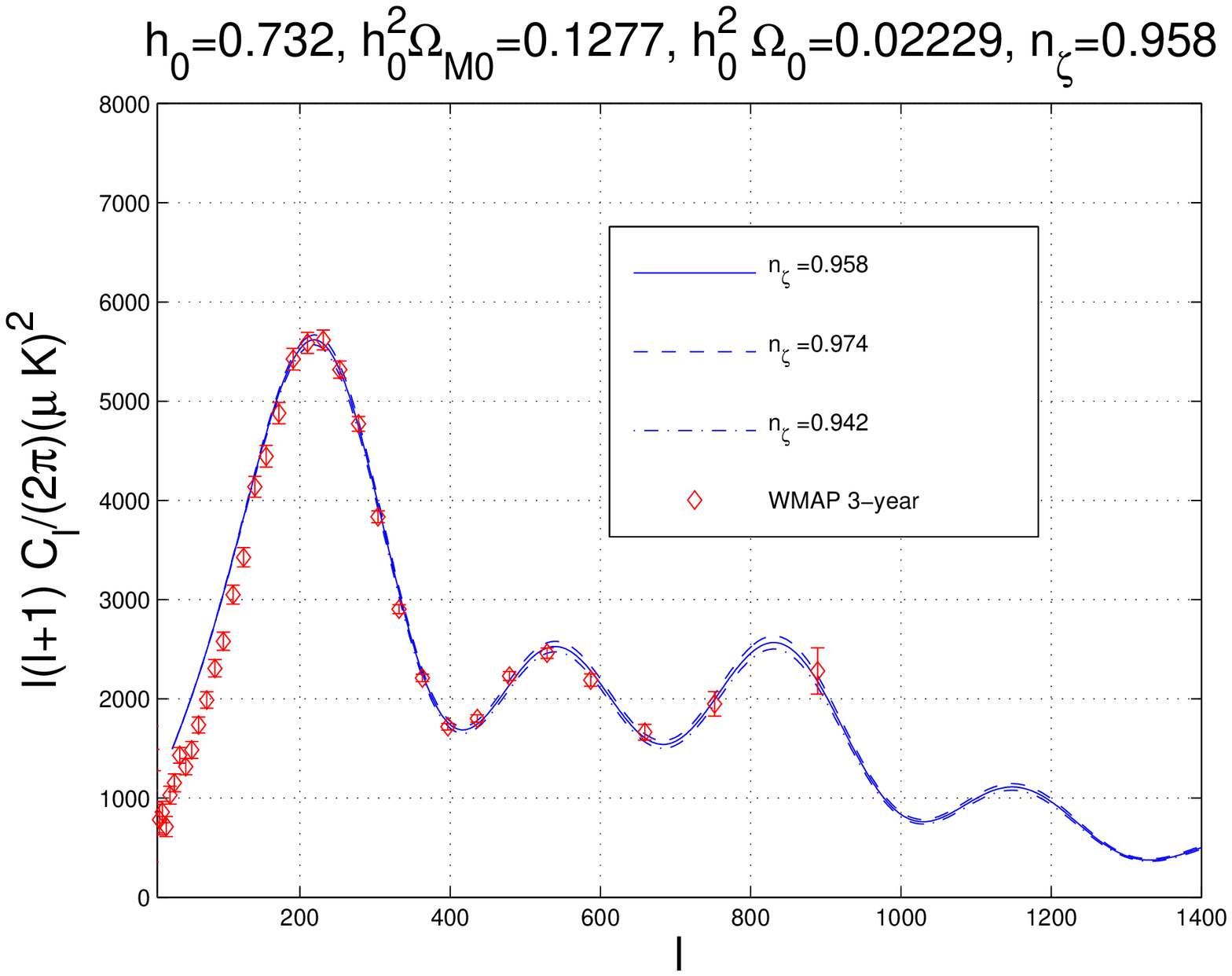}} &
      \hbox{\epsfxsize = 7.6 cm  \epsffile{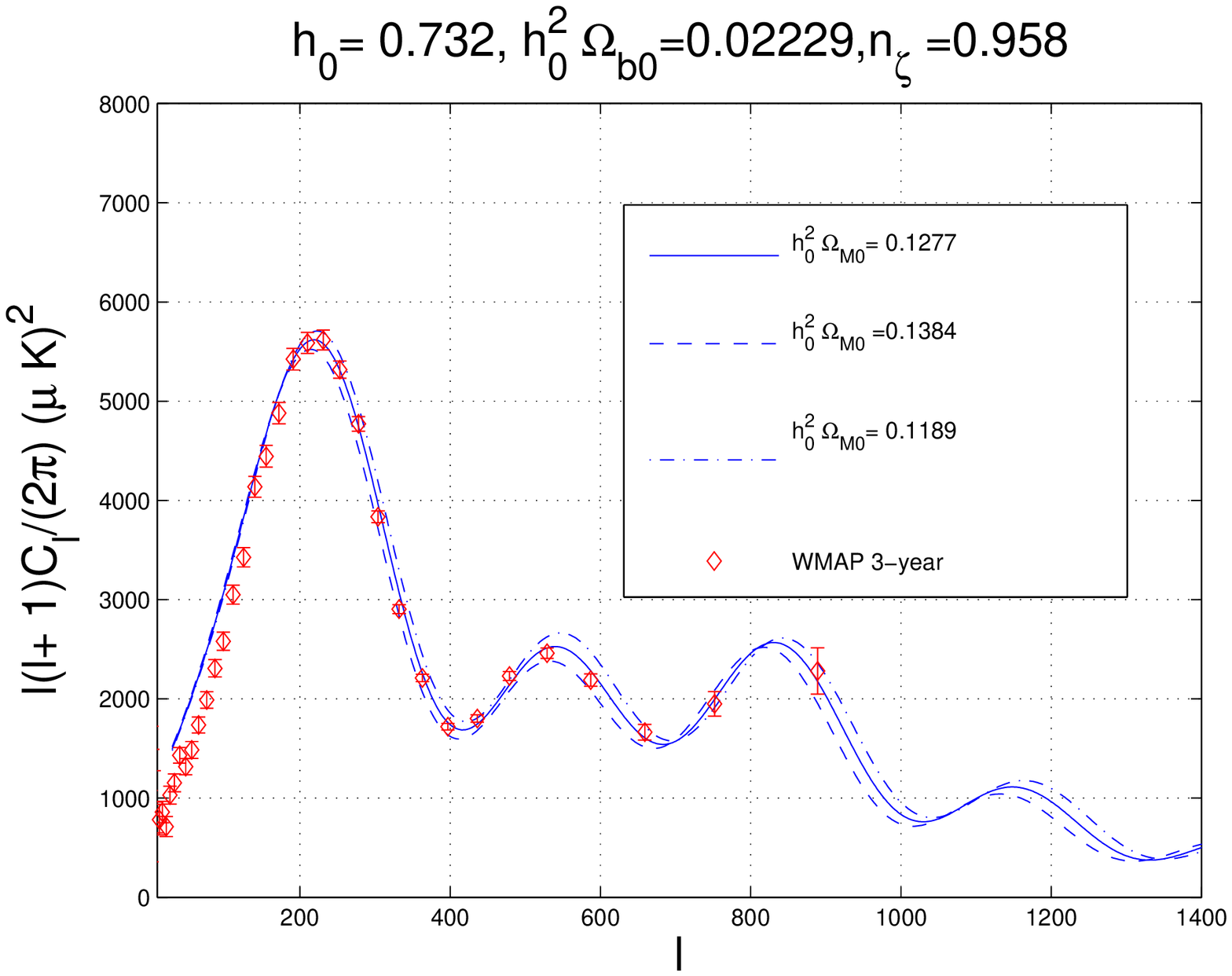}}\\
      \hline
\end{tabular}
\end{center}
\caption[a]{The temperature autocorrelations for a fiducial set 
of cosmological parameters chosen within a concordance model
and in the case $B_{{\mathrm{L}}}=0$.}
\label{Figure2}
\end{figure}
Consider, therefore, the case 
when the magnetic field vanishes (i.e. $B_{\mathrm{L}}=0$) in a  $\Lambda$CDM model with no tensors. 
In Fig. \ref{Figure1} the contribution of each of the 
integrals appearing in Eq. (\ref{CELLFIN}) is illustrated. 
The analytical form of these
integrals has been derived in Eqs. (\ref{C1}), (\ref{C2}), (\ref{C3}) and (\ref{C4}). In Fig. \ref{Figure1} (plot at the left) the separate contributions 
of $\ell(\ell+1) C_{1}(\ell)/(2\pi)$ and of $\ell(\ell+1) C_{2}(\ell)/(2\pi)$ 
have been reported for a fiducial set of parameters (i.e. $n_{\zeta} =0.958$, 
$h_{0}^2 \Omega_{\mathrm{M}0}=0.1277$ and $h_{0}^2 \Omega_{\mathrm{b}0}=0.0229$). 
This fiducial set of parameters corresponds to the best fit of the WMAP 3-year data alone \cite{WMAP1}.
As mentioned in Eq. (\ref{kscales}) the pivot wave-number is $k_{\mathrm{p}}
=0.002\, \mathrm{Mpc}^{-1}$. This is also the choice made by WMAP team.
In the plot at the right (always in Fig. \ref{Figure1}) the separate 
contributions of $\ell(\ell+1) C_{3}(\ell)/(2\pi)$ and of $\ell(\ell+1) C_{4}(\ell)/(2\pi)$ is illustrated for the same fiducial set of parameters (which is also described at the top of the plot).
The various contributions are expressed in units of $(\mu \mathrm{K})^2$ 
(i.e. $1\mu \mathrm{K} = 10^{-6} \mathrm{K}$) which are 
the appropriate ones for the comparison with the data. The normalization 
of the calculation is set by evaluating (analytically) the large-scale 
contribution for $\ell <30$ (see Eq. (\ref{SWP})) and by comparing it, in this region, with the 
WMAP 3-year data release. 

By summing up the four separate contributions illustrated in Fig. \ref{Figure1}, 
Eq. (\ref{CELLFIN}) allows to determine, for a given choice of cosmological parameters,
the full temperature autocorrelations.
The results, always 
in the absence of magnetized contribution, are reported in Fig. \ref{Figure2}.
In the plot at the left of Fig. \ref{Figure2} the critical fractions of matter and baryons, as well as 
$h_{0}$, are all fixed. The only quantity allowed to vary from one curve to the other is the 
scalar spectral index of curvature perturbations, i.e. $n_{\zeta}$. 
The full line denotes the pivot case $n_{\zeta}=0.958$ (corresponding to the central value for the 
spectral index as determined according to the WMAP data alone). The dashed and dot-dashed lines 
correspond, respectively, to $n_{\zeta}=0.974$ and $n_{\zeta}=0.942$ (which define the 
allowed range of $n_{\zeta}$ since $n_{\zeta} = 0.958 \pm 0.016$ \cite{WMAP1}). 

As already stressed, the regime $\ell < 100$ is only reasonably 
reproduced while the most interesting region, for the present purposes, is 
rather accurate (as the comparison with the WMAP data shows). 
The region of very large $\ell$ (i.e. $\ell > 1200$) is also beyond the treatment 
of diffusive effects adopted in the present paper.
In Figure \ref{Figure2} (plot at the right) the adiabatic spectral 
index is fixed (i.e. $n_{\zeta}= 0.958$) while the total (present) 
fraction of non-relativistic matter is allowed to vary ($h_{0}$ 
and $h_{0}^2 \Omega_{\mathrm{b}0}$ are, again, kept fixed). 
It can be observed that, according to Fig. \ref{Figure2}, 
the amplitude of the first peak increases as the total (dusty) matter fraction decreases.

\begin{figure}
\begin{center}
\begin{tabular}{|c|c|}
      \hline
      \hbox{\epsfxsize = 7.6 cm  \epsffile{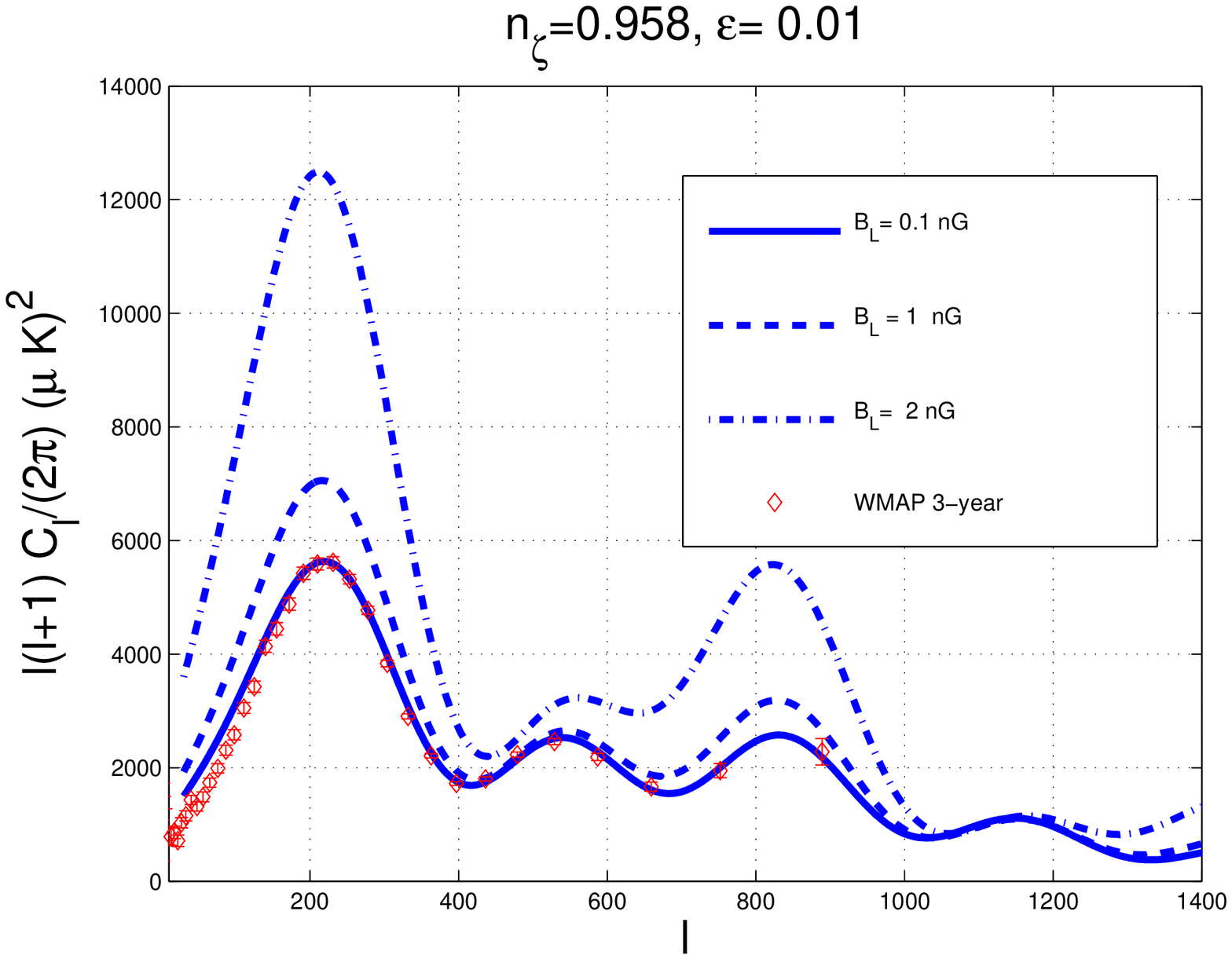}} &
      \hbox{\epsfxsize = 7.6 cm  \epsffile{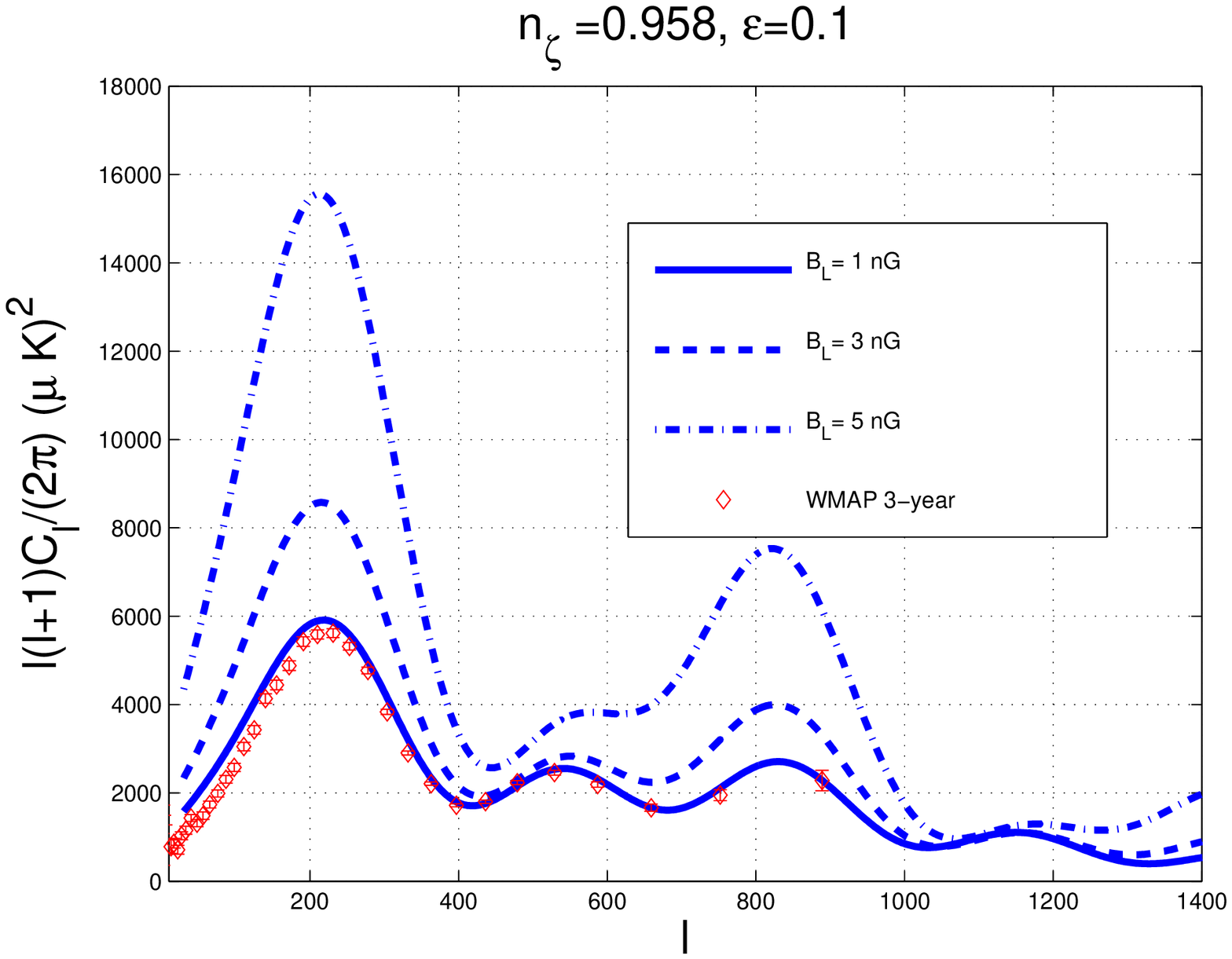}}\\
      \hline
\end{tabular}
\end{center}
\caption[a]{The inclusion of the effects of large-scale magnetic fields
in the case of nearly scale-invariant magnetic energy spectrum (i.e. $0< \epsilon <1$). 
The conventional adiabatic spectral index is fixed to the same value 
assumed in the right plot of Fig. \ref{Figure2}.}
\label{Figure3}
\end{figure}

\begin{figure}
\begin{center}
\begin{tabular}{|c|c|}
      \hline
      \hbox{\epsfxsize = 7.6 cm  \epsffile{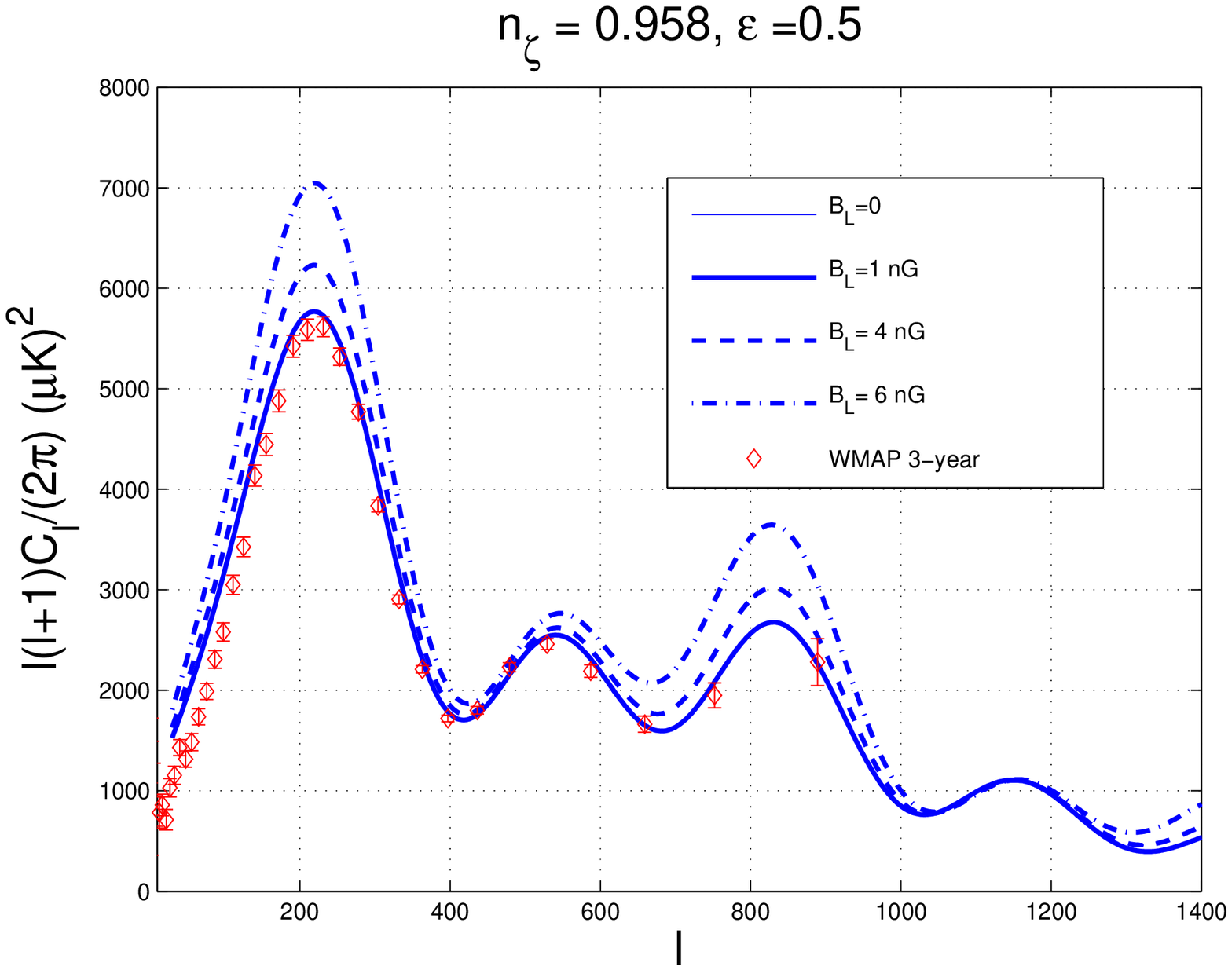}} &
      \hbox{\epsfxsize = 7.6 cm  \epsffile{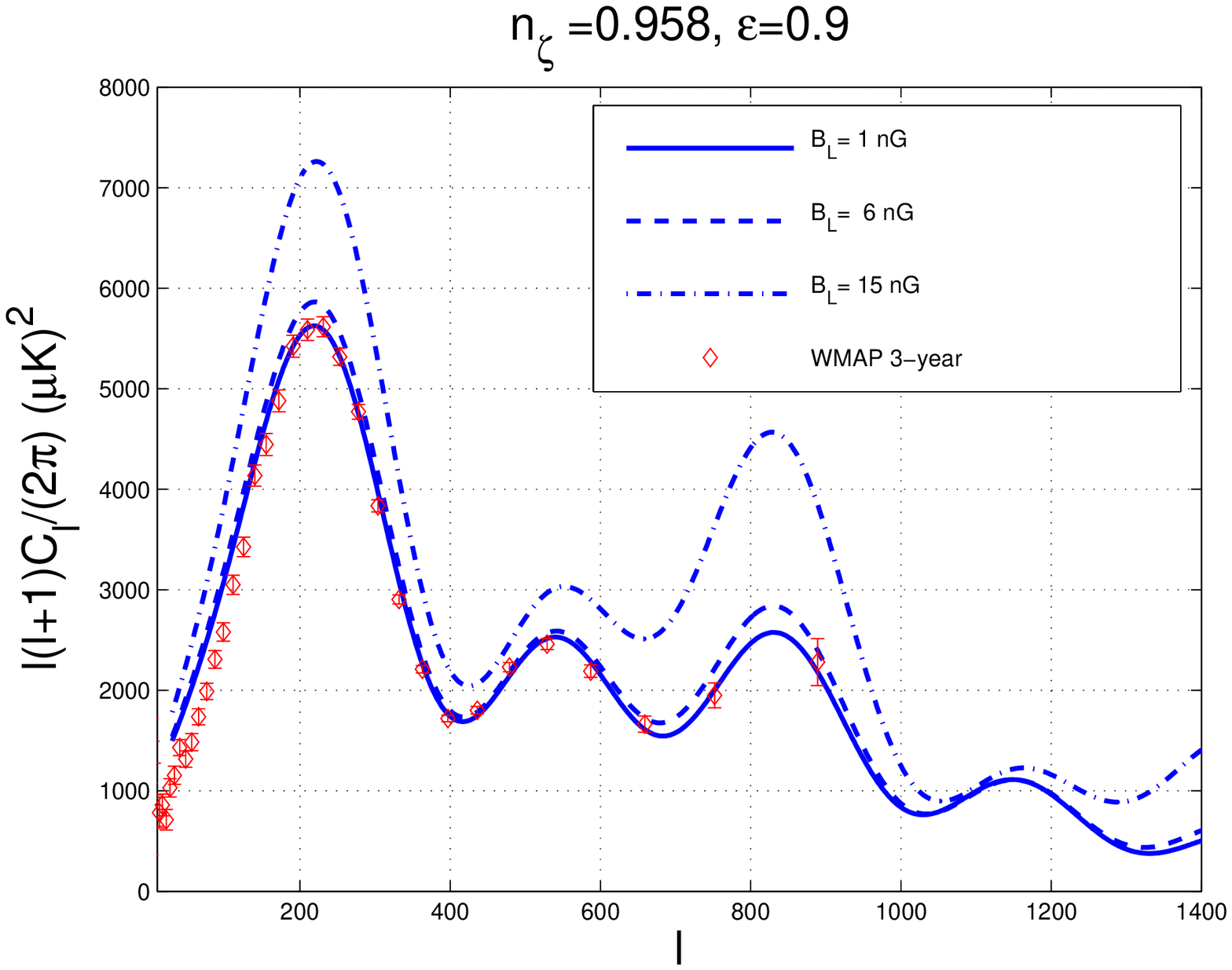}}\\
      \hline
\end{tabular}
\end{center}
\caption[a]{The inclusion of the effects related to laarge-scale 
magnetization in the case of a blue magnetic spectral index.}
\label{Figure4}
\end{figure}

The contribution of the magnetic fields will now be included both in the  Sachs-Wolfe 
region (as discussed in section \ref{sec3}) and in the Doppler region (as discussed in section \ref{sec4}). 
In Fig. \ref{Figure3} the temperature autocorrelations are computed 
in the presence of a magnetized background. The values of the relevant magnetic parameters 
(i.e. the smoothed amplitude of the field $B_{\mathrm{L}}$ and the spectral slope $\epsilon$)
are reported at the top of each plot and in the legends.  In the plot at the 
left of Fig. \ref{Figure3} the spectral slope is fixed as $\epsilon =0.01$ while $B_{\mathrm{L}}$ 
is allowed to vary. The other cosmological parameters are fixed 
to their concordance values stemming from the analysis of the WMAP 
3-year data and are essentially the ones already reported at the top 
of Fig. \ref{Figure2}. The diamonds are the WMAP 3-year data points.
In the plot at the right of Fig. \ref{Figure3} the spectral slope 
is still reasonably flat but, this time, $\epsilon = 0.1$. 
For a spectral slope $\epsilon =0.01$ the case $B_{\mathrm{L}}= 1 \mathrm{nG}$ is barely distinguishable (but not indistinguishable, as we shall see below) 
from the case $B_{\mathrm{L}}=0$. As soon as $B_{\mathrm{L}}$ increases
from $1$ to $5$ nG three different phenomena take place:
\begin{itemize}
\item{} the first Doppler peak increases dramatically and it reaches 
a value of the order of $1.2 \times10^{4}\, (\mu\mathrm{K})^2$
when $B_{\mathrm{L}}= 2$ nG;
\item{} already for $0.1 \mathrm{nG}< B_{\mathrm{L}} < 2 \,\mathrm{nG}$ 
the third peak increases while the second peak becomes less 
pronounced;
\item{} as soon as $B_{\mathrm{L}} \geq 2$ nG the second peak 
practically disappears and it is replaced by a sort of hump.
\end{itemize}
If the spectral slope increases a similar trend takes place as $B_{\mathrm{L}}$ increases.
However, the formation of the hump takes place for values of $B_{\mathrm{L}}$ 
which are comparatively larger than in the case of nearly scale-invariant magnetic energy spectrum. 
 In Fig. \ref{Figure3} (plot at the right) the magnetic spectral slope is $\epsilon = 0.1$ (while the adiabatic spectral slope is fixed to the concordance value, i.e. $n_{\zeta} = 0.984$). 
To observe the formation of the hump (which is of course excluded by experimental data) the values of $B_{\mathrm{L}}$ must be larger and in the range of $15$ to $20$ nG.
As soon as $\epsilon$ increases towards $1$ the minimal 
allowed $B_{\mathrm{L}}$ also increases. This is particularly evident from 
the two plots reported in Fig. \ref{Figure4} where the values of $\epsilon$ have been chosen to be $0.5$ (plot at the left) and $0.9$ (plot at the right).

\begin{figure}
\begin{center}
\begin{tabular}{|c|c|}
      \hline
      \hbox{\epsfxsize = 7.6 cm  \epsffile{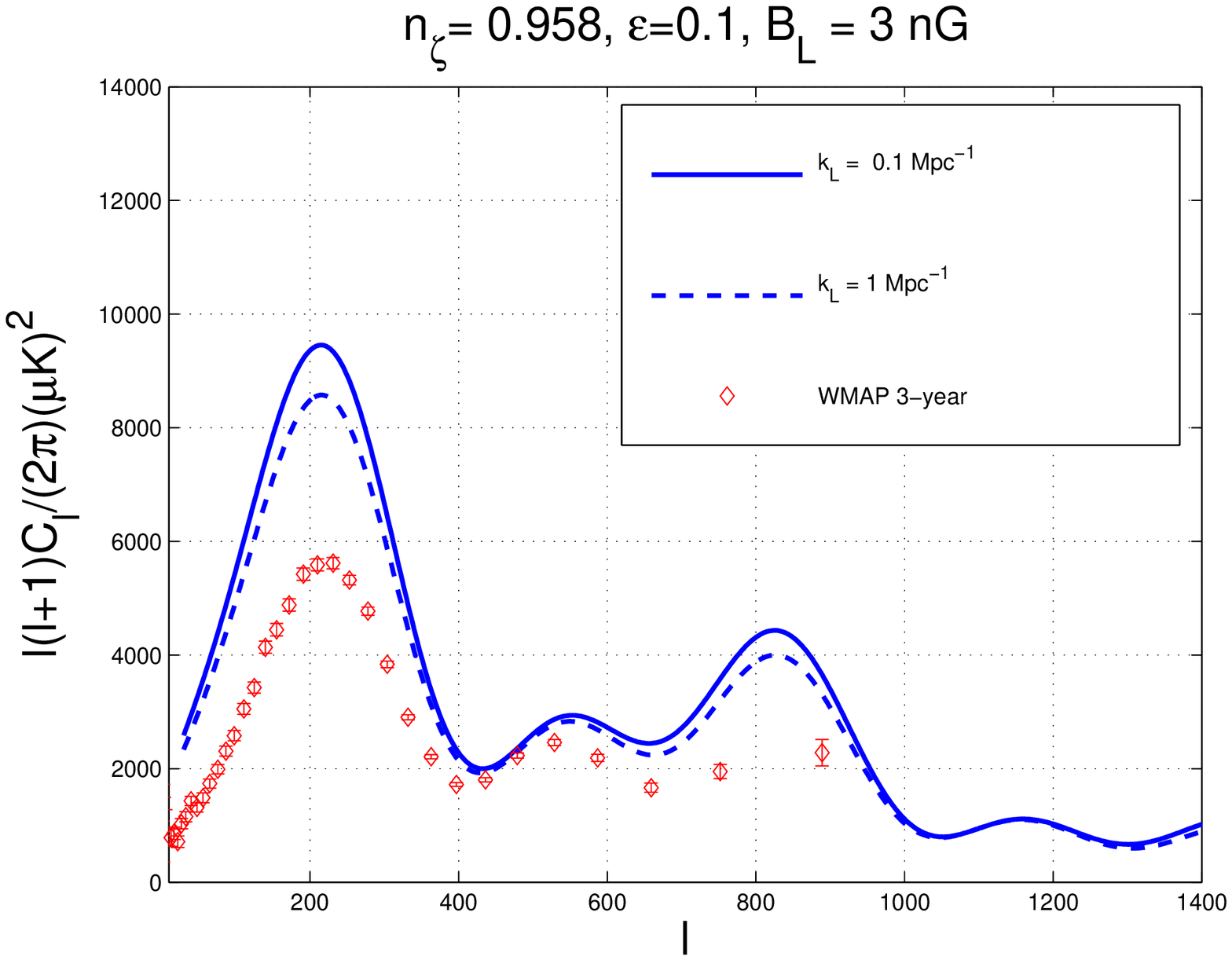}} &
      \hbox{\epsfxsize = 7.6 cm  \epsffile{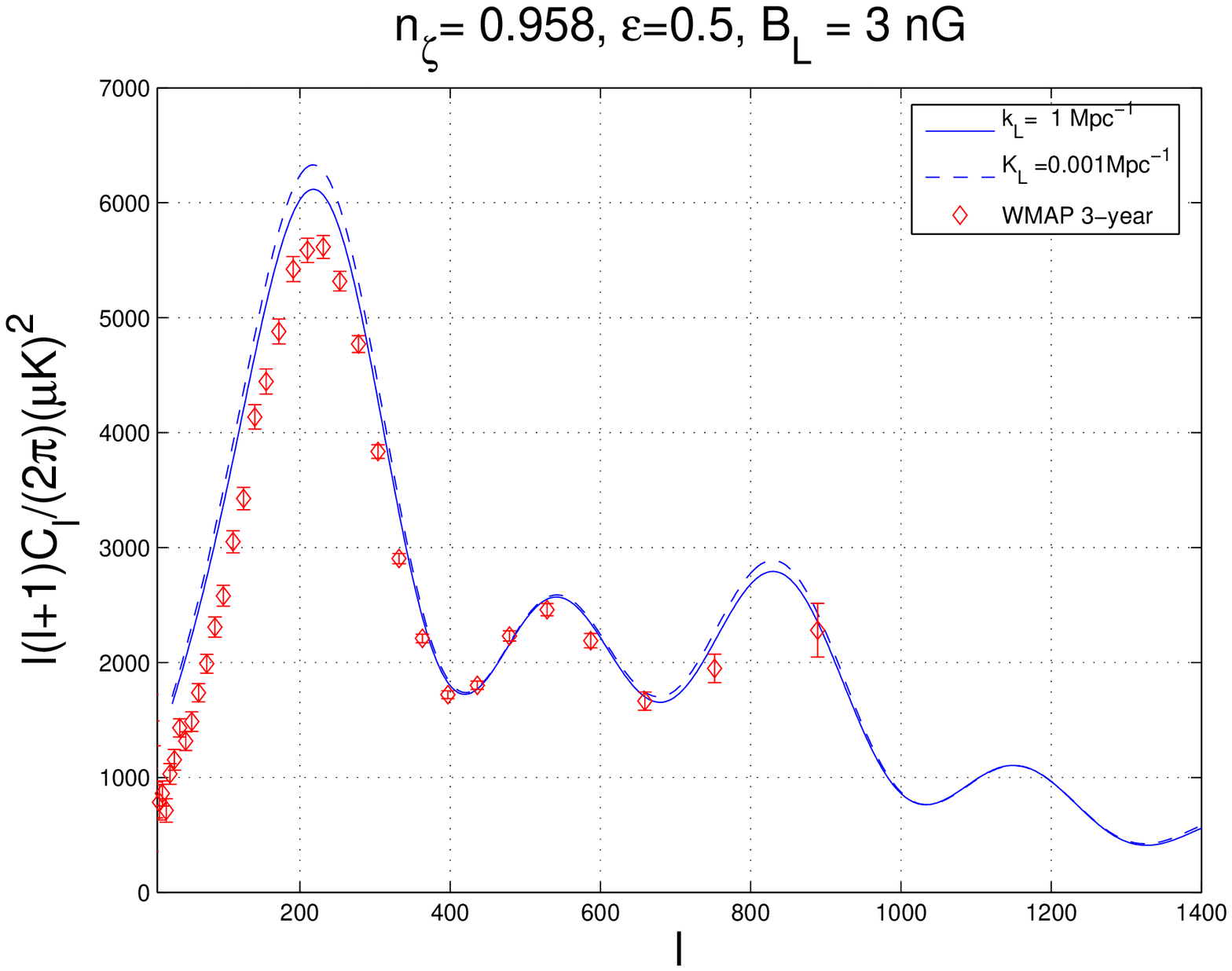}}\\
      \hline
\end{tabular}
\end{center}
\caption[a]{The variation of the magnetic pivot scale is illustrated for two different 
spectral slopes.}
\label{Figure5}
\end{figure}

In Fig. \ref{Figure4} the dashed curve in the plot at the right corresponds 
to $B_{\mathrm{L}} = 6\, \mathrm{nG}$. For this value of $B_{\mathrm{L}}$ 
the hump is not yet present, while for $\epsilon=0.01$ already for $B_{\mathrm{L}}= 2 \mathrm{nG}$ the second peak is completely 
destroyed. These differences are related to the fact that 
an increase in $\epsilon$ implies, indirectly, that the amplitude of the 
power spectrum of the magnetized background decreases at large
length-scales, i.e. for small wave-numbers.
 From Fig. \ref{Figure3} it can be argued, for instance,
that when the magnetic slope is nearly flat  (i.e. $\epsilon \simeq 0.01$), the allowed value 
of the smoothed field becomes $B_{\mathrm{L}} < 0.1 \mathrm{nG}$.
It should be remarked, to avoid confusion, that the scale invariant limit 
for the curvature perturbations, according to the conventions of the present paper is $n_{\zeta}\to 1$ while the scale invariant limit for the magnetic energy density fluctuations is $\epsilon \ll 1$.
Finally in Fig. \ref{Figure5} the effect of the variation of the magnetic pivot 
scale is illustrated. If $k_{\mathrm{L}}$ diminishes by one order of magnitude the temperature autocorrelations increase in a different way depending upon the value of $\epsilon$. By diminishing $k_{\mathrm{L}}$ 
the magnetic field is smoothed over a larger length-scale. The net effect
of this choice will be to increase the temperature autocorrelations
for the same values of $B_{\mathrm{L}}$ and $\epsilon$.

\begin{figure}
\begin{center}
\begin{tabular}{|c|c|}
      \hline
      \hbox{\epsfxsize = 7.6 cm  \epsffile{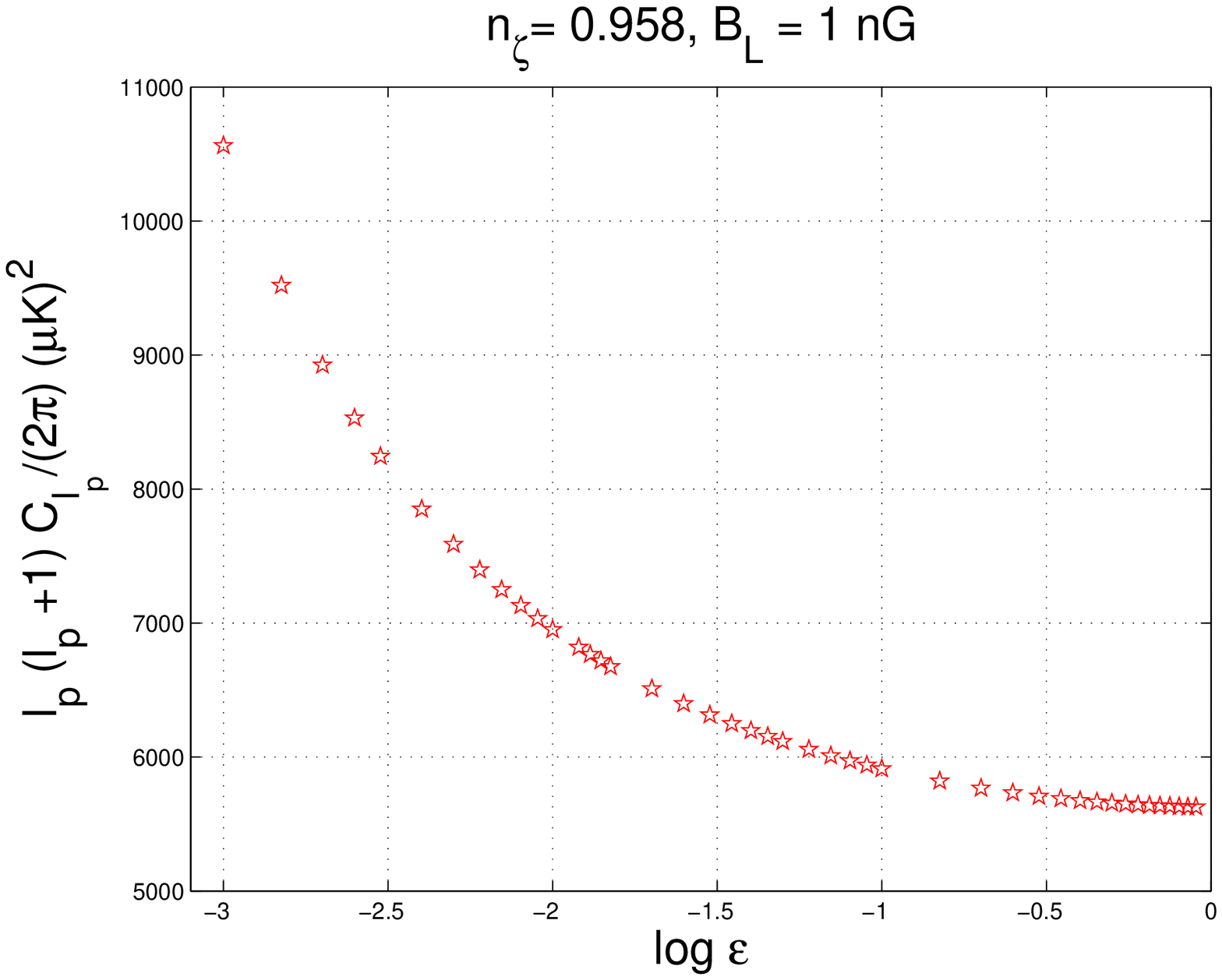}} &
      \hbox{\epsfxsize = 7.6 cm  \epsffile{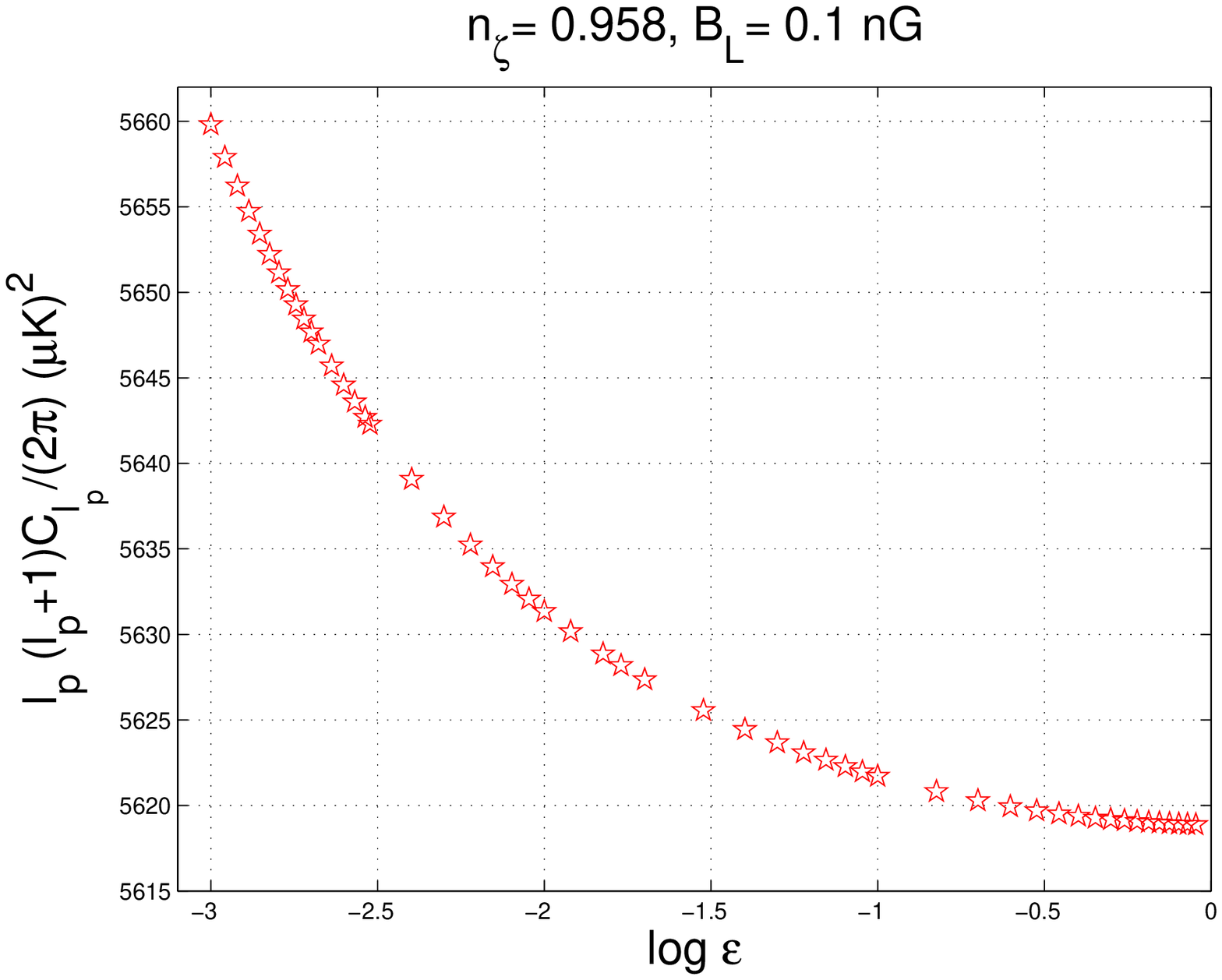}}\\
      \hline
\end{tabular}
\end{center}
\caption[a]{The stars represents the points obtained by numerical integration. On the vertical 
axis, in both plots, the (computed) value of the temperature autocorrelations at the first Doppler 
peak (i.e. $\ell = \ell_{\mathrm{p}}$) is reported as a function 
of the magnetic spectral slope for two values of the smoothed magnetic field intensity.}
\label{Figure6}
\end{figure}

\begin{figure}
\begin{center}
      \epsfxsize = 7.6 cm  \epsffile{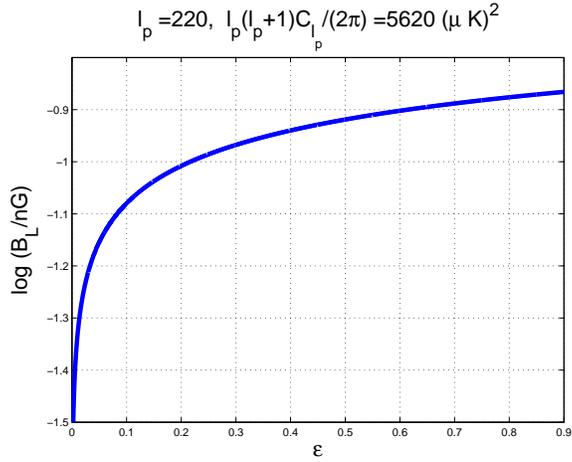}   
\end{center}
\caption[a]{Allowed region (below the thick curve) in the plane $(B_{\mathrm{L}},\epsilon)$ for $\ell_{\mathrm{p}}= 220$ and for $\Delta_{\mathrm{p}} =3 (\mu\mathrm{K})^2$.}
\label{Figure7}
\end{figure}

For $\ell=210$ the experimental value of the temperature  autocorrelations
is \cite{WMAP1,WMAP2,WMAP3} $5586 \pm 106.25 \, (\mu\mathrm{K})^2$, while for $\ell= 231$ the experimental value is $5616.35 \pm 99.94  \, (\mu\mathrm{K})^2$.
The next value, i.e. $\ell = 253$ implies $5318.06 \pm 86.19\, (\mu\mathrm{K})^2$.
By requiring that the addition of the magnetic field does not shift 
appreciably the height of the first Doppler peak it is possible to find, 
for each value of the spectral slope $\epsilon$ a maximal magnetic field 
which  approximately coincides, in the cases of Fig. \ref{Figure3} with the lowest 
curve of each plot.  
This argument is sharpened in Fig. \ref{Figure6} where the starred points represent the computed values of the temperature 
autocorrelations for two different values of $B_{\mathrm{L}}$ and for 
the interesting range of $\epsilon$. The value of $\ell_{\mathrm{p}}$, i.e. the multipole corresponding 
to the first Doppler peak, has been taken, according to \cite{WMAP1,WMAP2,WMAP3} to be 
$220$. If, according to experimental data, the following condition
\begin{equation}
\biggl[\frac{\ell_{\mathrm{p}} (\ell_{\mathrm{p}} +1) C_{\ell_{\mathrm{p}}}}{2\pi}\biggr]_{\mathrm{computed}} \leq 5260 (\mu \mathrm{K})^2
\label{BBOUND}
\end{equation}
is enforced, then, the the smoothed field intensity and the spectral slope will be bounded 
in terms of the position and height of the Doppler peak. This condition is indeed sufficient 
since, according to the numerical results reported in the previous figures, the distortion 
of the second and third peaks are always correlated with the increase of the first 
peak.  Already at a superficial level, it is clear that if $B_{\mathrm{L}} = 1\,\mathrm{nG}$ the 
only spectral slopes compatible with the requirement of Eq. (\ref{BBOUND}) 
are rather blue and, typically $\epsilon >  0.5$. 
The numerical values obtained with the method described in Fig. \ref{Figure6} are well 
represented by the following interpolating formula
\begin{equation}
\biggl[\frac{\ell_{\mathrm{p}} (\ell_{\mathrm{p}} +1) C_{\ell_{\mathrm{p}}}}{2\pi}\biggr]_{\mathrm{computed}}=
 \biggl[\biggl(\frac{B_{\mathrm{L}}}{\mathrm{nG}}\biggr)^2 \frac{c_1}{\log{\epsilon^{c_2} +1}} + 5617\biggr] (\mu \mathrm{K})^2,\qquad c_{1} = 46.71,\qquad c_{2}=0.55,
\label{BBOUND2}
\end{equation}
which holds for $B_{\mathrm{L}}\leq \mathrm{nG}$ a bit less accurate in the region 
$B_{\mathrm{L}} > \mathrm{nG}$ which is already excluded by inspection of the shape 
of the temperature autocorrelations.
By then comparing the value of the temperature autocorrelations in the location of the first
Doppler peak the amplitude of the magnetic field can therefore be bounded. In particular 
it is easy to show that 
\begin{equation}
\biggl(\frac{B_{\mathrm{L}}}{\mathrm{nG}}\biggr)^2 \leq \frac{1}{c_1}
\biggl(\frac{\sqrt{\Delta_{\mathrm{p}}}}{\mu \mathrm{K}}\biggr)^2 \log{[\epsilon^{c_2} +1]},
\label{BBOUND3}
\end{equation}
where 
\begin{equation}
\Delta_{\mathrm{p}} =\biggl[\frac{\ell_{\mathrm{p}} (\ell_{\mathrm{p}} +1) C_{\ell_{\mathrm{p}}}}{2\pi}\biggr]_{\mathrm{measured}} - 5617 (\mu \mathrm{K})^2.
\label{BBOUND4}
\end{equation}
The quantity $\Delta_{\mathrm{p}}$ is known once the experimental determination 
of the height of the peak is available. Consequently, 
by determining experimentally the value of the temperature autocorrelations 
at the first Doppler peak located for a multipole $\ell_{\mathrm{p}}$ the magnetic field 
intensity and the spectral slope will be bounded according to Eq. (\ref{BBOUND4}). 
If, as WMAP data suggest, we take $\Delta_{\mathrm{p}}= 3 (\mu\mathrm{K})^2$, the 
bounds on $B_{\mathrm{L}}$ and $\epsilon$ are illustrated in Fig. \ref{Figure7}.

Thus, according to the results described so far it is possible to say that to avoid gross distortion 
of the temperature autocorrelations attributed to large-scale magnetic fields we have to demand 
that the stochastic field satisfies
\begin{equation}
B_{\mathrm{L}} \leq 0.08 \mathrm{nG}, \qquad 0.001 \leq \epsilon <1
\end{equation}
If a magnetic field with smoothed amplitude $B_{\mathrm{L}} \leq 0.1 \mathrm{nG}$ is present 
before recombination the implication for the formation of magnetized structures are manifold. We 
recall that the value $B_{\mathrm{L}}$ is the smoothed magnetic field redshifted at the epoch 
of the gravitational collapse of the protogalaxy. We know that, during collapse, the 
freezing of magnetic flux justifies the compressional amplification of the pre-existing field that 
will be boosted by roughly four orders of magnitude during the collapse \cite{max1}. This will bring the amplitude 
of the field to the $\mu G$ level. It is however premature to speculate on these 
issues. There are, at the moment, two important steps to be undertaken:
\begin{itemize}
\item{} the forthcoming PLANCK explorer data will allow to strengthen the constraints derived in the present paper and, in particular, the formulae derived in the present section will allow to constraint directly the possible magnetized 
distortions stemming from the possible presence of large-scale magnetic fields;
\item{} another precious set of informations may come from the analysis of the magnetic fields 
in clusters and superclusters; it would be interesting to know, for instance, which is 
the spectral slope of the magnetic fields in galaxies, clusters and superclusters.
\end{itemize}
The other interesting suggestion of the present analysis is that the inclusion of a large-scale magnetic 
field as a fit parameter in an extended $\Lambda$CDM model is definitely plausible. 
The $\Lambda$CDM model has been extended to include, after all, different possibilities 
like the ones arising in the dark-energy sector. Here we have the possibility of adding the parameters 
of a magnetized background which are rather well justified on the physical ground. 
Notice, in particular, that interesting degeneracies can be foreseen. For instance, the increase 
of the first peak caused by a decrease in the dark-matter fraction can be combined with 
the presence of a magnetic field whose effect, as we demonstrated, is to shift the first Doppler peak upwards.
These issues are beyond the scopes of the present paper.
\renewcommand{\theequation}{6.\arabic{equation}}
\setcounter{equation}{0}
\section{Concluding remarks}
\label{sec6}
There are no compelling reasons why large-scale magnetic fields should not be present prior 
to recombination. In this paper, via a semi-analytical approach, the temperature 
autocorrelations induced by large-scale magnetic fields have been computed and 
confronted with the available experimental data. Of course the data analysis 
can be enriched by combining the WMAP data also with other data sets and by 
checking the corresponding effects of large-scale magnetic fields. The main spirit of this 
investigation was, however, not to discuss the analysis of data but to show 
that the effects of large-scale magnetic fields on the temperature 
autocorrelations can be brought at the same theoretical standard of 
the calculations that are usually performed in the absence of magnetic fields.

According to this perspective it is interesting to notice that, at the level of the pre-equality 
initial conditions, the presence of magnetic field induces a quasi-adiabatic mode. Depending 
on the features of the magnetic  spectrum (i.e. its smoothed amplitude $B_{\mathrm{L}}$ and 
its spectral slope $\epsilon$), 
possible distortions of the first and second peaks can jeopardize the shape 
of the observed temperature autocorrelations. In particular, for sufficiently strong 
magnetic backgrounds (i.e. $B_{\mathrm{L}} > 10 \mathrm{nG}$ and $\epsilon \leq 0.3$),
the second peak turns into a hump. From the analysis of these distortions it was possible 
to derive a bound that depends solely upon measurable quantities such as the location 
of the first peak and its height. The derived formulae will allow a swifter comparison of the possible effects 
of large-scale magnetic fields with the forthcoming experimental data such as the ones 
of PLANCK explorer. The available WMAP data suggest that 
$B_{\mathrm{L}} \leq 0.08 \mathrm{nG}$ for  $0.001 \leq \epsilon <1$. 
This range of parameters does not exclude that magnetic fields present prior to recombination 
could be the seeds of magnetized structures in the sky such as 
galaxies, clusters and superclusters. It is also interesting to remark that the allowed 
range of parameters does not exclude the possibility that the magnetic field of galaxies 
is produced from the pre-recombination field even without a strong dynamo action 
whose possible drawbacks and virtues are, at present, a subject of very interesting 
debates.

In recent years CMB data have been confronted with a variety of cosmological scenarios that 
take as a pivotal model the $\Lambda$CDM paradigm. Some of the parameters usually added 
encode informations stemming from effects that, even if extremely interesting, arise at very high 
energy and curvature scales. While it is certainly important to test any predictive cosmological scenario, we would 
like to stress that the purpose of the present work is, in some sense, more modest. We hope to learn 
from CMB not only what was the initial state of the Universe when the Hubble rate was only one millionth times 
smaller than the Planck (or string) mass scale; 
if possible we would like to learn from CMB how and why the largest
magnetized structures arose in the sky. 
Since we do see magnetic fields today it is definitely 
a well posed scientific question to know what were their effects 
prior to recombination.  It would be desirable, for instance, to find clear evidence of the absence 
of pre-recombination magnetic fields. It would be equally exciting to determine the possible 
presence of this natural component. 
It is therefore opinion of the author that the inclusion of a magnetized component 
in future experimental studies of CMB observables represents a physically motivated option which we do hope 
will be seriously considered by the various collaborations which are today active in experimental 
cosmology.
\newpage
\begin{appendix}
\renewcommand{\theequation}{A.\arabic{equation}}
\setcounter{equation}{0}
\section{Magnetized gravitational perturbations}
\label{APPA}
In this Appendix the evolution equations of the magnetized 
curvature perturbations will be presented in the conformally 
Newtonian gauge where the only two non vanishing 
components of the perturbed metric are 
\begin{equation}
\delta_{\mathrm{s}} g_{00}(\tau,\vec{x})= 2 a^2(\tau) \phi(\tau,\vec{x}),\qquad \delta_{\mathrm{s}} g_{ij}(\tau,\vec{x})= 2 a^2(\tau) \psi(\tau,\vec{x}) \delta_{ij},
\end{equation}
where $\delta_{\mathrm{s}}$ signifies the scalar nature of the fluctuation.
 While these equations are available in the literature \cite{mg1,mg3},
it seems appropriate  to give 
here an explicit and reasonably self-contained  treatment 
of some technical tools that constitute the basis of the 
results reported in the bulk of the paper. 

The magnetic fields are here treated in the magnetohydrodynamical (MHD) approximation where the 
displacement current is neglected and where the three 
dynamical fields of the problem (i.e., respectively, the 
magnetic field, the Ohmic electric field and the total 
Ohmic current) are all solenoidal.
The bulk velocity field, in this approach, is given by the centre 
of mass velocity of the electron-proton system. This 
is physically justified since electrons and protons are 
strongly coupled by Coulomb scattering. 
Photons and baryon are also strongly coupled 
by Thompson scattering, 
at least up to recombination which is the relevant time-scale 
for the effects of magnetic fields on temperature 
autocorrelations. The bulk velocity of the plasma 
can be separated into an irrotational part and into a rotational 
part which contributes to the evolution of the vector modes of the geometry \cite{mg2}. In the present investigation only the scalar modes are treated and, therefore, only the irrotational 
part of the velocity field will be relevant. 
In the MHD approach the magnetic fields enter, both, the perturbed Einstein equations and the Boltzmann hierarchy. 

\subsection{Perturbed Einstein equations}
The perturbed Einstein equations are affected by the 
various components of the (perturbed) 
energy-momentum. The contribution of the magnetic fields 
to the scalar fluctuations of the energy-momentum tensor are:
\begin{equation}
\delta_{\mathrm{s}} {\mathcal T}_{0}^{0} = \delta \rho_{\mathrm{B}} = \frac{B^{2} (\tau,\vec{x})}{8\pi a^4(\tau)},
\qquad \delta_{\mathrm{s}} {\mathcal T}_{i}^{j} = - \delta p_{\mathrm{B}} \delta_{i}^{j} + \tilde{\Pi}_{i}^{j},\qquad \delta p_{\mathrm{B}} =  \frac{\delta \rho_{\mathrm{B}}}{3},
\end{equation}
where 
\begin{equation}
\tilde{\Pi}_{i}^{j} = \frac{1}{4\pi a^4} \biggl[ B_{i}B^{j} - \frac{1}{3} B^2 \delta_{i}^{j} \biggr],
\end{equation}
is the magnetic anisotropic stress. Using the practical 
notation\footnote{In the conformally flat parametrization 
adopted in the present paper (see also Eq. (\ref{LEL})), 
$\nabla^2= \partial_{i}\partial^{i}$ is just the conventional Laplacian. If the spatial geometry would be 
curved, $\nabla^2$ will be defined in terms of the appropriate spatial geometry. The analysis 
of open or closed Universes is, however, not central for the present analysis (see first and second 
paragraph of section \ref{sec2}). Notice, furthermore, that, as in the bulk of the paper, the prime 
denotes a derivation with respect to the conformal time coordinate $\tau$.}
is  $\partial_{i} \partial^{j} \tilde{\Pi}_{j}^{i} = (p_{\gamma} + \rho_{\gamma}) \nabla^2
\sigma_{\mathrm{B}}$ the spatial (and traceless) components 
of the perturbed Einstein equations imply 
\begin{equation}
 \nabla^{4} (\phi - \psi) = 12 \pi G a^2 [ (p_{\nu} + \rho_{\nu}) \nabla^2 \sigma_{\nu} + (p_{\gamma} + \rho_{\gamma}) \nabla^2\sigma_{\mathrm{B}}],
 \label{anisstrein}
 \end{equation}
where $\sigma_{\nu}$ is the neutrino anisotropic stress. The Hamiltonian and momentum constraints can then be written as 
\begin{eqnarray}
&& \nabla^2 \psi - 3 {\mathcal H} ( {\mathcal H} \phi + \psi') = 4 \pi G a^2 ( \delta \rho_{\mathrm{t}} + \delta \rho_{\mathrm{B}}) ,
\label{HAM}\\
&& \nabla^2 ( {\mathcal H} \phi + \psi') = - 4\pi G a^2 (p_{\mathrm{t}} + \rho_{\mathrm{t}}) \theta_{\mathrm{t}},
\label{MOM}
\end{eqnarray}
where $\delta \rho_{\mathrm{t}} = \sum_{a} \delta \rho_{a} $ is the total density fluctuation (with the sum running over the four 
species of the plasma, i.e. photons, baryons, neutrinos and 
CDM particles). Equations (\ref{HAM}) and (\ref{MOM}) are simply derived from the 
perturbed components of the $(00)$ and $(0i)$ Einstein equations \cite{mg1,mg3} (see also 
\cite{MAXG1} for a comparison with the conventional situation where magnetic fields are absent).
Notice that the MHD Pointying vector has not been included in the momentum constraint.
The rationale for this approximation stems from the fact that this contribution is proportional
to $\vec{\nabla}\cdot(\vec{E}\times \vec{B})$ and it contains one electric field which is suppressed, in MHD,
by one power of $\sigma_{\mathrm{c}}$, i.e. the Ohmic conductivity.

In Eq. (\ref{MOM}) $\theta_{\mathrm{t}}$ is the three-divergence 
of the total peculiar velocity, i.e. 
\begin{equation}
(p_{\mathrm{t}} + \rho_{\mathrm{t}}) \theta_{\mathrm{t}} = 
\frac{4}{3} \rho_{\nu} \theta_{\nu} + \rho_{\mathrm{c}} \theta_{\mathrm{c}} + \frac{4}{3} \rho_{\gamma} ( 1 + R_{\mathrm{b}}) \theta_{\gamma\mathrm{b}},
\label{totth}
\end{equation}
where $\theta_{\gamma\mathrm{b}}$ is the baryon-photon 
velocity and $R_{\mathrm{b}}$ is baryon-photon ratio
defined in Eq. (\ref{Rbdef}). In particular, at recombination,
\begin{equation}
R_{\mathrm{b}}(z_{\mathrm{rec}}) = 0.664 \biggl(\frac{h_{0}^2 \Omega_{\mathrm{b}0}}{0.023}\biggr) \biggl(\frac{1051}{z_{\mathrm{rec}} +1  }\biggr).
\label{Rb}
\end{equation}
Finally from the spatial components of the perturbed Einstein equations we get 
\begin{equation}
 \psi'' + {\mathcal H} (\phi' + 2 \psi') + ( 2 {\mathcal H}' + {\mathcal H}^2) \phi + \frac{1}{3} \nabla^2 (\phi - \psi) = 4\pi G a^2 ( \delta p_{\mathrm{t}} + \delta p_{\mathrm{B}}).
\label{psidouble}
\end{equation}
The evolution equations of the metric fluctuations can be also usefully supplemented by the covariant 
conservation of the total density fluctuation of the fluid which can be written as 
\begin{equation}
\delta \rho_{\mathrm{t}}' + 3 {\mathcal H} ( \delta \rho_{\mathrm{t}} + \delta p_{\mathrm{t}}) + (p_{\mathrm{t}} + \rho_{\mathrm{t}}) \psi'   + (p_{\mathrm{t}} + \rho_{\mathrm{t}}) \theta_{\mathrm{t}}=0.
\label{TOTCONS1}
\end{equation}
This form of the total conservation equation allows to find rather swiftly the evolution equations 
of the gauge-invariant density contrast $\zeta$.
The evolution equation of the total velocity field of the mixture can also
be obtained from the covariant conservation of the total energy-meomentum tensor 
\begin{equation}
(p_{\mathrm{t}} + \rho_{\mathrm{t}}) \theta_{\mathrm{t}} + [ (p_{\mathrm{t}}' + \rho_{\mathrm{t}}') + 4 {\mathcal H}( p_{\mathrm{t}} + \rho_{\mathrm{t}})] 
+ \nabla^2 \delta p_{\mathrm{t}} + (p_{\mathrm{t}} + \rho_{\mathrm{t}}) \nabla^2 \phi + \frac{4}{3} \eta \nabla^2 \theta_{\mathrm{t}}=0,
\label{TOTCONS2}
\end{equation}
where $\eta$ denotes the shear viscosity coefficient which is particularly    
relevant for the baryon-photon system and which is related to the photon mean 
free path (see below in this Appendix). 
\subsection{Different fluids of the mixture}
The relevant equations will now be written directly 
in Fourier space omitting the explicit reference 
to the wave-number since the subscript may be confused 
with the other subscripts labeling each species of the fluid.
The evolution of CDM particles is rather simple since 
it is only sensitive to the fluctuation of the metric:
\begin{eqnarray}
&& \delta_{\mathrm{c}}' = 3 \psi' - \theta_{\mathrm{c}},
\label{CDM1}\\
&& \theta_{\mathrm{c}}' + {\mathcal H} \theta_{\mathrm{c}} = k^2 \phi.
\label{CDM2}
\end{eqnarray}
Defining the gauge-invariant density contrast of CDM, i.e.
$\zeta_{\mathrm{c}} = - \psi + \delta_{\mathrm{c}}/3$, it is 
immediate to combine Eqs. (\ref{CDM1}) and (\ref{CDM2})
and obtain:
\begin{equation}
\zeta_{\mathrm{c}}'' + {\mathcal H} \zeta_{\mathrm{c}} = - 
\frac{k^2}{3} \phi,
\label{CDM3}
\end{equation}
whose solution is 
\begin{equation}
\zeta_{\mathrm{c}} (\tau,k) = \zeta(\tau_{\mathrm{i}},k) - 
\frac{k^2}{3} \int_{\tau_{\mathrm{i}}}^{\tau} \frac{d\tau''}{a(\tau'')} \int_{\tau_{\mathrm{i}}}^{\tau''} 
\phi(k,\tau') a(\tau') d\tau',
\end{equation}
where $\tau_{\mathrm{i}}$ is the initial integration time.
In terms of the CDM density contrast and in the limit of 
vanishing anisotropic stress 
\begin{equation}
\delta_{\mathrm{c}}(\tau,k) = \delta_{\mathrm{c}}(\tau_{\mathrm{i}}, k) + 3 ( \phi(\tau,k) - \phi(\tau_{\mathrm{i}},k)) - 
k^2\int_{\tau_{\mathrm{i}}}^{\tau} \frac{d\tau''}{a(\tau'')} \int_{\tau_{\mathrm{i}}}^{\tau''} \phi(k,\tau') a(\tau') d\tau'
\end{equation}

The lowest multipoles of the neutrino hierarchy lead to the 
following set of equations where the contribution 
of the quadrupole (i.e. $2 \sigma_{\nu}$) and octupole (i.e. 
${\mathcal F}_{\nu3}$) have been explicitly included:
\begin{eqnarray}
&& \delta_{\nu}' = 4 \psi' - \frac{4}{3} \theta_{\nu},
\label{nueq1}\\
&& \theta_{\nu}' = \frac{k^2}{4} \delta_{\nu} - k^2 \sigma_{\nu} + k^2 \phi,
\label{nueq2}\\
&& \sigma_{\nu}' = \frac{4}{15} \theta_{\nu} - \frac{3}{10} {\mathcal F}_{\nu3}.
\label{nueq3}
\end{eqnarray}
Unlike neutrinos and CDM particles that feel the effect of the magnetic fields through 
the anisotropic stress and through the metric fluctuations, the photons, being tightly 
coupled with the baryons by Thompson scattering, are directly affected by the presence 
of large-scale magnetic fields.  Let us clarify this point by writing, separately, the relevant evolution 
equations for the photons and their counterpart for the baryons.
In real space the evolution equations of the photons can be written as:
\begin{eqnarray}
&& \delta_{\gamma}' = 4\psi' - \frac{4}{3} \theta_{\gamma},
\label{PH1}\\
&& \theta_{\gamma}' = -\frac{1}{4} \nabla^2\delta_{\gamma} - \nabla^2 \phi + \epsilon' (\theta_{\mathrm{b}} - \theta_{\gamma}),
\label{PH2}
\end{eqnarray}
where $\epsilon'$ is the inverse of the photon mean free path.
The baryon evolution equations are:
\begin{eqnarray}
&&  \delta_{\mathrm{b}}' = 3 \psi' - \theta_{\mathrm{b}},
\label{B1}\\
&& \theta_{\mathrm{b}}' + {\mathcal H} \theta_{\mathrm{b}} = -\nabla^2 \phi + \frac{4}{3} \frac{\rho_{\gamma}}{\rho_{\mathrm{b}}}
\epsilon' (\theta_{\gamma}- \theta_{\mathrm{b}} ) + \frac{\vec{\nabla}\cdot[\vec{J}\times \vec{B}]}{a^4 \rho_{\mathrm{b}}},
\label{B2}
\end{eqnarray}
where $\vec{J}\times \vec{B}$ is the MHD Lorentz force. 
Equations (\ref{PH2}) and (\ref{B2}) can now be summed and subtracted
after having multiplied Eq. (\ref{B2}) by $R_{\mathrm{b}}$.  By
subtracting the two aforementioned equations, we will obtain
an equation for $(\theta_{\gamma} -\theta_{\mathrm{b}})$ whose solution will imply a strong 
damping leading, in spite of the initial conditions, to $\theta_{\gamma} \simeq \theta_{\mathrm{b}} = \theta_{\gamma\mathrm{b}}$. From the sum of Eqs. (\ref{PH2}) and (\ref{B2}) the evolution equation 
of $\theta_{\gamma\mathrm{b}}$ will then be directly obtained.
In the tight coupling limit the evolution equations of the magnetized baryon-photon system is, therefore \cite{mg1}:
\begin{eqnarray}
&& \delta_{\gamma}' = 4 \psi' - \frac{4}{3} \theta_{\gamma\mathrm{b}},
\label{TC1}\\
&& \theta_{\gamma\mathrm{b}}' + \frac{{\mathcal H} R_{\mathrm{b}}}{1 + R_{\mathrm{b}}} \theta_{\gamma\mathrm{b}} + \frac{\eta}{\rho_{\gamma} (R_{\mathrm{b}} + 1)} k^2 \theta_{\gamma\mathrm{b}}= 
\frac{k^2}{4 ( 1 + R_{\mathrm{b}})} \delta_{\gamma} + k^2 \phi + \frac{k^2 (\Omega_{\mathrm{B}} - 4 \sigma_{\mathrm{B}})}{4 ( 1 + R_{\mathrm{b}})},
\label{TC2}\\
&& \delta_{\mathrm{b}}' = 3 \psi' - \theta_{\gamma\mathrm{b}},
\label{TC3}
\end{eqnarray}
where 
\begin{equation}
\eta = \frac{4}{15} \rho_{\gamma} \lambda_{\mathrm{T}},
\end{equation}
is the shear viscosity coefficient that leads to the Silk damping 
of the high harmonics in the CMB temperature autocorrelations. 
The important identity (heavily used in this algebra) is 
\begin{equation}
\frac{3}{4 a^4 \rho_{\gamma}} \vec{\nabla} \cdot[ \vec{J} \times \vec{B}] = \nabla^2 \sigma_{\mathrm{B}} - \frac{1}{4} \nabla^2 \Omega_{\mathrm{B}}.
\label{identity}
\end{equation}
This identity can be swiftly derived by recalling 
Eq. (\ref{anisstrein}) and by using two further vector identities:
\begin{equation}
\vec{\nabla}\cdot[\vec{J} \times \vec{B}] = \frac{1}{4\pi} \vec{\nabla}\cdot[(\vec{\nabla} \times \vec{B})\times \vec{B}] ,\qquad 
\partial_{i}B_{j}\partial^{j} B^{i} = \vec{\nabla}\cdot[(\vec{\nabla} \times \vec{B})\times \vec{B}] + \frac{1}{2} \nabla^2 B^2.
\label{identity2}
\end{equation}
The relation between the magnetic anisotropic stress and 
$\sigma_{\mathrm{B}}$ has been introduced before (i.e. 
after Eq. (\ref{anisstrein})) and it is simply $ \partial_{j}\partial^{i} 
\tilde{\Pi}_{i}^{j} = \nabla^2 \sigma_{\mathrm{B}}$.
Defining the gauge-invariant density contrast for the 
photons, i.e. $\zeta_{\gamma} = - \psi + \delta_{\gamma}/4$ 
and combining Eqs. (\ref{TC1}), (\ref{TC2}) and (\ref{TC3}) 
the following simple equation can be readily obtained:
\begin{equation}
\zeta_{\gamma}''  + \frac{{\mathcal H} R_{\mathrm{b}}}{R_{\mathrm{b}} +1} \zeta_{\gamma}' + 
\frac{4}{15} k^2 \frac{\lambda_{\mathrm{T}}}{R_{\mathrm{b}} +1} \zeta_{\gamma}' + \frac{k^2}{3 (R_{\mathrm{b}}+1)} \zeta_{\gamma} = - \frac{k^2}{3}\biggl[ \phi + \frac{\psi}{R_{\mathrm{b}} +1}\biggr] + \frac{k^2}{12(R_{\mathrm{b}} +1)} ( 4 \sigma_{\mathrm{B}} - \Omega_{\mathrm{B}}).
\label{TC4}
\end{equation}
By now defining the photon-baryon sound speed $c_{\mathrm{sb}}$ we have 
\begin{equation}
c_{\mathrm{sb}} = \frac{1}{\sqrt{3(R_{\mathrm{b}} +1)}}, \qquad 
\frac{(c_{\mathrm{sb}}^2)'}{c_{\mathrm{sb}}^2} = - 
\frac{{\mathcal H} R_{\mathrm{b}}}{R_{\mathrm{b}}+1},
\end{equation}
which also implies, when inserted into Eq. (\ref{TC4}), that 
\begin{equation}
\zeta_{\gamma}'' - \frac{(c_{\mathrm{sb}}^2)'}{c_{\mathrm{sb}}^2} \zeta_{\gamma}' + \frac{4}{5} k^2 c_{\mathrm{sb}}^2 \lambda_{\mathrm{T}} \zeta_{\gamma}' + k^2 c_{\mathrm{sb}}^2 = - \frac{k^2}{3} ( \phi + 3 c_{\mathrm{sb}}^2 \psi)+ 
\frac{k^2}{4} c_{\mathrm{sb}}^2( 4 \sigma_{\mathrm{B}} - 
\Omega_{\mathrm{B}}).
\label{TC5}
\end{equation}
By now changing variable from the conformal time 
coordinate $\tau$ to $ dq = c_{\mathrm{sb}}^2 d\tau$, 
Eq. (\ref{TC5}) becomes:
\begin{equation}
\frac{d^2 \zeta_{\gamma}}{d q^2} + \frac{4}{5} k^2 \lambda_{\mathrm{T}} \frac{d \zeta_{\gamma}}{d q} + \frac{k^2}{c_{\mathrm{sb}}^2} \zeta_{\gamma} = - \frac{k^2}{3 c_{\mathrm{sb}}^4} ( \phi + 3 c_{\mathrm{sb}}^2 \psi) + 
\frac{k^2}{ 4 c_{\mathrm{sb}}^2} ( 4 \sigma_{\mathrm{B}} - \Omega_{\mathrm{B}}).
\label{TC6}
\end{equation}
The solution of the homogeneous equation can be simply obtained in the WKB approximation 
\cite{pavel1,hu1,hu2,muk} and it is 
\begin{equation}
\zeta_{\gamma}(\tau,k) = \frac{1}{\sqrt{3(R_{\mathrm{sb}} +1)}} \{ C_1 \cos{[\alpha(\tau,k)]} + C_2 \sin{[\alpha(\tau,k)]}\} e^{- \frac{k^2}{k^2_{\mathrm{D}}}}, 
\label{TC7}
\end{equation}
where the quantities appearing in Eq. (\ref{TC7}) are:
\begin{eqnarray}
\frac{1}{k^2_{\mathrm{D}}(\tau)} &=& \frac{2}{5} \int_{0}^{\tau} c_{\mathrm{sb}}(\tau') \frac{ a_{0} d\tau'}{a(\tau')\,\,x_{\mathrm{e}} n_{\mathrm{e}} \sigma_{\mathrm{T}}}, 
\label{TC7a}\\
\alpha(\tau,k) &=& k\int^{\tau} c_{\mathrm{sb}}(\tau') d\tau'.
\label{TC7b}
\end{eqnarray} 
Since $\delta_{\gamma} = 4 (\zeta_{\gamma} + \psi)$, 
Eqs. (\ref{TC6}), (\ref{TC7}) can also be used 
to determine the photon density contrast which is 
a key ingredient for the estimate of the temperature 
autocorrelations both at large and small angular scales.

\subsection{Evolution of the brightness perturbations}
The evolution equations of the  brightness 
perturbations can be easily derived within
the set of conventions employed in the 
present paper.  Recalling that $\theta_{\mathrm{b}}$ is the divergence of the peculiar 
velocity field of the baryons, it is convenient to define, for notational convenience, $v_{\mathrm{b}}$ 
i.e.
\begin{equation}
v_{\mathrm{b}} = \frac{\theta_{\mathrm{b}}}{ik} \simeq \frac{\theta_{\gamma\mathrm{b}}}{i k}, 
\end{equation}
where the second equality holds in the tight coupling approximation. 
From Eq. (\ref{B2}) it then follows that the evolution of $v_{\mathrm{b}}$ is simply given by
\begin{equation}
 v_{\rm b}' + {\cal H} v_{\rm b} + i k\phi  + \frac{\epsilon'}{R_{\mathrm{b}}}
 \biggl( 3 i \Delta_{{\rm I}1} + v_{b} \biggr) = \frac{i k}{4 R_{\mathrm{b}}} ( \Omega_{\mathrm{B}} - 4 \sigma_{\mathrm{B}}),
\label{vb}
\end{equation}
having used that $\theta_{\gamma} = 3 k \Delta_{\mathrm{I}1}$ which simply reflects the occurrence 
that the dipole of the intensity of the perturbed radiation field is related to the peculiar velocity 
of the photons once the fluctuation of the intensity is expanded in multipoles as 
\begin{equation}
\Delta_{\mathrm{I}}( \vec{k}, \hat{n},\tau) = \sum_{\ell} (-i)^{\ell} ( 2 \ell +1) \Delta_{\mathrm{I}\ell}(k, \tau)
P_{\ell}(\mu),
\label{expansion}
\end{equation}
where $ \hat{n}$ is the direction of the momentum of the photon, $\mu = \hat{n}\cdot\hat{k}$ and 
$P_{\ell}(\mu)$ are the Legendre polynomials of index $\ell$ and argument $\mu$.
Analog expansion hold for the brightness perturbations related with the other two Stokes 
parameters, i.e. $Q$ and $U$. It should be remarked, incidentally, that the $\ell$-dependent factors 
appearing in Eq. (\ref{expansion}) are conventional. If different conventions in the 
expansion are adopted, the various expressions of the multipoles will change accordingly.
With these necessary specifications we have that
\begin{eqnarray}
&& \Delta_{\rm I}' + (i k\mu + \epsilon')  \Delta_{\rm I}  = \psi' - i k\mu \phi +
\epsilon' \biggl[ \Delta_{{\rm I}0 } +  \mu v_{b} - 
\frac{1}{2} P_{2}(\mu) S_{\rm Q}\biggr],
\label{BRI}\\
&& \Delta_{\rm Q}' + (i k\mu + \epsilon') \Delta_{\rm Q}  =   
\frac{\epsilon'}{2} [1- P_{2}(\mu)] S_{\rm Q}\biggr\},
\label{BRQ}\\
&& \Delta_{\rm U}' + (i k\mu + \epsilon' )\Delta_{\rm U}  = 0 ,
\label{BRU}
\end{eqnarray}
where we defined, for notational convenience 
\begin{equation}
S_{\rm Q}=  \Delta_{{\rm I}2} + \Delta_{{\rm Q}0} + \Delta_{{\rm Q}2}.
\end{equation}
In Eqs. (\ref{BRQ})--(\ref{BRU}), $P_{2}(\mu) = (3 \mu^2 -1)/2$ is the Legendre 
polynomial of second order, which appears in the collision operator of the Boltzmann 
equation for the photons 
due to the directional nature of Thompson scattering. 
Since we shall be chiefly concerned with the temperature 
autocorrelations, let us remind that, using the technique 
of integration along the line of sight, Eq. (\ref{BRI}) 
can be solved, after integration by parts, as
\begin{equation}
\Delta_{\mathrm{I}}(\vec{k},\tau_{0}) = \int_{0}^{\tau_{0}} 
e^{-\epsilon(\tau,\tau_{0})} (\psi' + \phi') + 
\int_{0}^{\tau_{0}} {\mathcal K}(\tau) \biggl[ \Delta_{\mathrm{I}0} + \phi + \mu v_{\mathrm{b}} - \frac{1}{2} P_{2}(\mu) S_{\mathrm{Q}}\biggr] e^{- i \mu k (\tau_{0} - \tau)},
\label{BRI2}
\end{equation}
where ${\mathcal K} = \epsilon' e^{-\epsilon}$ is the visibility function. In the sudden recombination 
approximation, relevant for wavelengths larger 
that the Hubble radius we will have that, according 
to Eq. (\ref{BRI2}) the Sachs-Wolfe contribution 
and the integrated Sachs-Wolfe contribution 
are simply given by 
\begin{equation}
\Delta^{(\mathrm{SW})}_{\mathrm{I}}(\vec{k},\tau_{0}) = 
\Delta_{\mathrm{I}0}(\vec{k},\tau_{\mathrm{rec}}) + 
\phi(\vec{k},\tau_{\mathrm{rec}}),\qquad \Delta^{(\mathrm{SW})}_{\mathrm{I}}(\vec{k},\tau_{0}) = \int_{0}^{\tau_{0}} (\phi' + \psi').
\label{SWISW}
\end{equation}
At smaller angular scales the finite thickness 
of the last scattering surface cannot be neglected anymore 
and the general expression for the $C_{\ell}$ coefficients becomes then:
\begin{equation}
C_{\ell} = \frac{2}{\pi} \int k^3 d\ln{k} |\Delta_{\mathrm{I}\ell}(k,\tau_{0})|^2,
\label{CL1}
\end{equation}
where 
\begin{equation}
\Delta_{\mathrm{I}\ell}(k,\tau_{0}) = {\mathcal K}(\tau_{\mathrm{rec}}) 
\biggl[ (\Delta_{\mathrm{I}0} + \phi)_{\tau_{\mathrm{rec}}} 
j_{\ell}(x_0) + i v_{\mathrm{b}} \frac{d j_{\ell}(x_{0})}{d x_{0}}\biggr].
\label{CL2}
\end{equation}
In Eq. (\ref{CL2}) we the visibility function is approximated by a Gaussian in a way similar to what 
has been done in \cite{pavel1,pavel2,seljak,muk}:
\begin{equation}
{\mathcal K}(\tau_{\mathrm{rec}}) = e^{- \sigma^2 \tau_{\mathrm{rec}}^2 k^2}, \qquad \sigma= \frac{1}{\alpha_{1}
{\mathcal H}_{\mathrm{rec}}\tau_{\mathrm{rec}}} = 0.0148 \biggl( \frac{\sqrt{ z_{\mathrm{rec}} + z_{\mathrm{eq}}}   + \sqrt{z_{\mathrm{rec}}} }{\sqrt{ z_{\mathrm{rec}} + z_{\mathrm{eq}}} }\biggr).
\end{equation}
where $\alpha_1= 33.59$. 
\end{appendix}

\end{document}